\documentclass[12pt,a4paper,english,twoside]{article}
\pdfoutput=1

\usepackage{booktabs}
\usepackage{mathtools}
\usepackage[english]{babel}
\usepackage{amsmath,amssymb,amsbsy,amstext, amsthm}
\usepackage{array}
\usepackage{url}
\usepackage{slashed}
\usepackage{multicol}
\usepackage{blkarray}
\usepackage{enumerate}
\usepackage{graphicx}
\usepackage{amsfonts}
\usepackage{amssymb}
\usepackage{empheq}
\usepackage{colortbl}
\usepackage{float}
\usepackage[utf8]{inputenc}
\usepackage{subcaption}
\usepackage{hhline}
\usepackage[pdftex,hyperref,x11names]{xcolor}
\usepackage[pdftex,colorlinks=true,
pdfstartview=FitV,
pdfnewwindow=true,
linktoc = page,
linkcolor= RoyalBlue3,
citecolor= Red3,
urlcolor= RoyalBlue3,
hyperindex=true,
hyperfigures=false]{hyperref}
\usepackage{setspace}
\usepackage{a4wide}
\usepackage{makecell}
\usepackage{cancel}
\usepackage[normalem]{ulem}

\usepackage[all]{xy}

\usepackage{cite}

\allowdisplaybreaks[1]

\makeatletter
\newlength{\apb@width}
\makeatother

\numberwithin{equation}{section}

\def\be{\begin{equation}}
\def\ee{\end{equation}}
\def\bea{\begin{eqnarray}}
\def\eea{\end{eqnarray}}
\def\lp{\left(}
\def\rp{\right)}

\def\d{{\rm d}}
\def\nn{\nonumber}
\def\del{\partial}

\newcommand*\widefbox[1]{\fbox{\hspace{2em}#1\hspace{2em}}}

\newcommand{\eq}[1]{\begin{equation}#1\end{equation}}
\newcommand{\spl}[1]{\begin{split}#1\end{split}}

\newcommand{\Pkin}{P_{\rm kin}^{\pm}}
\newcommand{\Pkinp}{P_{\rm kin}^{+}}
\newcommand{\Pkinm}{P_{\rm kin}^{-}}
\newcommand{\Pk}{P_k}
\newcommand{\Pkp}{P_{k\,\phi}}
\newcommand{\Pp}{P_{\phi}}
\newcommand{\Pmp}{P_{m\,\phi}}
\newcommand{\Pm}{P_m}
\newcommand{\Prp}{P_{r\,\phi}}
\newcommand{\Prr}{P_r}

\begin{document}
\numberwithin{equation}{section}

\begin{titlepage}
\begin{center}

\phantom{DRAFT}

\vspace{-0.1in}

{\LARGE \bf{Exponential Quintessence: \\ \vspace{0.1in} curved, steep and stringy?}}\\

\vspace{2.0 cm} {\Large David Andriot$^{1}$, Susha Parameswaran$^{2}$, Dimitrios Tsimpis$^{3}$,}\\
\vspace{0.3 cm} {\Large Timm Wrase$^{4}$, Ivonne Zavala$^{5}$}\\

\vspace{0.9cm} {\upshape\ttfamily andriot@lapth.cnrs.fr; susha.parameswaran@liverpool.ac.uk; tsimpis@ipnl.in2p3.fr; timm.wrase@lehigh.edu; e.i.zavalacarrasco@swansea.ac.uk}\\

\vspace{2.2cm}

{\bf Abstract}
\vspace{0.1cm}
\end{center}
\begin{quotation}
We explore the possibility that our universe's current accelerated expansion is explained by a quintessence model with an exponential scalar potential, $V =V_0\, e^{-\lambda\, \phi}$, keeping an eye towards $\lambda \geq \sqrt{2}$ and an open universe, favorable to a string theory realisation and with no cosmological horizon. We work out the full cosmology of the model, including matter, radiation, and optionally negative spatial curvature, for all $\lambda>0$, performing an extensive analysis of the dynamical system and its phase space. The minimal physical requirements of a past epoch of radiation domination and an accelerated expansion today lead to an upper bound $\lambda \lesssim \sqrt{3}$, which is driven slightly up in the presence of observationally allowed spatial curvature. Cosmological solutions start universally in a kination epoch, go through radiation and matter dominated phases and enter an epoch of acceleration, which is only transient for $\lambda>\sqrt{2}$. Field distances traversed between BBN and today are sub-Planckian. We discuss possible string theory origins and phenomenological challenges, such as time variation of fundamental constants. We provide theoretical predictions for the model parameters to be fitted to data, most notably the varying dark energy equation of state parameter, in light of recent results from DES-Y5 and DESI.
\end{quotation}

\newpage

\vspace{0.9 cm} {\small\slshape $^1$ Laboratoire d'Annecy-le-Vieux de Physique Th\'eorique (LAPTh),\\
CNRS, Universit\'e Savoie Mont Blanc (USMB), UMR 5108,\\
9 Chemin de Bellevue, 74940 Annecy, France}\\

\vspace{0.2 cm} {\small\slshape $^2$ Department of Mathematical Sciences, University of Liverpool,\\
Liverpool, L69 7ZL, United Kingdom}\\

\vspace{0.2 cm} {\small\slshape $^3$ Institut de Physique des Deux Infinis de Lyon (iP2i),\\
Universit\'{e} de Lyon, UCBL, CNRS/IN2P3, UMR 5822,\\
4 rue Enrico Fermi, 69622 Villeurbanne Cedex, France}\\

\vspace{0.2 cm} {\small\slshape $^4$ Department of Physics, Lehigh University,\\
16 Memorial Drive East, Bethlehem, PA 18018, USA}\\

\vspace{0.2 cm} {\small\slshape $^5$ Physics Department, Swansea University,\\ Swansea, SA2 8PP, United Kingdom}\\

\end{titlepage}

\tableofcontents

\section{Introduction and results summary}\label{sec:intro}

\begin{figure}[H]
\begin{center}
\begin{subfigure}[H]{0.48\textwidth}
\includegraphics[width=\textwidth]{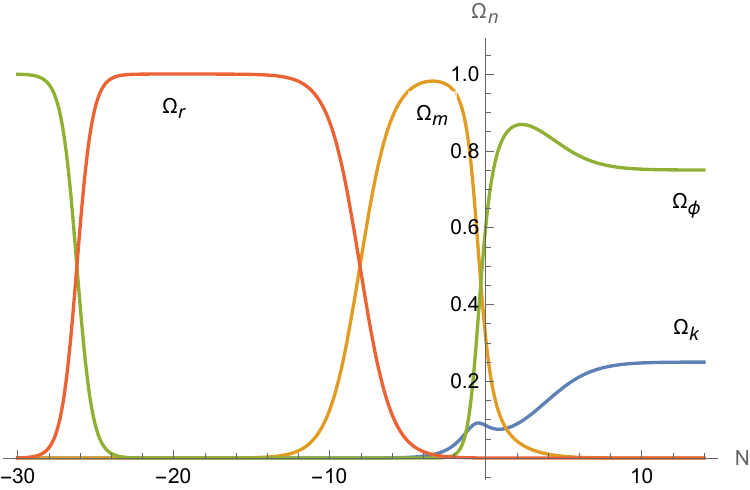}\caption{$\Omega_n$}\label{fig:SolphiOintro}
\end{subfigure}\quad
\begin{subfigure}[H]{0.48\textwidth}
\includegraphics[width=\textwidth]{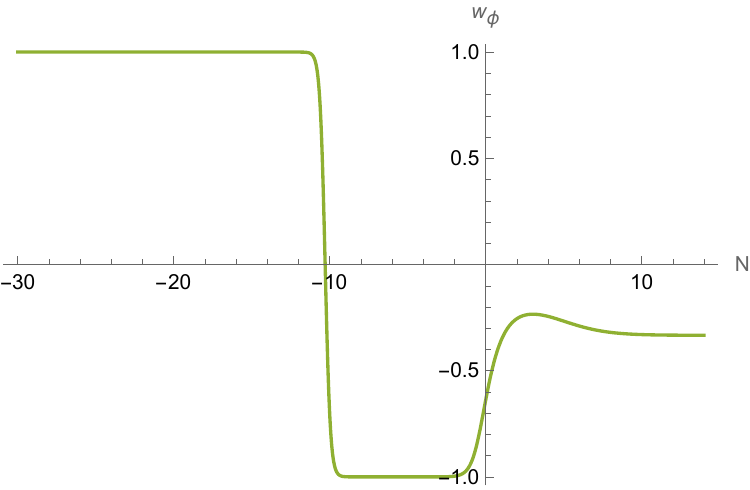}\caption{$w_{\phi}$}\label{fig:Solphiwphiintro}
\end{subfigure}
\caption{Evolution of the density parameters, $\Omega_n$, and of the dark energy equation of state parameter, $w_{\phi}$, in terms of the number of e-folds $N$, for a possibly realistic cosmological solution in the exponential quintessence model. Here $\lambda=\sqrt{\frac{8}{3}}$, with spatial curvature (open universe), the amount of which is taken to be near the limits of expected observational bounds. The universe today corresponds to $N=0$. We refer to figure \ref{fig:Solphi} for more details.}\label{fig:intro}
\end{center}
\end{figure}

The universe, as observed today, is not only expanding but this expansion is currently accelerating. The energy responsible for this acceleration is named dark energy; understanding its nature or origin is among the most fundamental questions in contemporary physics. A well-established, simple cosmological model, $\Lambda$CDM, postulates that dark energy is due a (positive) cosmological constant $\Lambda$; this implies that the equation of state parameter for dark energy is constant, with $w_{{\rm DE}}=-1$. While this model is generally in good agreement with observations, different options have been put forward, most notably dynamical dark energy in the form of quintessence models \cite{Ratra:1987rm,Peebles:1987ek,Caldwell:1997ii}.
In its simplest realisation, quintessence considers
 a 4-dimensional (4d) theory of a single scalar  field $\phi$, minimally coupled to gravity, together with a scalar potential $V(\phi)$
\be
{\cal S} = \int \d^4 x \sqrt{|g_4|} \left( \frac{M_p^2}{2} {\cal R}_4 - \frac{1}{2} (\del \phi)^2 - V(\phi) \right) \ , \label{action}
\ee
where $M_p$ stands for the reduced 4d Planck mass, and the scalar field is canonically normalized. In such quintessence models, a homogeneous field $\phi(t)$ can be rolling down (or up) the potential. In that case, dark energy receives contributions from both the potential and the kinetic energy, $\frac{1}{2} \dot{\phi}^2$, where $\dot{\phi}\equiv \del_t \phi$. The equation of state parameter becomes
\be
w_{{\rm DE}} = w_{\phi} =\frac{\frac{1}{2} \dot{\phi}^2 - V(\phi) }{\frac{1}{2} \dot{\phi}^2 + V(\phi)} \ ,
\ee
and, importantly, it can vary over time (see figure \ref{fig:Solphiwphiintro} for an illustration).

Beyond various tensions between observational data that seem to emerge when fitting to $\Lambda$CDM (see \cite{Abdalla:2022yfr} for a review), which may call for some extensions of the cosmological model, the motivation for considering such quintessence models is twofold. Firstly, recent observations have exhibited possible agreements with a varying dark energy equation of state parameter, e.g.~\cite{DES:2024tys, DESI:2024mwx}, and upcoming missions such as Euclid will provide even better constraints on this parameter. It is then important to understand possible theory expectations for the evolution of dark energy's equation of state. Secondly, in recent years, significant effort has been made to construct solutions in string theory with a positive cosmological constant (de Sitter solutions) that are well-controlled. The latter refers to the trustability of the various approximations made, or equivalently, to the possible corrections that could arise and spoil the solution. Such constructions turn out to be very difficult, and to date, it is fair to say that no de Sitter solution from string theory can be claimed to be unequivocally under control, despite many attempts such as the recent ones \cite{Leedom:2022zdm, ValeixoBento:2023nbv, Andriot:2024cct}.\footnote{See also \cite{Cicoli:2023opf} for a recent review on string cosmology.} As a consequence, some recent string theory activity has been devoted to constructing or constraining quintessence models instead: see for instance \cite{Olguin-Trejo:2018zun, Cicoli:2020cfj, ValeixoBento:2020ujr, Cicoli:2020noz, Cicoli:2021skd, Russo:2022pgo, Brinkmann:2022oxy, Rudelius:2022gbz, Andriot:2022xjh, Calderon-Infante:2022nxb, Marconnet:2022fmx, Danielsson:2022lsl, Shiu:2023nph, Shiu:2023fhb, Cremonini:2023suw, Hebecker:2023qke, Freigang:2023ogu, Andriot:2023wvg, VanRiet:2023cca, Revello:2023hro, Seo:2024fki, Gallego:2024gay, Seo:2024qzf, Arapoglu:2024umz}. The possibility that string theory may prefer dynamical dark energy and quintessence models over a cosmological constant is among the motivations for this work.

A related line of investigation has aimed at characterising genuine scalar potentials obtained in effective theories of the type \eqref{action} from string theory. The question there is whether positive scalar potentials can admit a critical point (a de Sitter solution) or, if not, what is their slope? Within the swampland program, this has led to the trans-Planckian censorship conjecture (TCC) and the strong de Sitter conjecture \cite{Bedroya:2019snp, Rudelius:2021oaz}, which characterize positive scalar potentials from string theory in the asymptotics of field space ($\phi \rightarrow \infty$). The claim is that those should obey (in Planckian units $M_p=1$), for $V>0$,
\be
\frac{\nabla V}{V} \geq \sqrt{2} \ ,
\ee
in such asymptotics. Here, $(\nabla V)^2 = g^{ij} \del_{\phi^i} V \del_{\phi^j} V $ in a general multifield case, with field space metric $g_{ij}$. Importantly, to this day there is no known counter-example to this claim.

When considering a field space asymptotic $\phi \rightarrow \infty$, scalar potentials from string theory often boil down to an exponential for the canonically normalised field; this is e.g.~the case of the 4d fields arising from diagonal metric components of the extra dimensions (radions or K\"ahler moduli). This motivates us to focus on an exponential quintessence model
\be
V(\phi) = V_0 \ e^{-\lambda \, \phi} \ , \label{exppotintro}
\ee
where $V_0, \lambda >0$. The above strong de Sitter conjecture then becomes a condition on the exponential rate: $\lambda \geq \sqrt{2}$. Note that we leave aside the multifield case; it would be interesting to see how the results of this paper would change there. For now, the question we would like to address is whether such an exponential quintessence model can provide a realistic description of  dark energy in our universe, especially with $\lambda \geq \sqrt{2}$.\\

Answers to this question have been proposed using a dynamical system approach, at first in the case of a spatially flat universe ($\Omega_k=0$) \cite{Copeland:1997et, Ferreira:1997hj, Bahamonde:2017ize, SavasArapoglu:2017pyh, Rudelius:2022gbz, Andriot:2022xjh, Shiu:2023nph, Shiu:2023fhb} (see in particular \cite{Bahamonde:2017ize} for a review of dynamical systems applied to cosmology). Since the above statements are made in the asymptotics (of field space but also of time), it is natural to consider the asymptotics of cosmological solutions, in other words the attractor fixed point of the system. The result of such an analysis is that the stable fixed point only allows for an accelerating universe ($\ddot{a}>0$) when $\lambda < \sqrt{2}$. Therefore, the simplest string theory realisations of the exponential potential in a flat universe, which have $\lambda>\sqrt{2}$, do not allow for asymptotic acceleration.  Moreover, even if transient acceleration is possible when $\lambda>\sqrt{2}$, observational constraints on $\lambda$ in a flat universe tend to prefer values $\lambda<\sqrt{2}$ \cite{Agrawal:2018own, Akrami:2018ylq, Raveri:2018ddi}, outside the favoured regime of string theory.  In this paper, we study how this changes when including spatial curvature.

In particular, we are motivated in the present paper to focus on the case of an open universe, which has negatively curved spatial slices and $\Omega_k>0$.  In the context of string theory models, negative spatial curvature has been argued to be a natural outcome of Coleman and de Luccia tunneling in the string landscape \cite{Freivogel:2005vv}; even if the most recent investigations also allow for other possibilities  (see for example \cite{Cespedes:2020xpn, Cespedes:2023jdk}) negative spatial curvature can certainly arise or be imposed as an extra condition on string compactifications. Moreover, it is known that in a system with pure quintessence and curvature $\Omega_k>0$, a new attractor fixed point appears for $\lambda > \sqrt{2}$ \cite{Halliwell:1986ja, Chen:2003dca, Andersson:2006du, Marconnet:2022fmx, Andriot:2023wvg}, whose corresponding solution is on the threshold of acceleration ($\ddot{a}=0$).  Intriguingly, cosmological solutions can approach this attractor while accelerating; such solutions then exhibit an ``asymptotic acceleration'', in a framework favorable to a string theory realisation.  More intriguingly, as emphasized in \cite{Andriot:2023wvg} (see also \cite{Boya:2002mv}), these asymptotically accelerating solutions do not admit a cosmological event horizon. This could be a hint of a {\sl no cosmological horizon conjecture} \cite{Andriot:2023wvg} (see also \cite{Townsend:2003qv}), stating that cosmological solutions obtained from a quantum gravity theory do not allow for such horizons, in line with a possible absence of stable de Sitter solutions and power-law accelerating solutions.  A natural question is then to what extent these asymptotically accelerating solutions can be realistic when comparing to observations of our own universe. At the same time, it is also important to note the existence of cosmological solutions with a transient phase of acceleration, which could also be candidates for a realistic quintessence.

Asking for a realistic exponential quintessence model requires the inclusion of matter and radiation to properly describe the past and present of our universe; early related works include \cite{Ratra:1987rm, Wetterich:1994bg, Copeland:1997et, Cline:2001nq, Kolda:2001ex}. The starting point of this work is therefore to provide a complete dynamical system analysis of exponential quintessence including radiation, matter and curvature (each of these ingredients being possibly turned off). A dynamical system analysis with a single barotropic fluid with general equation of state, together with curvature, was studied in \cite{vandenHoogen:1999qq}, while in \cite{Gosenca:2015qha} quintessence, curvature and a combined fluid of matter and radiation was analysed. Furthermore, \cite{Marconnet:2022fmx, Andriot:2023wvg} considered quintessence plus curvature, \cite{SavasArapoglu:2017pyh} included quintessence plus matter and radiation, and \cite{Bahamonde:2017ize} presented various other relevant subsystems. We perform our analysis of the complete system in 4d in section \ref{sec:dynsys}, appendix \ref{ap:analytic} and \ref{ap:asymptotics}, and in arbitrary spacetime dimensions in appendix \ref{ap:dynsys}. Beyond the fixed points and their stability, this analysis provides a first overview of all possible cosmological solutions. We obtain in particular analytic solutions at, and close to, the fixed points, as well as on specific subspaces; the rest of the solutions are obtained numerically. An outcome is that all cosmological solutions (except those at a fixed point) start around the same unstable fixed points $\Pkin$, which correspond to an initial kination phase, with the energy budget dominated by the kinetic energy of the scalar field. The solutions end at the unique stable fixed point, which -- depending on the value of $\lambda$ and the presence or absence of spatial curvature -- may be scalar field dominated, $\Pp$, scaling with matter, $\Pmp$, or scaling with curvature, $\Pkp$. Along the way, the solutions may pass close to some of the saddle fixed points. Of interest for a realistic solution are the radiation dominated fixed point, $\Prr$, and the matter dominated one, $\Pm$. To what extent the solutions pass close to them in the past will be a relevant question.

Among all possible solutions, we identify those that stand a chance of describing our universe. Remarkably, we are able to achieve a quite advanced understanding of the parameter space using a combination of analytical and empirical study of the phase space. We first do this analysis without spatial curvature in section \ref{sec:solutions}. As we discuss in subsection \ref{sec:radmatdom}, the minimal cosmological requirement of a past epoch of radiation domination inevitably implies that the universe subsequently passes through an epoch of matter domination of amplitude and duration similar to that of $\Lambda$CDM. In addition, we ask -- for each $\lambda$ value -- that the solutions considered pass by a point corresponding to our universe today, defined in terms of values for today's density parameters, $\Omega_{n0}$ ($n=m,r,k,\phi$), and today's equation of state parameter, $w_{\phi0}$. For a given $\Omega_{n0}$, we determine numerically in \eqref{wphi0values} a correspondence between $\lambda$ and $w_{\phi0}$: the smaller $\lambda$ is, the closer to $-1$ is $w_{\phi0}$. Requiring furthermore that the universe is accelerating today leads finally to an upper bound on $\lambda$; we show this in subsection \ref{sec:upperbound}, where we find \eqref{lambdaboundacc}:
\be
\lambda \lesssim \sqrt{3} \ .\label{upperboundintro}
\ee
Equation \eqref{upperboundintro} is an important constraint for string theory model building. At the same time, observational constraints impose stronger bounds than this theoretical one.  Those have been worked out for the flat case in \cite{Agrawal:2018own, Akrami:2018ylq, Raveri:2018ddi, Schoneberg:2023lun}, and we review these results in subsection \ref{sec:obs}, together with our theoretical ones.  Moreover, in subsection \ref{sec:w0wa} we provide theoretical predictions for the so-called CPL or $w_0w_a$ parametrisation \cite{Chevallier:2000qy, Linder:2002et}, which is often used as a fiducial model to fit to observational data. Having established candidate realistic solutions, we summarise the characteristics (start and duration) of their acceleration phase in subsection \ref{sec:accphase}; indeed, in the case without curvature, when $\lambda > \sqrt{2}$ these solutions admit only a transient acceleration phase which includes the universe today.

We include spatial curvature in section \ref{sec:curvature}. Observations constrain the amount of curvature to be relatively small today, with $\Omega_{k0}\lesssim\mathcal{O}(10^{-1}-10^{-3})$, and thus even more subdominant in the past.  Consequently, curvature leads to no significant difference in the past of candidate realistic solutions.  In contrast, the future can be very different.  Although this is less relevant to observations, it still has significant conceptual implications.  In contrast to the case without matter and radiation, which allowed eternal acceleration when $\lambda>\sqrt{2}$, the solutions considered here have only transient acceleration.  We discuss all this in subsection \ref{sec:pastfuture}. Moreover, there are interesting  quantitative differences for the universe today, which we work out in subsection \ref{sec:today}. If turning on curvature, with $\Omega_{k0} > 0$, is done by correspondingly lowering the dark energy density parameter today, $\Omega_{\phi0}$, we show that it could be favorable to string theory models. Indeed, for a given value of $w_{\phi0}$, this allows one to reach a higher value for $\lambda$, as shown in \eqref{wphi0valuesk}. Similarly, the theoretical upper bound \eqref{upperboundintro} on $\lambda$ for viable models becomes higher as in \eqref{lambdaboundacccurv}, and we expect something similar for the observational bounds. We obtain in addition theoretical predictions for the $w_0w_a+\Omega_k$ model often used in cosmological analyses. A complete cosmological fit of the curved exponential quintessence model with observational data, beyond the scope of this work, \cite{BBMPTZ, Alestas:2024gxe}, would determine whether spatial curvature can make values $\lambda \geq \sqrt{2}$  observationally viable; our results indicate that including curvature could help in this direction. We illustrate in figure \ref{fig:intro} a candidate realistic cosmological solution, with $\lambda>\sqrt{2}$ and negatively curved spatial slices.

We finally discuss in section \ref{sec:stringtheory} the concrete realisation of string theory models with such possibly realistic cosmological solutions, in particular one example using a well-controlled type II supergravity compactification on a negatively curved 6d compact Einstein manifold, with the quintessence scalar corresponding to the 6d volume modulus, giving $\lambda= \sqrt{\frac{8}{3}}$ and $k=-1$ \cite{Townsend:2003fx, Andersson:2006du, Marconnet:2022fmx, Andriot:2023wvg}. In such time dependent compactifications, we discuss the question of the time variation of fundamental constants, like masses of fundamental particles and coupling constants, as well as fifth forces. In particular, we observe how the epochs of radiation and matter domination in the universe inevitably lead to a freezing of the quintessence field; fundamental constants do not change during this time and the field distances traversed are easily sub-Planckian from BBN until today.  On the other hand, following further backwards in time, all cosmological solutions eventually reach an early epoch of kination as they approach the only fully unstable fixed points $\Pkin$.  This non-standard early universe epoch could have interesting physical consequences (see e.g. \cite{Conlon:2022pnx}), although it may be that the effective field theory breaks down before these can be studied as the field traverses super-Planckian distances. It would be interesting to then patch our cosmological solution to an earlier inflation model. \\

Let us close this summary by remarking that we have not addressed the cosmological constant problem.  String theory dark energy scenarios at the asymptotics of field space may go some way towards resolving this as string loop and $\alpha'$ corrections should be small there.  However, to properly conquer the problem, the cancellation of the Standard Model's contributions to the vacuum energy must be explained. This would require to go beyond the classical analysis performed here, within a complete model including a string realisation of matter.

\section{4-dimensional dynamical system}\label{sec:dynsys}

In this section, we analyse a 4-dimensional (4d) cosmology with radiation, matter, curvature and a scalar field, as a dynamical system. We describe the fixed points, the phase space and take a first look at cosmological solutions. A complementary analysis in $d$ dimensions is provided in appendix \ref{ap:dynsys}, while solutions in the vicinity of the fixed points are presented in appendix \ref{ap:asymptotics}.

\subsection{4d cosmology}\label{sec:4dcosmo}

We consider a 4d FLRW universe with metric
\be\label{metric}
\d s^2 = -\d t^2 +a^2(t) \left( \frac{\d r^2}{1-kr^2}+r^2 \d\Omega_{2}^2\right)\,,
\ee
with a scale factor $a(t)>0$, and where $k=0,\pm1$ characterizes the 3d spatial curvature. This universe is filled with a canonically normalised scalar field $\phi$, with scalar potential $V(\phi)$, and a set of perfect fluids labeled by $i$, which obey the equation of state $p_i=w_i\, \rho_i$; for us those components will be radiation and matter, $i=r,m$. In this case, the equations of motion (e.o.m.) can be written as follows
\begin{subequations}
 \begin{align}
      H^2 &= \frac{1}{3} \sum_n\rho_n \,,\label{eq:Fried}\\
    \dot H &= -\frac{1}{2} \sum_n\left(\rho_n +p_n \right)\,,  \label{eq:Hdot}\\
    \ddot \phi &= -3H\dot \phi - \del_\phi V \,,  \label{eq:phi}
 \end{align}
\end{subequations}
where we set the reduced Planck mass $M_p=(8\pi G)^{-1/2}=1$. The dot stands for $\del_t$ and $H=\dot a /a$ and we  assume in the following that $H \neq0$. We also used a general perfect fluid notation with label $n$, whose components are detailed in table \ref{tab:n}. For radiation, matter and curvature, the only time variation of $\rho_n$ is assumed to be through its dependence in $a(t)$.

\begin{table}[H]
\begin{center}
\centering
\begin{tabular}{| l | c || c | c | c |}
\hline
\cellcolor[gray]{0.9} & \cellcolor[gray]{0.9} & \cellcolor[gray]{0.9} & \cellcolor[gray]{0.9} & \cellcolor[gray]{0.9}\\[-8pt]
\cellcolor[gray]{0.9} $\!\! n$ & \cellcolor[gray]{0.9} component &  \cellcolor[gray]{0.9} $\rho_n$ & \cellcolor[gray]{0.9} $p_n$ & \cellcolor[gray]{0.9} $w_n$ \\[5pt]
\hline
&&&&\\[-8pt]
$r$ & radiation & $\propto a^{-4}$  & $\propto a^{-4}$ & $\frac{1}{3}$ \\[3pt]
\hline
&&&&\\[-8pt]
$m$ & matter & $\propto a^{-3}$  & $\propto a^{-3}$ & $0$ \\[3pt]
\hline
&&&&\\[-8pt]
$k$ & curvature & $-\frac{3\,k}{a^2}$  & $\frac{k}{a^2}$ & $-\frac{1}{3}$ \\[6pt]
\hline
&&&&\\[-8pt]
$\phi$ & scalar field & $\frac{\dot\phi^2}{2} + V(\phi)$  & $\frac{\dot\phi^2}{2} - V(\phi)$ & $w_{\phi}$ \\[6pt]
\hline
\end{tabular}
\end{center}
\caption{Perfect fluid notation with, for each component $n$, the energy density $\rho_n$, the pressure $p_n$, and the equation of state parameter $w_n=\frac{p_n}{\rho_n}$.}
\label{tab:n}
\end{table}

For each component $n$, we can define the density parameters  in the universe as $\Omega_n=\frac{\rho_n}{3H^2}$. The first Friedmann equation \eqref{eq:Fried} can then be rewritten as
\be\label{eq:friO}
1 = \sum_n\Omega_n \quad \Leftrightarrow \quad 1-\Omega_k =\Omega_T \,,
\ee
where
\be
\Omega_T = \Omega_\phi + \Omega_m + \Omega_r  \,,\qquad \Omega_k = - \frac{k}{a^2 H^2} \,.\label{OT}
\ee
We see from \eqref{eq:friO} the standard result that in an open universe ($k=-1$), $\Omega_T<1$, while in a closed universe ($k=1$), $\Omega_T>1$, and in a flat universe $\Omega_T=1$.

Finally, we can  define  $\rho_{\rm eff} = \sum_n \rho_n$ and $p_{\rm eff} = \sum_n p_n$, to obtain an {\sl effective equation of state} of the full system as $p_{\rm eff}=w_{\rm eff}\, \rho_{\rm eff}$. We then find
\be
w_{\rm eff} = \frac{\sum_n p_n}{\sum_n \rho_n}= \frac{\sum_n w_n \rho_n}{\sum_n \rho_n} = \frac{\sum_n w_n \Omega_n}{\sum_n \Omega_n} = \sum_n w_n \Omega_n = w_{\phi} \Omega_{\phi} - \frac{1}{3} \Omega_k + \frac{1}{3} \Omega_r \ .\label{weffO}
\ee

We will be interested in accelerating solutions, meaning $\ddot a>0$. This is quantified through the $\epsilon$ parameter defined as follows
\be
\frac{\ddot a}{a} = H^2 \lp1-\epsilon\rp \ ,\qquad \epsilon\equiv  -\frac{\dot H}{H^2}\,,
\ee
with acceleration amounting to $\epsilon<1$. Using \eqref{eq:Hdot} and the definitions above, we can write $\epsilon$ as
\be\label{epsilon}
\epsilon = \frac{3}{2} \sum_n (1+w_n)\, \Omega_n = \frac{3}{2} ( 1 + w_{\rm eff}) \,.
\ee
The condition for acceleration becomes
\be
\text{Acceleration:}\quad \epsilon< 1 \quad \Leftrightarrow \quad w_{\rm eff} <-\frac{1}{3} \,, \label{weffaccel}
\ee
as is the case for a single fluid component.

Finally, in correspondence with \eqref{OT}, let us introduce the following notation
\be
\rho_T = \rho_{\phi} + \rho_m + \rho_r \ ,\quad p_T = p_{\phi} + p_m + p_r \ ,\quad w_T= \frac{p_T}{\rho_T} \ .
\ee
The Friedmann equations \eqref{eq:Fried} and \eqref{eq:Hdot} can then be rewritten as follows
\be
H^2 + \frac{k}{a^2} = \frac{1}{3} \rho_T \ ,\qquad  \frac{\ddot{a}}{a}= -\frac{1}{2} \left(\frac{1}{3} + w_T \right) \rho_T \ . \label{eomsT}
\ee
The acceleration condition can then also be written  in terms of $w_T$ as $w_T<-1/3$. While $w_T$ is a priori different from $w_{\rm eff}$, the condition takes the same form as \eqref{weffaccel} because $w_k = -1/3$. \\

We end this subsection with a note on de Sitter in an open universe, $w_T$ and $w_{\rm{eff}}$.  In absence of curvature ($k=0$), a de Sitter spacetime admits a constant $H=H_0$, meaning an exponential scale factor $a(t)$. From \eqref{eomsT}, we then obtain a constant $\rho_T$, together with $w_T=-1$. The same holds true for $k=-1$: the metric \eqref{metric} of a de Sitter spacetime in hyperbolic slicing admits the scale factor
\be
a(t)=l\, \sinh\left(\frac{t}{l}\right) \ ,\label{adesitter}
\ee
where the de Sitter radius $l$ also appears in the scalar curvature ${\cal R}_4 = 12/l^2$. From \eqref{eomsT}, we obtain the constant $\rho_T=3/l^2 = - p_T $, i.e.~once again
\eq{
\text{de~Sitter}\ \Leftrightarrow \ w_T=-1~.
}
However, one has a priori $w_{\rm eff}\neq -1$ for de Sitter. Take for instance a realisation with $\Omega_r=\Omega_m =0$: in that case, $w_T= w_{\phi} = -1$.\footnote{One may ask whether de Sitter could be realised differently, namely with a non-zero matter or radiation contribution: indeed, requiring a constant $\rho_T$, together with a varying $\rho_m$ or $\rho_r$, would imply a very specific varying $\rho_{\phi}$.} In addition, $\Omega_{\phi}=1-\Omega_k$ so we read from \eqref{weffO} that $w_{\rm eff} = -1 +\frac{2}{3} \Omega_k $. With $k=-1$, we indeed have $w_{\rm eff}\neq -1$.

\subsection{Dynamical system}\label{sec:dynsysintro}

We now describe the  cosmology and its equations as a dynamical system. To that end, we restrict ourselves to the case $k\leq 0$, i.e.~$k=0$ or $k=-1$. We define the following dynamical variables:
\be\label{eq:variables}
   x= \frac{\phi'}{\sqrt{6}} \,, \qquad
   y = \frac{\sqrt{V}}{\sqrt{3} H}\,,    \qquad
   z = \frac{\sqrt{-k}}{a H}\,,  \qquad
   u = \frac{\sqrt{\rho_r}}{\sqrt{3} H}\,,  \qquad
   \lambda = -\frac{\del_\phi V}{V} \,,
   \ee
where the prime $'$ denotes the derivative with respect to the number $N$ of e-folds, $\d N= H \d t$ and $N=\ln \frac{a}{a_0}$ with $a_0$ taken to be the scale factor today. We recall that $a > 0,\, H \neq0$, and assume that $V \geq 0$. Strictly speaking, defining $\lambda$ requires $V \neq0$; we will come back to the case $V=0$. In terms of these variables, one has
\bea
&& \Omega_{\phi}=x^2+y^2\ ,\quad w_{\phi}=\frac{x^2 - y^2}{x^2+y^2}  \ ,\quad \Omega_k=z^2\ ,\quad \Omega_r=u^2 \,, \label{varOx}\\
\nonumber\\
\Leftrightarrow && x^2= \frac{1}{2}\Omega_{\phi} (1+w_{\phi})\ ,  \quad  y^2= \frac{1}{2}\Omega_{\phi} (1-w_{\phi}) \ ,\quad z^2=\Omega_k\ ,\quad u^2=\Omega_r \,,
\eea
and the first Friedmann equation \eqref{eq:Fried} can be written as
\be\label{Omegaxyzu}
 \Omega_m=1-  x^2- y^2- z^2 -u^2\,.
\ee
This equation will be a constraint to the system in the following.

We derive the dynamical system equations using only the second Friedmann equation \eqref{eq:Hdot} and the field e.o.m.~\eqref{eq:phi}. Equations are expressed in terms of the variables $x,y,z,u,\lambda$, together with $\Omega_m$ and derivatives of the potential. We obtain the system
\begin{subequations}\label{eq:system}
 \begin{empheq}[box=\widefbox]{align}
    x'&= \sqrt{\frac{3}{2}}\, y^2\,\lambda + x \left( 3\,(x^2-1)  +  z^2 +\frac32\Omega_m +2u^2\right) \,,\\
  y' &= y \left(- \sqrt{\frac{3}{2}}\, x \,\lambda + 3\,x^2 + z^2 +\frac32 \Omega_m +2u^2\right) \,,\\
  z' &= \,z\left( z^2-1 +3 \, x^2 + \frac32 \Omega_m +2u^2\right)\,,\\
  u' &= \,u\left( z^2-2 +3 \, x^2 + \frac32 \Omega_m +2u^2\right)\,,\\
  \lambda' &= -\sqrt{6}\,x\, \left(\frac{\del_{\phi}^2 V}{V} -\frac{(\del_{\phi}V)^2}{V^2}\right)\,,
\end{empheq}
\end{subequations}
and  the additional constraint \eqref{Omegaxyzu}.

Given that $\Omega_m$ can be expressed in terms of the other variables thanks to \eqref{Omegaxyzu}, whether the system is autonomous only depends on the last equation: it requires $\del_{\phi}^2 V/V$ to be expressed in terms of $x,y,z,\lambda$. This will hold from now on as we restrict ourselves to an exponential scalar potential
\be
V(\phi) = V_0\ e^{-\lambda \phi} \ . \label{potexp}
\ee
The previous variable $\lambda$ now matches the constant exponential rate, and we take $\lambda>0$, $V_0\geq0$. This allows us to extend the previous equations and variables to cases where $V = V_0 = 0$. In the system \eqref{eq:system}, the equation for $\lambda$ is trivially satisfied; we are left with the other equations and variables.

Note that we can also write previously defined quantities in terms of the dynamical variables: we obtain in particular from \eqref{weffO}
\be\label{weff}
w_{\rm eff} = x^2 -y^2 -\frac{z^2}{3} +\frac{u^2}{3}\,.
\ee

\subsection{Fixed points and physical interpretation}\label{sec:fixedpoints4d}

Finding the fixed points for the system \eqref{eq:system} with the constraint \eqref{Omegaxyzu} and the exponential potential \eqref{potexp} can now be done: we detail this derivation in appendix \ref{ap:dynsys} for general $d$. We summarize here in table \ref{tab:fixedpoints} the fixed points for $d=4$. For each of them we compute $w_{\rm eff}$ using \eqref{weff}: this determines whether the fixed point solution is accelerating. We see that only $\Pp$ offers this possibility, provided $\lambda< \sqrt{2}$, as is well-known \cite{Copeland:1997et, Bahamonde:2017ize, Shiu:2023nph, Shiu:2023fhb}.\\

\begin{table}[h]
\begin{center}
\centering
\begin{tabular}{| l | c | c | c | }
\hline
\cellcolor[gray]{0.9} & \cellcolor[gray]{0.9} & \cellcolor[gray]{0.9} & \cellcolor[gray]{0.9}\\[-8pt]
\cellcolor[gray]{0.9} \hskip 1.4cm $(x,y,z,u)$ & \cellcolor[gray]{0.9} $\Omega_m$ &  \cellcolor[gray]{0.9} Existence & \cellcolor[gray]{0.9} $w_{\rm eff}$ \\[5pt]
\hline
&&&\\[-8pt]
$\Pkin= $ $(\pm 1,0,0,0) $ & $0$  & $\forall\, \lambda$ & $1$ \\[3pt]
\hline
&&&\\[-8pt]
$\Pk= $ $(0,0,\pm1,0) $ & $0$  & $ \forall\, \lambda$ & $ -\frac{1}{3}$ \\[3pt]
\hline
&&&\\[-8pt]
$\Pkp= $ $\lp \frac{1}{\lambda}\sqrt{\frac{2}{3}},\pm\frac{2}{\lambda\sqrt{3}} ,\pm \sqrt{1-\frac{2}{\lambda^2}},0\rp $ & $0$ & $ \lambda>\sqrt{2}$ & $ -\frac{1}{3}$ \\[8pt]
\hline
&&&\\[-8pt]
$\Pp= $ $\lp \frac{\lambda}{\sqrt{6}}, \pm \frac{\sqrt{6-\lambda^2}}{\sqrt{6}}, 0 , 0\rp $ & $0$ & $  \lambda < \sqrt{6}$ & $ \frac{\lambda^2}{3}-1$ \\[8pt]
\hline
&&&\\[-8pt]
$\Pmp= $ $\lp \frac{1}{\lambda} \sqrt{\frac{3}{2}},\pm\frac{1}{\lambda} \sqrt{\frac{3}{2}},0,0\rp $ & $1-\frac{3}{\lambda^2}$ & $  \lambda > \sqrt{3}$  & $0$\\[8pt]
\hline
&&&\\[-8pt]
$\Pm= $ $(0,0,0,0) $ & $1$ & $ \forall\, \lambda$ & $0$ \\[3pt]
\hline
&&&\\[-8pt]
$\Prr= $ $(0,0,0,\pm 1)$ & $0$ & $ \forall\, \lambda$ & $\frac{1}{3}$ \\[3pt]
\hline
&&&\\[-8pt]
$\Prp= $ $\left( \frac{1}{\lambda} \sqrt{\frac{8}{3}},\pm \frac{2}{\lambda\, \sqrt{3}},0, \pm \sqrt{1-\frac{4}{\lambda^2}} \right)$ & $0$ & $\lambda > 2$ & $\frac{1}{3}$ \\[8pt]
\hline
\end{tabular}
\end{center}
\caption{Fixed points for the system \eqref{eq:system} with the constraint \eqref{Omegaxyzu} and the exponential potential \eqref{potexp}. The nomenclature used comes from the fluid components present at each fixed point: we detail it in the main text. The existence condition for $\Pmp$ comes from the requirement $\Omega_m \geq 0$ that we impose; we make the inequality strict to distinguish it from $\Pp$. The points $\Pkin,\Pk,\Pkp,\Pp$ have $\Omega_m=\Omega_r=0$ and thus correspond to those found in \cite{Andriot:2023wvg}, denoted there  by $P_{\pm},P_0,P_1,P_2$, respectively. The points $\Pkin,\Pp,\Pmp,\Pm$ have $\Omega_k=\Omega_r=0$ and already appeared in the literature under the respective names $A_{\pm},C, B, O$ \cite{Bahamonde:2017ize}.}
\label{tab:fixedpoints}
\end{table}

The fixed points have the following physical interpretations
\begin{itemize}
    \item {\sl Kinetic domination}: $\Pkin$ are dominated by the kinetic energy of the scalar field, with $x^2=1$, and have thus $w_{\rm eff} =1$. From the phenomenological point of view, these points could be of interest for an early kination epoch.

    \item {\sl Curvature domination}: $\Pk$ is a curvature dominated point with $\Omega_k=1$ and $w_{\rm eff} =-1/3$. It corresponds to no (de)acceleration: $\ddot{a}=0$.

    \item {\sl Curvature scaling}: at $\Pkp$, the universe evolves under the influence of both the curvature and the scalar field, with $\Omega_k=\frac{\lambda^2-2}{\lambda^2}>0$, $\Omega_\phi=\frac{2}{\lambda^2}$. However, it has $w_{\rm eff} =-1/3$,  so that the expansion mimics pure curvature domination and corresponds to no (de)acceleration: $\ddot{a}=0$. It merges with $\Pp$ for $\lambda\rightarrow \sqrt{2}$ and $\Pk$ for $\lambda \rightarrow \infty$.

    \item {\sl Scalar domination}: $\Pp$ is dominated by the scalar field with $\Omega_\phi=1$; this is a standard point in quintessence models of late dark energy \cite{Bahamonde:2017ize}. It is the only point that allows acceleration, with $\lambda<\sqrt{2}$ giving $w_{\rm eff}<-1/3$. It merges with $\Pkinp$ for $\lambda\rightarrow \sqrt{6}$, and corresponds to a pure de Sitter universe for $\lambda \rightarrow 0$.

    \item {\sl Matter scaling}: in $\Pmp$, the universe evolves under the influence of both matter and the scalar field, with $\Omega_m=1-\frac{3}{\lambda^2} > 0$ and $\Omega_\phi=\frac{3}{\lambda^2}$. But similar to the above, $w_{\rm eff}=0$, means that it expands as if it were completely matter dominated. It merges with $\Pp$ for $\lambda\rightarrow \sqrt{3}$ and $\Pm$ for $\lambda \rightarrow \infty$.

    \item {\sl Matter domination}: $\Pm$ is dominated by matter with $\Omega_m=1$ and has $w_{\rm eff}=0$. This point is of obvious phenomenological interest.

    \item {\sl Radiation domination}: $\Prr$ is dominated by radiation with $\Omega_r=1$ and has $w_{\rm eff}=\frac{1}{3}$. This point is also of phenomenological interest.

    \item {\sl Radiation scaling}: in $\Prp$, the universe evolves under the radiation and the scalar field influence, with $\Omega_r=1-\frac{4}{\lambda^2} > 0$ and $\Omega_{\phi}=\frac{4}{\lambda^2}$. It has $w_{\rm eff}=\frac{1}{3}$, which mimics a pure radiation domination. It merges with $\Pp$ for $\lambda\rightarrow 2$ and $\Prr$ for $\lambda \rightarrow \infty$.

\end{itemize}

\subsection{Stability}\label{sec:stab}

To find the stability of the fixed points listed in table \ref{tab:fixedpoints}, we first write the system \eqref{eq:system}, without the $\lambda'$ equation, and in which we replace $\Omega_m$ using \eqref{Omegaxyzu}, as follows
\be
x' = f(x,y,z,u) \ ,\ y'=g(x,y,z,u) \ ,\ z'=h(x,y,z,u) \ ,\ u'=v(x,y,z,u) \ .
\ee
The stability of the fixed points can be read off from the eigenvalues of the Jacobian
\be \label{Jacobian}
M= \left( \begin{array}{cccc} \del_x f & \del_y f & \del_z f & \del_u f \\
\del_x g & \del_y g & \del_z g & \del_u g \\
\del_x h & \del_y h & \del_z h & \del_u h \\
\del_x v & \del_y v & \del_z v & \del_u v
\end{array} \right) \ ,
\ee
evaluated at each fixed point, with all negative real parts indicating a stable (or ``attractor'') direction and positive real parts indicating an unstable (or ``repeller'') direction.  We list these eigenvalues for each fixed point in table \ref{tab:fixedpointstab}, and indicate the corresponding stability.

\begin{table}[H]
\begin{center}
\centering
\begin{tabular}{| l | c | c | c |}
\hline
\cellcolor[gray]{0.9} & \cellcolor[gray]{0.9} & \cellcolor[gray]{0.9} &  \cellcolor[gray]{0.9} \\[-8pt]
\cellcolor[gray]{0.9} Point & \cellcolor[gray]{0.9} Eigenvalues &  \cellcolor[gray]{0.9} Stability &  \cellcolor[gray]{0.9} Existence \\[5pt]
\hline
&&&\\[-10pt]
&& $\Pkinm:$ Fully unstable \phantom{for $\lambda \leq \sqrt{6}$} & \\
$\Pkin $ & $ \lp 3,2,3 \mp \lambda \sqrt{\frac32}, 1 \rp$  & $\Pkinp:$ Fully unstable for $\lambda \leq \sqrt{6}$ & - \\
&& $\Pkinp:$ Saddle for $\lambda >\sqrt{6}$ \phantom{Fully i} &\\[6pt]
\hline
&&&\\[-8pt]
$\Pk $ & $(-2,-1,1,-1)$  & Saddle & - \\[4pt]
\hline
&&&\\[-10pt]
$\Pkp$ & $\lp -1, -1 -\frac{\sqrt{8-3\lambda^2}}{\lambda},-1 +\frac{\sqrt{8-3\lambda^2}}{\lambda},-1\rp $  & Stable & $\lambda > \sqrt{2}$ \\[6pt]
\hline
&&&\\[-7pt]
 &  & Stable for $\lambda \leq \sqrt{2}$ & \\[-8pt]
$\Pp$ & $ \lp \frac{\lambda^2}{2}-3, \lambda^2-3,\frac{\lambda^2}{2}-1 , \frac{\lambda^2}{2}-2\rp$ && $\lambda < \sqrt{6}$ \\[-10pt]
&& Saddle for $\lambda > \sqrt{2}$ & \\[6pt]
\hline
&&&\\[-10pt]
$\Pmp $ & $\lp \frac12,-\frac{3(\lambda+\sqrt{24-7\lambda^2})}{4\lambda},-\frac{3(\lambda -\sqrt{24-7\lambda^2})}{4\lambda}, -\frac12 \rp $  & Saddle & $\lambda > \sqrt{3} $ \\[6pt]
\hline
&&&\\[-8pt]
$\Pm $ & $ (-\frac32,\frac32,\frac12,-\frac12)$  & Saddle & - \\[5pt]
\hline
&&&\\[-8pt]
$\Prr $ & $ (-1,2,1,1)$  & Saddle & - \\[5pt]
\hline
&&&\\[-8pt]
$\Prp $ & $ (1,1, -\frac{\lambda+\sqrt{64-15 \lambda^2}}{2\lambda}, -\frac{\lambda-\sqrt{64-15 \lambda^2}}{2\lambda} )$  & Saddle & $\lambda > 2 $ \\[5pt]
\hline
\end{tabular}
\end{center}
\caption {Stability of the fixed points in table \ref{tab:fixedpoints}, read from the eigenvalues of the Jacobian \eqref{Jacobian}. We also recall restrictions on the existence of each fixed point.}
\label{tab:fixedpointstab}
\end{table}

An important information that can be read from table \ref{tab:fixedpointstab} is that all solutions start at $\Pkin$, the fully unstable point(s), and end at the attractor $\Pkp$ ($\lambda> \sqrt{2}$) or $\Pp$ ($\lambda \leq \sqrt{2}$). Physically, this implies that all solutions start in a kination epoch. Whether they subsequently pass nearby other points, which are saddles, is a relevant question for phenomenology. Note also that approaching $\Pkp, \Pmp, \Prp$ can be partly done as a spiral instead of a standard node, when the value of $\lambda$ is such that the eigenvalues have non-trivial imaginary parts.\\

Before studying solutions in more detail, it is interesting to focus on important subcases. In the system \eqref{eq:system}, it is clear that $u=0$ (no radiation) or $z=0$ (no curvature) provide solutions: indeed, setting $u$ or $z$ to zero allows one to discard the corresponding equations in $u'$ or $z'$. Doing so also provides consistent subsystems (called ``invariant manifolds'' in dynamical system terminology, or invariant subspaces in the following), where we ignore from the start the variable $u$ or $z$. When considering such subsystems, the fixed points remain the same, as long as they are part of the solutions selected by $u=0$ or $z=0$. However, their stability may change: indeed, one or several eigenvalues are removed, and the stability can thus be altered. We present the corresponding stability for some of these invariant subspaces in tables  \ref{tab:fixedpointstabwithoutr} and \ref{tab:fixedpointstabwithoutrk}.\\

\begin{table}[h]
\begin{center}
\centering
\begin{tabular}{| l | c | c | c |}
\hline
\cellcolor[gray]{0.9} & \cellcolor[gray]{0.9} & \cellcolor[gray]{0.9} &  \cellcolor[gray]{0.9} \\[-8pt]
\cellcolor[gray]{0.9} Point & \cellcolor[gray]{0.9} Eigenvalues &  \cellcolor[gray]{0.9} Stability &  \cellcolor[gray]{0.9} Existence \\[5pt]
\hline
&&&\\[-10pt]
&& $\Pkinm:$ Fully unstable \phantom{for $\lambda \leq \sqrt{6}$} & \\
$\Pkin $ & $ \lp 3,2,3 \mp \lambda \sqrt{\frac32} \rp$  & $\Pkinp:$ Fully unstable for $\lambda \leq \sqrt{6}$ & - \\
&& $\Pkinp:$ Saddle for $\lambda >\sqrt{6}$ \phantom{Fully i} &\\[6pt]
\hline
&&&\\[-8pt]
$\Pk $ & $(-2,-1,1)$  & Saddle & - \\[4pt]
\hline
&&&\\[-10pt]
$\Pkp$ & $\lp -1, -1 -\frac{\sqrt{8-3\lambda^2}}{\lambda},-1 +\frac{\sqrt{8-3\lambda^2}}{\lambda}\rp $  & Stable & $\lambda > \sqrt{2}$ \\[6pt]
\hline
&&&\\[-7pt]
 &  & Stable for $\lambda \leq \sqrt{2}$ & \\[-8pt]
$\Pp$ & $ \lp \frac{\lambda^2}{2}-3, \lambda^2-3,\frac{\lambda^2}{2}-1 \rp$ && $\lambda < \sqrt{6}$ \\[-10pt]
&& Saddle for $\lambda > \sqrt{2}$ & \\[6pt]
\hline
&&&\\[-10pt]
$\Pmp $ & $\lp \frac12,-\frac{3(\lambda+\sqrt{24-7\lambda^2})}{4\lambda},-\frac{3(\lambda -\sqrt{24-7\lambda^2})}{4\lambda} \rp $  & Saddle & $\lambda > \sqrt{3} $ \\[6pt]
\hline
&&&\\[-8pt]
$\Pm $ & $ (-\frac32,\frac32,\frac12)$  & Saddle & - \\[5pt]
\hline
\end{tabular}
\end{center}
\caption {Stability of the fixed points in the invariant subspace without radiation ($u=0$): with respect to table \ref{tab:fixedpoints} and \ref{tab:fixedpointstab}, $\Prr, \Prp$ are not present anymore. We refer to the main text for more details.}
\label{tab:fixedpointstabwithoutr}
\end{table}

\begin{table}[h]
\begin{center}
\centering
\begin{tabular}{| l | c | c | c |}
\hline
\cellcolor[gray]{0.9} & \cellcolor[gray]{0.9} & \cellcolor[gray]{0.9} &  \cellcolor[gray]{0.9} \\[-8pt]
\cellcolor[gray]{0.9} Point & \cellcolor[gray]{0.9} Eigenvalues &  \cellcolor[gray]{0.9} Stability &  \cellcolor[gray]{0.9} Existence \\[5pt]
\hline
&&&\\[-10pt]
&& $\Pkinm:$ Fully unstable \phantom{for $\lambda \leq \sqrt{6}$} & \\
$\Pkin $ & $ \lp 3,3 \mp \lambda \sqrt{\frac32} \rp$  & $\Pkinp:$ Fully unstable for $\lambda \leq \sqrt{6}$ & - \\
&& $\Pkinp:$ Saddle for $\lambda >\sqrt{6}$ \phantom{Fully i} &\\[6pt]
\hline
&&&\\[-7pt]
 &  & Stable for $\lambda \leq \sqrt{3}$ & \\[-8pt]
$\Pp$ & $ \lp \frac{\lambda^2}{2}-3, \lambda^2-3 \rp$ && $\lambda < \sqrt{6}$ \\[-10pt]
&& Saddle for $\lambda > \sqrt{3}$ & \\[6pt]
\hline
&&&\\[-10pt]
$\Pmp $ & $\lp -\frac{3(\lambda+\sqrt{24-7\lambda^2})}{4\lambda},-\frac{3(\lambda -\sqrt{24-7\lambda^2})}{4\lambda} \rp $  & Stable & $\lambda > \sqrt{3} $ \\[6pt]
\hline
&&&\\[-8pt]
$\Pm $ & $ (-\frac32,\frac32)$  & Saddle & - \\[5pt]
\hline
\end{tabular}
\end{center}
\caption {Stability of the fixed points in the invariant subspace without radiation and curvature ($z=u=0$): with respect to table \ref{tab:fixedpoints} and \ref{tab:fixedpointstab}, $\Pk, \Pkp,\Prr, \Prp$ are not present anymore. We refer to the main text for more details. The results presented here match those of \cite[Tab. 5]{Bahamonde:2017ize}, with $\Pkin= A_{\pm}$, $\Pp=C$, $\Pmp=B$, $\Pm=O$.}
\label{tab:fixedpointstabwithoutrk}
\end{table}

A further invariant subspace is given by the results from \cite{Andriot:2023wvg} where one considers no matter and radiation ($u=0, x^2+y^2+z^2=1$). It is less straightforward to see this is an invariant subspace: one should compare the system \eqref{eq:system} and constraint \eqref{Omegaxyzu} (with $\Omega_m=0, u=0$) to \cite[(2.7)]{Andriot:2023wvg}.\footnote{While equations $x',y'$ match, we see from the $z'$ equation here that the fixed point solutions should obey $z(2x^2-y^2)=0$. One verifies that this holds for $\Pkin,\Pk,\Pkp,\Pp$.} The fixed points are $\Pkin,\Pk,\Pkp,\Pp$ from table \ref{tab:fixedpoints}, and they can be expressed in terms of $(x,y)$ only. Their stability is given in table \ref{tab:fixedpointstabwithoutrm}.\footnote{The results in table \ref{tab:fixedpointstabwithoutrm} reproduce those of \cite[Sec. 2.3]{Andriot:2023wvg}, up to a mistake we note in \cite[(2.23)]{Andriot:2023wvg}: the second eigenvalue for $\Pkin$ is $d-1 \mp \sqrt{(d-1)(d-2)}\, \lambda/2 $ (where $\lambda$ here stands for $\gamma$ there); the second term was missed.}\\

\begin{table}[h]
\begin{center}
\centering
\begin{tabular}{| l | c | c | c |}
\hline
\cellcolor[gray]{0.9} & \cellcolor[gray]{0.9} & \cellcolor[gray]{0.9} &  \cellcolor[gray]{0.9} \\[-8pt]
\cellcolor[gray]{0.9} Point & \cellcolor[gray]{0.9} Eigenvalues &  \cellcolor[gray]{0.9} Stability &  \cellcolor[gray]{0.9} Existence \\[5pt]
\hline
&&&\\[-10pt]
&& $\Pkinm:$ Fully unstable \phantom{for $\lambda \leq \sqrt{6}$} & \\
$\Pkin $ & $ \lp 4,3 \mp \lambda \sqrt{\frac32} \rp$  & $\Pkinp:$ Fully unstable for $\lambda \leq \sqrt{6}$ & - \\
&& $\Pkinp:$ Saddle for $\lambda >\sqrt{6}$ \phantom{Fully i} &\\[6pt]
\hline
&&&\\[-8pt]
$\Pk $ & $(-2,1)$  & Saddle & - \\[4pt]
\hline
&&&\\[-10pt]
$\Pkp$ & $\lp -1 -\frac{\sqrt{8-3\lambda^2}}{\lambda},-1 +\frac{\sqrt{8-3\lambda^2}}{\lambda}\rp $  & Stable & $\lambda > \sqrt{2}$ \\[6pt]
\hline
&&&\\[-7pt]
 &  & Stable for $\lambda \leq \sqrt{2}$ & \\[-8pt]
$\Pp$ & $ \lp \frac{\lambda^2}{2}-3, \lambda^2-2 \rp$ && $\lambda < \sqrt{6}$ \\[-10pt]
&& Saddle for $\lambda > \sqrt{2}$ & \\[6pt]
\hline
\end{tabular}
\end{center}
\caption{Stability of the fixed points in the invariant subspace without radiation and matter: with respect to table \ref{tab:fixedpoints} and \ref{tab:fixedpointstab}, $\Pmp,\Pm,\Prr, \Prp$ are not present anymore. We refer to the main text for more details. These results are consistent with those of \cite[Sec. 2.3]{Andriot:2023wvg}.}
\label{tab:fixedpointstabwithoutrm}
\end{table}

The main change of stability that occurs when restricting to an invariant subspace happens when comparing the situation with or without curvature.  Consider e.g.~table \ref{tab:fixedpointstabwithoutr} and \ref{tab:fixedpointstabwithoutrk}, for which radiation is turned off and there is, respectively, curvature and no curvature.  Since $\Pkp$ cannot exist anymore without curvature, its role of attractor in the presence of curvature (for $\lambda > \sqrt{2}$) is taken over by $\Pp$ (for $\lambda\leq \sqrt{3}$) or $\Pmp$ (for $\lambda > \sqrt{3}$) when curvature is turned off. Note also that the value of $\lambda$ at which $\Pp$ transitions from stable to saddle changes from $\sqrt{2}$ with curvature to $\sqrt{3}$ without curvature. This will have an impact on the solutions we consider below.

\subsection{Cosmological solutions at the fixed points}\label{sec:fixedpointsol}

In this subsection we translate the fixed points that we found in table \ref{tab:fixedpoints} back to the original variables, namely the scale factor $a(t)$ and the scalar field $\phi(t)$. We also express the necessary conditions for the fixed points to exist in terms of the curvature $k$, $V_0$, $\Omega_r$ and $\Omega_m$. This allows one to immediately identify some of the fixed points as corresponding to a cosmology with only curvature ($\Pk)$, only matter ($\Pm$) or only radiation ($\Prr$). We summarize the results in table \ref{tab:fixedpointsfields}, where we present only the results for expanding cosmologies, i.e., for $H>0$.\\

It is also interesting to see how the solutions in phase space can leave or approach the fixed points. This can be determined by expanding $a(t)$ and $\phi(t)$ around the solutions in table \ref{tab:fixedpointsfields} and solving the equations of motion. The results are slightly lengthy and therefore we list them in appendix \ref{ap:asymptotics}. Furthermore, appendix \ref{ap:analytic} discusses fixed curves and surfaces in the $(x,y,z,u)$ parameter space.
We also derive closed-form expressions for fully analytic solutions of  $a(t)$ and $\phi(t)$ that exist within a subspace of the full parameter space.

\renewcommand{\arraystretch}{2}
\begin{table}[H]
\begin{center}
\centering
\begin{tabular}{| l | c | c | c |}
\hline
\cellcolor[gray]{0.9} Point & \cellcolor[gray]{0.9} $a(t)$ &  \cellcolor[gray]{0.9} $\phi(t)$ &  \cellcolor[gray]{0.9} Conditions \\[4pt]
\hline
$\Pkin $ & $a(t)=a_0\, t^{\frac13}$  & $\phi(t)=\phi_0\pm\sqrt{\frac{2}{3}}\ln t$ & \makecell{$k=V_0=\Omega_m=\Omega_r=0$, \\[2pt] $a_0 ,\phi_0$ arbitrary} \\[7pt]
\hline
$\Pk$ &  $a(t)=a_0 \,t$  & $\phi(t)=\phi_0$ & \makecell{$k=-1$, $V_0=\Omega_m=\Omega_r=0$,\\[2pt] $a_0=1$, $\phi_0$ arbitrary} \\[7pt]
\hline
$\Pkp$ & $a(t)=a_0\, t$  & $\phi(t)=\phi_0+\frac{2}{\lambda}\ln t$ & \makecell{$k=-1$, $\Omega_m=\Omega_r=0$, \\[2pt] $a_0=\frac{\lambda}{\sqrt{\lambda^2-2}}$, $\phi_0=\frac{1}{\lambda}\ln\frac{\lambda^2V_0}{4}$ } \\[7pt]
\hline
$\Pp$ & $a(t)=a_0\, t^{\frac{2}{\lambda^2}}$  & $\phi(t)=\phi_0+\frac{2}{\lambda}\ln t$ & \makecell{$k=\Omega_m=\Omega_r=0$,\\[2pt]
$\phi_0=\frac{1}{\lambda}\ln\left(\frac{\lambda^4V_0}{2(6-\lambda^2)}\right)$, $V_0, a_0$ arbitrary} \\[7pt]
\hline
$\Pmp$ & $a(t)=a_0\, t^{\frac{2}{3}}$  & $\phi(t)=\phi_0+\frac{2}{\lambda}\ln t$ & \makecell{$k=\Omega_r=0$, $\Omega_m=\frac{\lambda^2-3}{\lambda^2}$,\\[2pt]  $\phi_0=\frac{1}{\lambda}\ln\left(\frac{\lambda^2V_0}{2}\right)$, $V_0, a_0$ arbitrary} \\[7pt]
\hline
$\Pm$ & $a(t)=a_0\, t^{\frac{2}{3}}$  & $\phi(t)=\phi_0$ & \makecell{$k=V_0=\Omega_r=0$,
 $\Omega_m=1$,\\[2pt] $a_0, \phi_0$ arbitrary} \\[7pt]
\hline
$\Prr$ & $a(t)=a_0 \, t^{\frac12}$  & $\phi(t)=\phi_0$ & \makecell{$k=V_0=\Omega_m=0$,
 $\Omega_r=1$,\\[2pt] $a_0, \phi_0$ arbitrary} \\[7pt]
\hline
$\Prp$ & $a(t)=a_0 \, t^{\frac12}$  & $\phi(t)=\phi_0+\frac{2}{\lambda}\ln t$ &  \makecell{$k=\Omega_m=0$, $\Omega_r=\frac{\lambda^2-4}{\lambda^2}$,\\[2pt]  $\phi_0=\frac{1}{\lambda}\ln (\lambda^2 V_0)$, $V_0, a_0$ arbitrary}  \\[7pt]
\hline
\end{tabular}
\end{center}
\caption {Behavior of $a(t)$ and $\phi(t)$ at the fixed points from table \ref{tab:fixedpoints}. Whenever there were two sign choices $\pm$ possible then we restricted to the one corresponding to expansion. We have used the shift symmetry in the time coordinate to set $a(t=0)=0$. The last column lists restrictions on the parameters.}
\label{tab:fixedpointsfields}
\end{table}
\renewcommand{\arraystretch}{1}

\subsection{Graphical summary}\label{sec:graphsum}

We provide here various illustrations of the phase space and its fixed points. Fixing the variable $\lambda$, which is constant for an exponential potential, we have a $4$-dimensional phase space in terms of $(x,y,z,u)$. A graphical representation requires us to restrict to 3 dimensions.\\

We first do so by turning off the radiation, i.e., by setting $u=0$. In addition, we trade $z$ for $\sqrt{\Omega_m}$ using the constraint \eqref{Omegaxyzu}. We then plot the 3d phase space with coordinates $(x,y,\sqrt{\Omega_m})$. This allows us to extend the illustrations of \cite{Marconnet:2022fmx, Andriot:2023wvg} that were restricted to the plane $(x,y)$: the third dimension directly accounts for the inclusion of matter. Since $\sqrt{\Omega_m} \geq 0$, only the upper half of the 3d space is meaningful.

Given the constraint \eqref{Omegaxyzu} for $u=0$, $z^2 = 1- x^2 -y^2 - \Omega_m$, we deduce that $k=0=z$ corresponds, in the $(x, y,\sqrt{\Omega_m})$ space, to the upper half sphere of radius $1$, while $k=-1$ is its interior, i.e.~the upper half ball of unit radius. In other words, the distance to the sphere measures how large $\Omega_k=z^2$ is. Restricting to the $(x,y)$ horizontal plane ($\Omega_m=0$), we recover the $k=0$ circle and its inside disk as depicted in \cite{Marconnet:2022fmx, Andriot:2023wvg}.

An expanding universe, of interest here, requires $y\geq 0$ and $z \geq0$ (see \eqref{eq:variables}). We can then restrict to the upper quarter of the ball corresponding to $y, \sqrt{\Omega_m}\geq 0$. Note that the system \eqref{eq:system} is invariant under the three symmetries $y \rightarrow -y$, $z \rightarrow -z $, and $x \rightarrow -x ,\ \lambda \rightarrow -\lambda $. We broke the last one by restricting ourselves to $\lambda>0$.

\begin{figure}[H]
\centering
\includegraphics[width=0.6\textwidth]{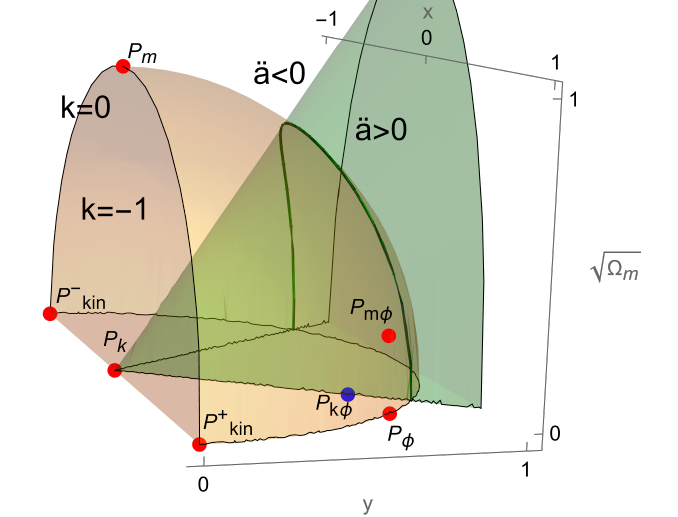}
\caption{Phase space for the system \eqref{eq:system} with constraint \eqref{Omegaxyzu} and exponential potential \eqref{potexp}, turning off the radiation ($u=0$). The three coordinates are given by the system variables $(x,y,\sqrt{\Omega_m})$. We consider the case of an expanding universe, restricting the illustration to a quarter of the 3d space. The interior of the ball of radius 1 is the region with $k=-1$, while its spherical boundary corresponds to $k=0$. The green region is that of accelerating solutions. The fixed points of table \ref{tab:fixedpoints} with $u=0$ are depicted, for $\lambda=\sqrt{3}+0.05$. In red are the saddle or unstable ones, while in blue is the stable one (attractor). Recall that the existence of the fixed points, as well as their stability properties, depend on $\lambda$; the value chosen here allows to depict all fixed points of the stability table \ref{tab:fixedpointstabwithoutr}. The horizontal $(x,y)$ plane reproduces the phase space depicted in \cite{Marconnet:2022fmx, Andriot:2023wvg}. We refer to the main text for more details.}\label{fig:PhaseSpace}
\end{figure}

Finally, another region of interest is that of acceleration. Reformulating the requirement $w_{\rm eff} < - \frac{1}{3}$ with \eqref{weff} and $u=0$, we deduce that the acceleration zone corresponds to $
\Omega_m < 2 y^2 - 4 x^2$. This region is a 3d cone, and is depicted in green in the phase space illustration given in figure \ref{fig:PhaseSpace}.\\

\begin{figure}[H]
\centering
\includegraphics[width=0.6\textwidth]{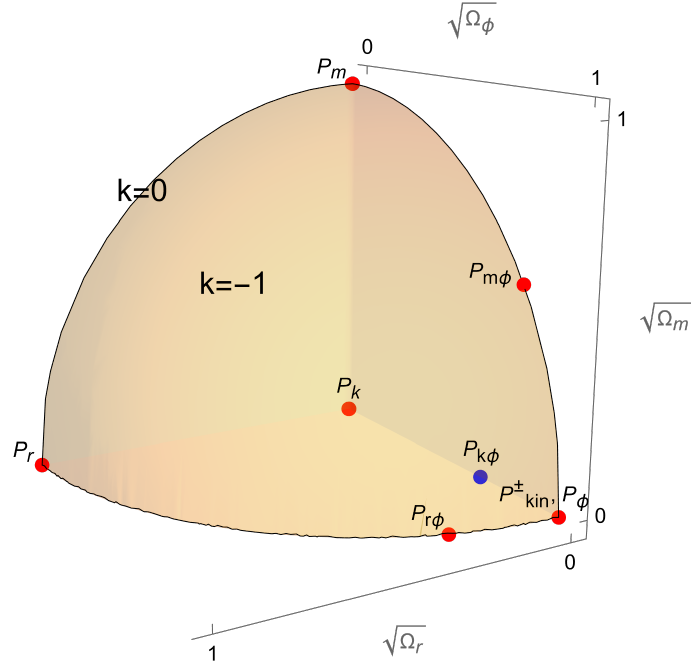}
\caption{Phase space for the system \eqref{eq:system} with constraint \eqref{Omegaxyzu} and exponential potential \eqref{potexp}, ignoring the parameter $w_{\phi}$. The three coordinates are $(\sqrt{\Omega_{r}}, \sqrt{\Omega_{\phi}}, \sqrt{\Omega_{m}})$. Considering an expanding universe, the illustration is restricted to an eighth of the 3d space. The ball of radius 1 is the region with $k=-1$, while its spherical boundary corresponds to $k=0$. The accelerating region cannot be represented. The fixed points of table \ref{tab:fixedpoints} are depicted, for $\lambda=2+0.1$; in this representation, $\Pkin$ and $\Pp$ are merged. In red are the saddle or unstable ones, while in blue is the stable one (attractor), as described by table \ref{tab:fixedpointstab}. We refer to the main text for more details.}\label{fig:PhaseSpacer}
\end{figure}

As a second 3d illustration, we take the combination of $x$ and $y$ into the single variable $\Omega_{\phi}=x^2+y^2$. The freedom given by the other variable $w_{\phi}$ \eqref{varOx} is not represented. We then plot the 3d phase space with coordinates $(\sqrt{\Omega_{r}}, \sqrt{\Omega_{\phi}}, \sqrt{\Omega_{m}})= (u, \sqrt{x^2+y^2}, \sqrt{\Omega_{m}}) $, where again we restrict ourselves to an expanding universe ($u \geq 0$). Since all three coordinates are positive, only one eighth of the 3d space is necessary. The constraint is phrased as
\be
\Omega_k = 1 - \Omega_r - \Omega_{\phi} - \Omega_m \ ,
\ee
so we can use similar to before the eighth of a ball corresponding to $k=-1$, together with its spherical boundary for $k=0$. The acceleration region cannot be depicted on such an illustration, as $w_{\rm eff}$ depends on $w_{\phi}$ \eqref{weffO}. This phase space illustration is given in figure \ref{fig:PhaseSpacer}.

\subsection{A first glimpse at cosmological solutions}\label{sec:firstsol}

Having determined analytically cosmological solutions at the fixed points in subsection \ref{sec:fixedpointsol}, as well as in their vicinity in appendix \ref{ap:asymptotics}, we now take a look at complete cosmological solutions obtained numerically. Before focusing on specific solutions of interest, we first provide an overall picture of possible solutions, using the phase space illustrations presented in subsection \ref{sec:graphsum}.

\begin{figure}[H]
\begin{center}
\begin{subfigure}[H]{0.48\textwidth}
\includegraphics[width=\textwidth]{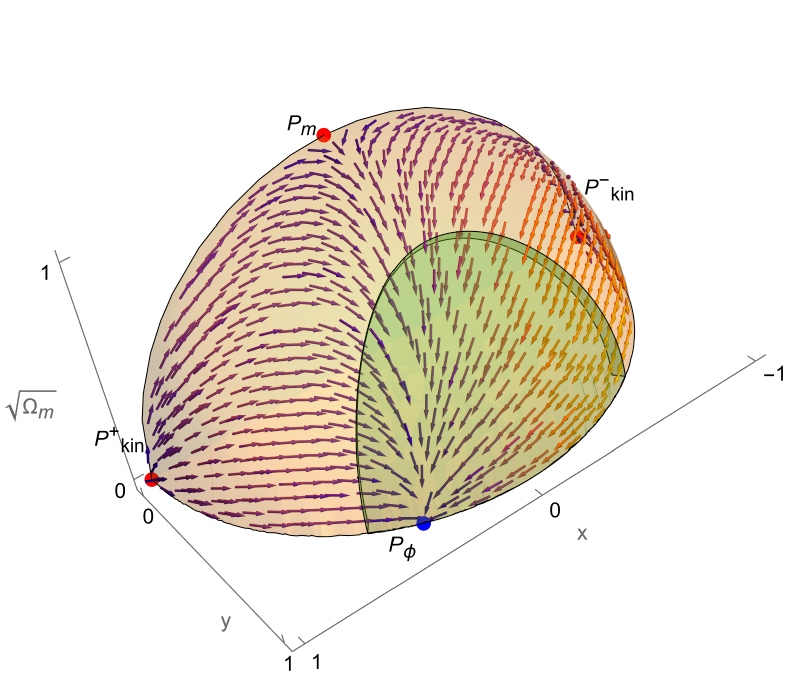}\caption{$\lambda=1$}\label{fig:PhaseSpaceL0}
\end{subfigure}\quad
\begin{subfigure}[H]{0.48\textwidth}
\includegraphics[width=\textwidth]{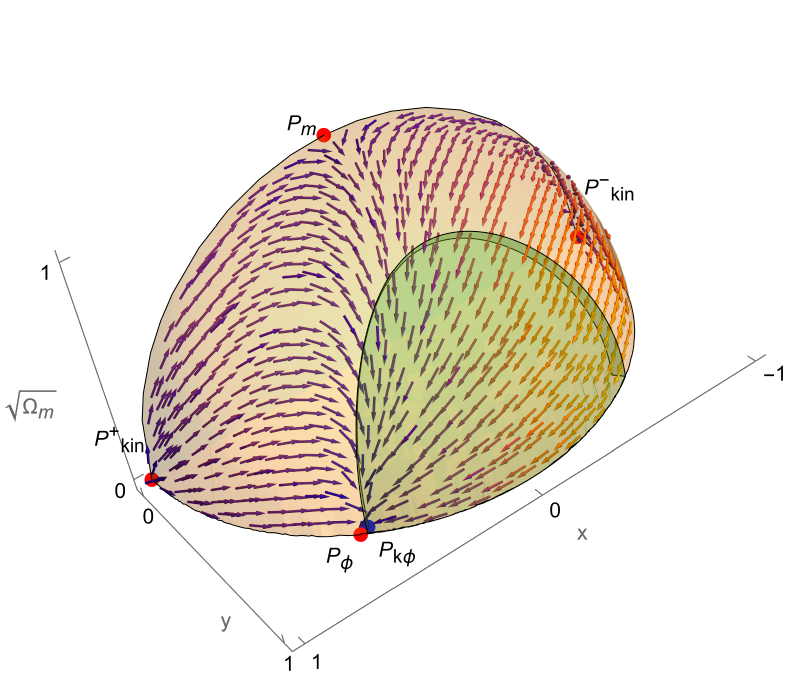}\caption{$\lambda=\sqrt{2}+0.05$}\label{fig:PhaseSpaceL1}
\end{subfigure}
\caption{Cosmological solutions as vector fields in the phase space of figure \ref{fig:PhaseSpace}, where radiation is turned off ($u=0$). Only parts of solutions within the slice $0\leq \Omega_k \leq 0.1$ (close to the spherical boundary) are represented. Different values of $\lambda$ are chosen, with which we show the change in the solutions and in the fixed points that exist in each case. We refer to table \ref{tab:fixedpoints} for the fixed point existence conditions, and to table \ref{tab:fixedpointstabwithoutr} for their stability. We do not necessarily represent all fixed points but only the relevant ones, for a better readability. As in figure \ref{fig:PhaseSpace}, the color of the fixed points correspond to their stability, while the green zone corresponds to the accelerating region. We refer to the main text for physical comments on the solutions. Further illustrations are provided in figure \ref{fig:PhaseSpaceLbis}.} \label{fig:PhaseSpaceL}
\end{center}
\end{figure}

\begin{figure}[H]
\begin{center}
\begin{subfigure}[H]{0.48\textwidth}
\includegraphics[width=\textwidth]{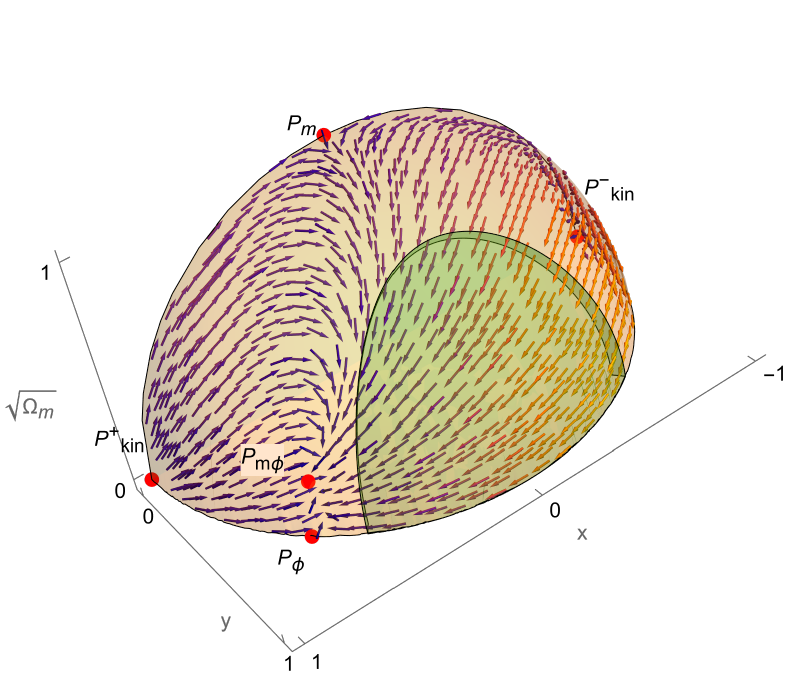}\caption{$\lambda=\sqrt{3}+0.05$}
\label{fig:PhaseSpaceL2}
\end{subfigure}\\
\begin{subfigure}[H]{0.48\textwidth}
\includegraphics[width=\textwidth]{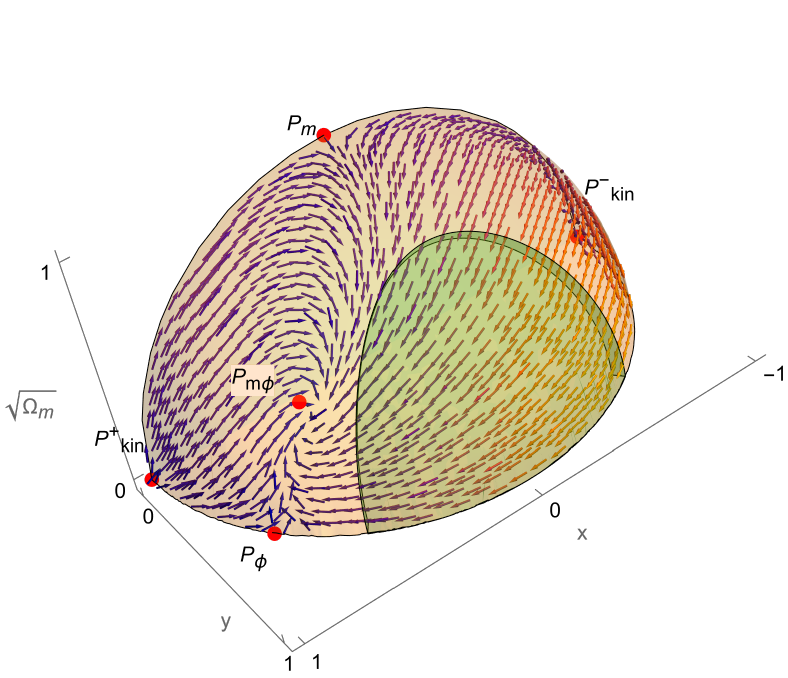}\caption{$\lambda=2$}\label{fig:PhaseSpaceL3}
\end{subfigure}\quad
\begin{subfigure}[H]{0.48\textwidth}
\includegraphics[width=\textwidth]{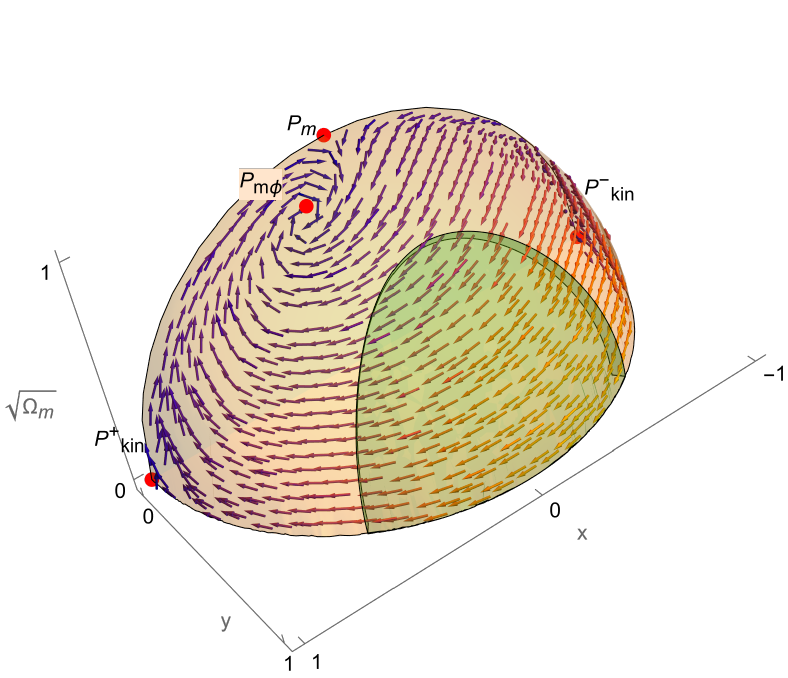}\caption{$\lambda=5$}\label{fig:PhaseSpaceL4}
\end{subfigure}
\caption{Illustrations of cosmological solutions as vector fields in the phase space of figure \ref{fig:PhaseSpace}: see figure \ref{fig:PhaseSpaceL} for a detailed description.}\label{fig:PhaseSpaceLbis}
\end{center}
\end{figure}

Having an overall view of solutions can be achieved by depicting solution vector fields, on slices of the multidimensional phase space. Since the system \eqref{eq:system} has different dependencies on $x,y$, the phase space illustration of figure \ref{fig:PhaseSpacer} is not suited to a vector field approach, and we rather use the phase space of figure \ref{fig:PhaseSpace}, for which radiation is turned off ($u=0$). There, we pick a slice where $0 \leq \Omega_k \leq 0.1$,
which will be of interest later. That is, we pick  a thin spherical shell of radius $\sqrt{1-\Omega_k}$ in the $x,y,\sqrt{\Omega_m}$ space, recalling $z^2 = 1- x^2 -y^2 - \Omega_m$. The corresponding solution vector fields are depicted in figure \ref{fig:PhaseSpaceL} and \ref{fig:PhaseSpaceLbis} for different values of $\lambda$. Note that other slices in $\Omega_k$ can be depicted, but we do not learn much more from them.

Several observations can be made about the solutions in figure \ref{fig:PhaseSpaceL} and \ref{fig:PhaseSpaceLbis} and how they change with $\lambda$. We first note that all solutions start at $\Pkin$, as expected since those are the fully unstable points for most $\lambda$ values; only for $\lambda=5$ $(>\sqrt{6}$), solutions start at $\Pkinm$ while we observe the saddle behaviour of $\Pkinp$. Depending on $\lambda$, the solutions either end at the stable point when it is present in the slice ($\Pkp, \Pp$), or we stop seeing the solutions in this slice when they are close to $\Pmp$: they then go inside the ball towards $\Pkp$. We also note the change in stability properties of $\Pmp$, which exhibits an inspiraling approach for $\lambda > \sqrt{\frac{24}{7}}$, as can be seen in figure \ref{fig:PhaseSpaceL3} and \ref{fig:PhaseSpaceL4}.

Another observation to be emphasized is that from the start of the solutions until either their end point, or until $\Pmp$, they remain within this specific slice: physically this means that having a low value of $\Omega_k$ at some point in the solution implies to always have a low $\Omega_k$ in the past. As will be discussed, this can be understood from the dependence on the scale factor $a$ of the different energy density contributions; this will be important.

Finally, we can focus on solutions passing close to $\Pm$, meaning having enough matter domination, and then through the acceleration region, as our universe does. Such solutions do not seem to exist for high $\lambda$ values. Indeed, they seem barely existent at $\lambda=2$ on figure \ref{fig:PhaseSpaceL3}, while they exist for lower $\lambda$ values. We will make this observation more precise in the following, as we turn to realistic solutions.

\section{Realistic cosmological solutions without curvature}\label{sec:solutions}

A question motivating this work is whether cosmological solutions to the previous dynamical system, for a given $\lambda$ value, can reproduce the known evolution of our universe. As a warm-up, we focus in this section on solutions without spatial curvature, $\Omega_k = z^2 =0$ (see e.g. \cite{SavasArapoglu:2017pyh} for a past study of this system); we include curvature in the next section. Mathematically, for a given $\lambda$, a solution is fully determined by specifying one of its points $(x,y,z,u)$ at a given time $t=t_0$: these are ``initial conditions''. As seen from \eqref{varOx}, a point $(x,y,z,u)$ can also be expressed in terms of the $(\Omega_\phi, \Omega_k, \Omega_r)$ and $w_{\phi}$ (with $\Omega_m=1-\Omega_\phi - \Omega_k -  \Omega_r$).   If, for a given $\lambda$, we can find a solution passing through a point $\Omega_{n}, w_{\phi}$ that is consistent with cosmological observations today (denoted $\Omega_{n0}, w_{\phi0}$ with $t_0=0$ today), then we stand a chance of having a realistic solution.

There are however a couple of immediate issues with this strategy. Firstly, observational constraints on $\Omega_{n0}, w_{\phi0}$ are generally model dependent (and in particular here, $\lambda$ dependent); their values are usually inferred by fitting a given cosmological model to the observational data.  Take for instance the flat $\Lambda$CDM model: it corresponds in our framework to the limit $\lambda=0$ and to solutions  that have $z=0$, together with $x=0 \, \Leftrightarrow\, w_{\phi}=-1$. For example, the Planck Collaboration's  best fits and 68\% confidence levels for $\Lambda$CDM model, using CMB data, are $\Omega_{m0} = 0.3153 \pm 0.0073$ and $ \Omega_\Lambda = 0.6847 \pm 0.0073$ \cite{Planck:2018vyg}. Turning to the exponential quintessence model with $\lambda\neq0$ and $\Omega_k=0$, we will review observational analyses in the literature in subsection \ref{sec:obs}, where we will see that constraints on the $\Omega_{n0}$ are relatively tight (a few $1\%$) \cite[Fig.3]{Schoneberg:2023lun}, whilst constraints on $w_{\phi0}$ are quite loose. Therefore, in the following we take as fiducial values for the flat exponential quintessence model:
\be
\Omega_{{\rm \phi0}} = 0.6850 \ ,\ \Omega_{m0} = 0.3149 \ ,\ \Omega_{r0} =0.0001 \ ,\ \Omega_{k} =0 \ , \label{OmegaLCDM}
\ee
and allow $w_{\phi0}$ to vary as required. Although our qualitative results will not change when varying the fiducial values, the precise numerical bounds we quote will be sensitive to them.

A second worry with the above strategy to find realistic solutions is whether a solution is guaranteed to be realistic in the past, with the requisite epochs of radiation and matter domination, once we fix the values for $\Omega_{n0}$ today.  Although this is automatic for the $\Lambda$CDM model, the quintessence model will have many different histories possible for a given $\lambda$ and $\Omega_{n0}$ today, depending on $w_{\phi0}$. In particular, there will be cosmological solutions that start in a kination epoch, at one of the unstable fixed points $\Pkin$, and reach the attractor fixed point, $P_\phi$ or $P_{m\phi}$, passing through today's density parameters $\Omega_{n0}$ but without ever approaching radiation or matter domination, i.e.~without ever passing through a region close to $P_r$ or $P_m$.

In the following, we will turn these two problems around: for a given $\lambda$, we will first ensure a realistic past (radiation and matter domination), together with observationally appropriate values for $\Omega_{n0}$. Proceeding in this way will fix the value of $w_{\phi0}$, as we will explain in section \ref{sec:radmatdom}. Whether that value is in agreement with observations requires a full cosmological fit with the data \cite{BBMPTZ, Alestas:2024gxe}, which is beyond the scope of the present paper, but we will for now give a minimal criterion on this question by requiring acceleration today. In turn, this will constrain admissible values of $\lambda$; indeed, as we will show in section \ref{sec:upperbound}, we will find an upper bound for $\lambda$ from the minimal constraints of past radiation and matter domination and an accelerated expansion today. In section \ref{sec:obs}, we will summarize the candidate realistic solutions for this quintessence model, and compare them to observational constraints from the literature in absence of curvature. In section \ref{sec:accphase}, we discuss the acceleration phase in these solutions.  This will prepare the ground to include curvature.

\subsection{Radiation and matter domination, and $w_{\phi0}$}\label{sec:radmatdom}

As motivated above, we use as a starting point the following minimal requirements:
\be
\text{``Realistic'' solution:}\qquad \exists\, t<0 \ \text{s.t.} \ \Omega_r(t) \geq 0.5 \ ,\quad \text{and}\ \Omega_{n0} \ \checkmark \ \text{obs.} \ . \label{realsol}
\ee
In other words, we require from a candidate realistic cosmological solution that it achieved some radiation domination in the past and that its values for $\Omega_n$ today appear in agreement with observational data. For the latter, we will take, as argued above, the fiducial values \eqref{OmegaLCDM}; as already emphasised, our qualitative results will not depend on the precise values of the $\Omega_{n0}$ although our numerical predictions do.  For the epoch of radiation domination, one might impose that it starts at least around 20 e-folds before today, in time for BBN with $z_{BBN} \approx 4 \cdot 10^8\sim e^{20}$; however, in a first approximation we will not need to specify the precise amplitude of the domination, nor its duration.

Interestingly, whilst the requirements in \eqref{realsol} appear minimal, they completely fix two important features in the solutions. Firstly, requiring some degree of radiation domination in the past, $\Omega_r(t) \geq 0.5$, turns out to imply a subsequent epoch of matter domination, with amplitude and duration automatically similar to that of $\Lambda$CDM.  Secondly, having fixed $\Omega_{n0}$, finding solutions with past radiation domination automatically fine tunes the value of $w_{\phi0}$ (and increasing the fine-tuning increases the amplitude and duration of domination). We spend the rest of this subsection explaining these two points.\\

That matter domination is automatically obtained is a simple consequence of how the radiation, matter and dark energy densities evolve with the scale factor $a$. Let us make this more precise, and reproduce the evolution of the $\Omega_n$, following e.g.~\cite{Jaffe:2012}. For future purposes, we include curvature in this derivation.

Given that $\rho_r \propto a^{-4}$, $\rho_m \propto a^{-3}$, $\rho_k \propto a^{-2}$, with proportionality factors assumed constant, one can rewrite these quantities at any time in terms of those today
\be
\rho_r =  \rho_{r0} \left(\frac{a_0}{a} \right)^4 \ ,\ \rho_m =  \rho_{m0} \left(\frac{a_0}{a} \right)^3 \ ,\ \rho_k =  \rho_{k0} \left(\frac{a_0}{a} \right)^2 \ .\label{rhos}
\ee
Using this and the $\Omega_{n0}$, the first Friedmann equation \eqref{eq:Fried} can be rewritten as follows
\be
\frac{H^2}{H_0^2} = \left(\Omega_{r0} \left(\frac{a_0}{a} \right)^4 + \Omega_{m0} \left(\frac{a_0}{a} \right)^3  + \Omega_{k0} \left(\frac{a_0}{a} \right)^2  + \Omega_{\phi0} \frac{\rho_{\phi}}{\rho_{\phi0}} \right) \ \Leftrightarrow \ H^2 = H_0^2\, (\dots) \ ,
\ee
where we use the latter parentheses in the following. Combining the above equations, the various $\Omega_n$ at any time can be expressed as follows
\be
\Omega_r =  \frac{\Omega_{r0}}{(\dots)} \left(\frac{a_0}{a} \right)^4 \ ,\ \Omega_m =  \frac{\Omega_{m0}}{(\dots)} \left(\frac{a_0}{a} \right)^3 \ ,\ \Omega_k =  \frac{\Omega_{k0}}{(\dots)} \left(\frac{a_0}{a} \right)^2 \ , \ \Omega_{\phi} = \frac{\Omega_{\phi 0}}{(\dots)} \frac{\rho_{\phi}}{\rho_{\phi0}}  \ .\label{Omegas}
\ee

\begin{figure}[h]
\centering
\includegraphics[width=0.55\textwidth]{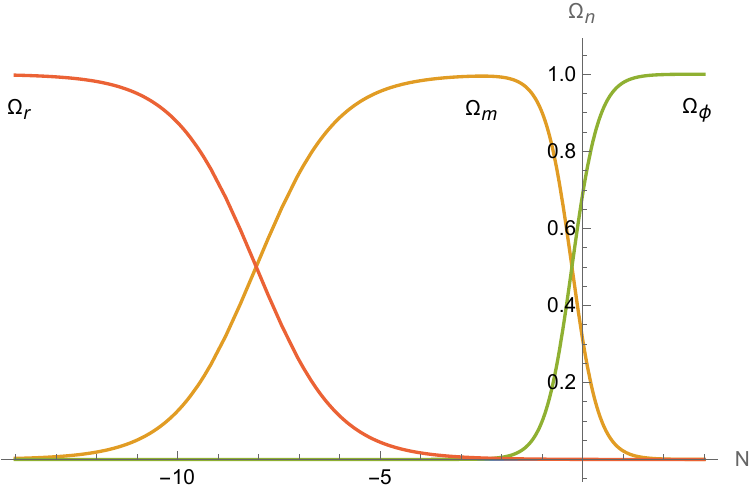}
\caption{Evolution of the $\Omega_n$ in terms of $N= \ln \frac{a}{a_0}$, for $\Lambda$CDM, where $\Omega_{\phi}$ stands for the dark energy component $\Omega_{\Lambda}$. The values \eqref{OmegaLCDM} for today's universe are those at $N=0$.}\label{fig:EvolutionLCDM}
\end{figure}

We now restrict ourselves to $\Omega_k = \Omega_{k0}=0$. To provide a first illustration of the evolution of the $\Omega_n$, let us consider $\Lambda$CDM, setting $\frac{\rho_{\phi}}{\rho_{\phi0}} =1$ to reproduce the cosmological constant contribution. The evolution as functions of $N=\ln \frac{a}{a_0}$ is then depicted in Figure \ref{fig:EvolutionLCDM}. We see there the well-known three successive domination phases.

Moving away from $\Lambda$CDM to quintessence, the evolution of $\rho_{\phi}$ and $\Omega_{\phi}$ are more complex (including phases like kination,\footnote{\label{foot:kin}Recall that $w_{\phi}=1$ for kination (see table \ref{tab:n})  and in general $\rho \sim a^{-3(1+w)}$.} with $\rho_{\phi} \sim a^{-6}$, frozen quintessence, with $\rho_{\phi} \sim a^0$, or something in between). With a few assumptions, however, we can see that the main characteristics of the matter domination phase are still fixed. We imposed in \eqref{realsol} that the realistic solutions admit a radiation domination phase in the past. Let us assume that during and shortly after this epoch, $\Omega_{\phi}$ remains negligible; this will be explicitly verified in the solutions considered. Then radiation domination has to be followed by matter domination, and the moment at which the equality $\Omega_r =\Omega_m$ is obtained, is given by
\be
N_{r=m} = \ln \frac{\Omega_{r0}}{\Omega_{m0}} \approx -8.1 \ .
\ee
We can similarly estimate the moment at which the equality $\Omega_m = \Omega_{\phi}$ is obtained, since by this time $\Omega_r$ will be negligible (given its $\sim a^{-4}$ decay). We then obtain
\be
N_{m=\phi} = \frac{1}{3}\ln \frac{\Omega_{m0}}{\Omega_{\phi0}}\frac{\rho_{\phi0}}{\rho_{\phi}}\,,
\ee
where $\rho_\phi$ above should be evaluated at the time when $\Omega_m=\Omega_\phi$. For $\Lambda$CDM, $\rho_{\phi}=\rho_{\phi0}$, giving the value $N_{m=\phi} =-0.26$. In our numerical quintessence solutions, we will find similar values for this moment. To conclude, the start of the matter domination in the realistic solutions \eqref{realsol} is {\sl automatically} fixed to a value close to that of $\Lambda$CDM, and we will  find similar values for its duration.

We may also determine the maximum of $\Omega_m$ during the matter domination phase. Doing so requires us to evaluate $\frac{\rho_{\phi}}{\rho_{\phi0}}$ and its derivative with respect to $a$ at this maximum. If we can set $\frac{\rho_{\phi}}{\rho_{\phi0}} \approx 1$ and neglect its derivative at this moment, as in $\Lambda$CDM, then we obtain the following expressions and value for the maximum
\be
\!\!\! N_{m\, {\rm max}} = \ln\frac{ a_{{\rm max}}}{ a_0} = \frac{1}{4} \ln \frac{\Omega_{r0}}{3 \Omega_{\phi 0}} \approx -2.5 \, ,\ \ \Omega_{m\, {\rm max}}= \frac{\Omega_{m0}}{ \Omega_{m0}  + 4 \Omega_{\phi 0} \left(\frac{\Omega_{r0}}{3 \Omega_{\phi 0}} \right)^{\frac{3}{4}} } \approx 0.995 \ .\label{Omegammax}
\ee
In fact, during matter domination, we can expect the scalar field to be frozen by Hubble friction so that indeed $\frac{\rho_{\phi}}{\rho_{\phi0}}$ remains roughly constant; we will also see empirically that the maximum reached is $\Omega_{m\, {\rm max}} \approx 1$.

\begin{figure}[H]
\begin{center}
\begin{subfigure}[H]{0.48\textwidth}
\includegraphics[width=\textwidth]{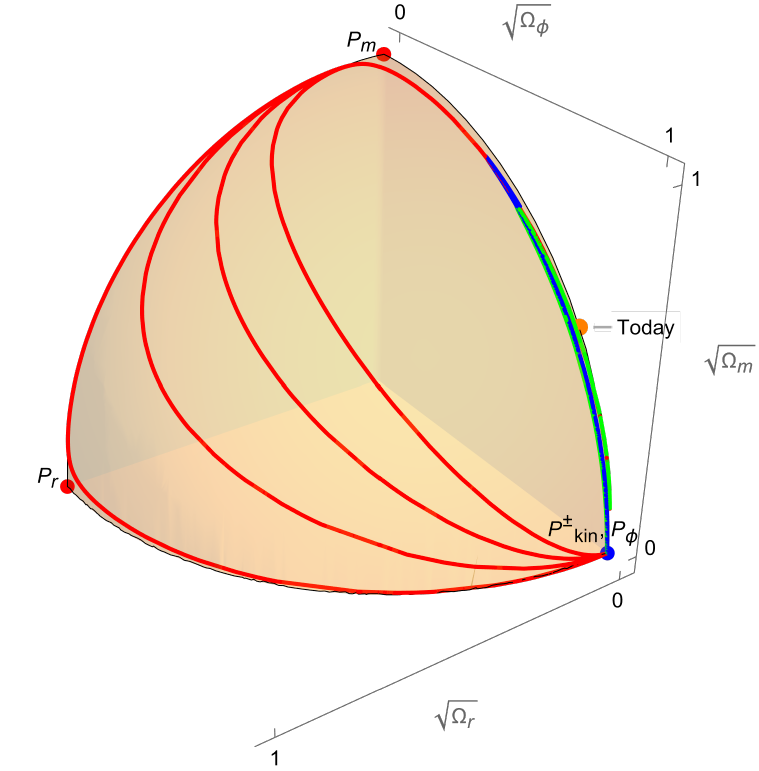}\caption{}\label{fig:Solr0}
\end{subfigure}\quad
\begin{subfigure}[H]{0.48\textwidth}
\includegraphics[width=\textwidth]{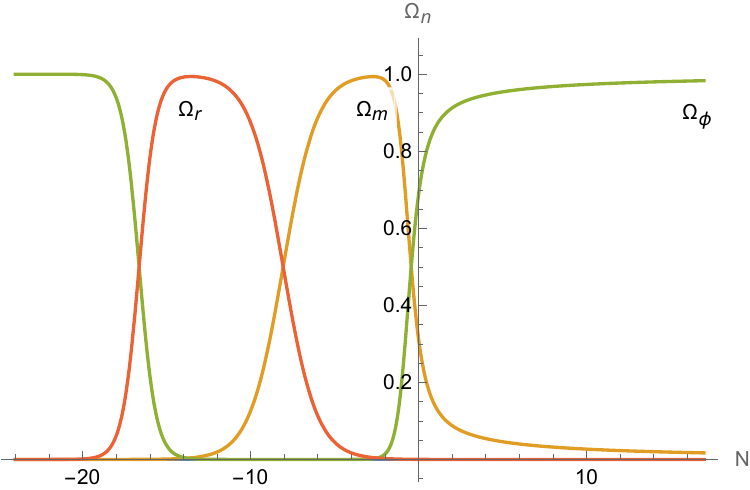}\caption{$w_{\phi0}=-0.51073606$}\label{fig:SolrO1}
\end{subfigure}\\
\begin{subfigure}[H]{0.48\textwidth}
\includegraphics[width=\textwidth]{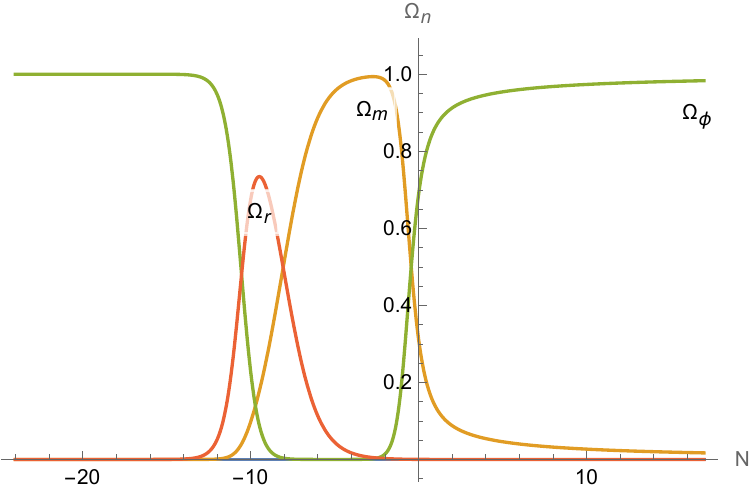}\caption{$w_{\phi0}=-0.510737$}\label{fig:SolrO2}
\end{subfigure}\quad
\begin{subfigure}[H]{0.48\textwidth}
\includegraphics[width=\textwidth]{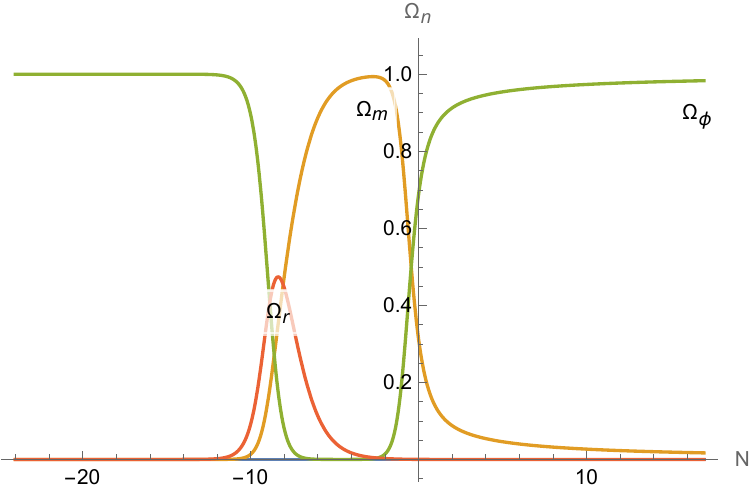}\caption{$w_{\phi0}=-0.510731$}\label{fig:SolrO3}
\end{subfigure}
\caption{Cosmological solutions in the phase space illustration of figure \ref{fig:PhaseSpacer} for $\lambda=\sqrt{3}$, together with their $\Omega_n$. The solutions start at $\Pkin$ and end at $\Pp$, which are merged in this illustration of phase space. We require the solutions to pass by a point (indicated as ``Today'' in figure \ref{fig:Solr0}) corresponding to today's universe, defined by the $\Omega_{n0}$ values  \eqref{OmegaLCDM}. The solutions differ by the value given to $w_{\phi0}$: the green and blue solutions in figure \ref{fig:Solr0} have respectively $w_{\phi0}=-0.4$ and $-0.7$, and barely allow for any radiation, nor reach a matter domination with very high amplitude. On the contrary, the red solutions allow for radiation (and matter) domination. Among these solutions, the one very close to $\Prr$ has $w_{\phi0}=-0.51073606$, while moving away from it, we find solutions respectively with $w_{\phi0}=-0.510737$, $-0.510731$, $-0.510720$. We give the evolution of the $\Omega_n$ for the first three red solutions, where $N=0$ corresponds to today's universe. We refer to the main text for further comments on the solutions.}\label{fig:Solr}
\end{center}
\end{figure}

Having provided some analytic arguments as to why considering the solutions \eqref{realsol} to the quintessence model {\sl automatically} provides a matter domination phase with appropriate duration and amplitude, we now verify this numerically. This is illustrated in figure \ref{fig:Solr}.

In figure \ref{fig:Solr0} we show  the solutions, in red, that pass through an epoch of radiation domination, each with a different amplitude and duration.  All these solutions subsequently pass very close to $\Pm$; hence, we infer that the radiation domination phase is always followed by a matter domination, whose amplitude reaches $\Omega_{m\, {\rm max}} \approx 1$. This is confirmed in the $\Omega_n$ curves in figure \ref{fig:Solr}: any kind of radiation domination in the past apparently leads to the same matter domination phase, in terms of duration and amplitude. The analytic explanation provided above thus seems to apply here: among the assumptions mentioned, we see in particular that $\Omega_{\phi}$ is negligible at the time of radiation - matter equality.

\begin{figure}[H]
\begin{center}
\begin{subfigure}[H]{0.7\textwidth}
\includegraphics[width=\textwidth]{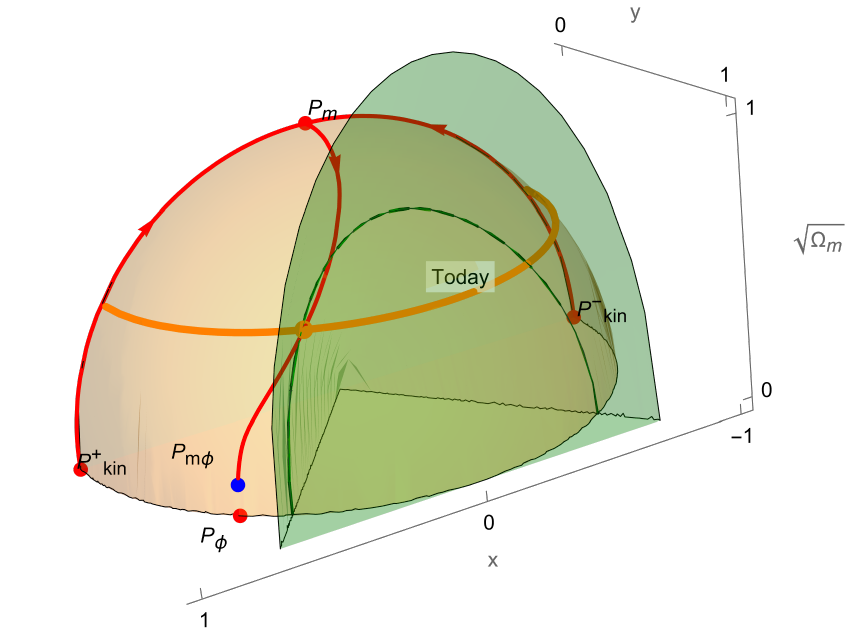}\caption{}\label{fig:Solm0}
\end{subfigure}\\
\begin{subfigure}[H]{0.48\textwidth}
\includegraphics[width=\textwidth]{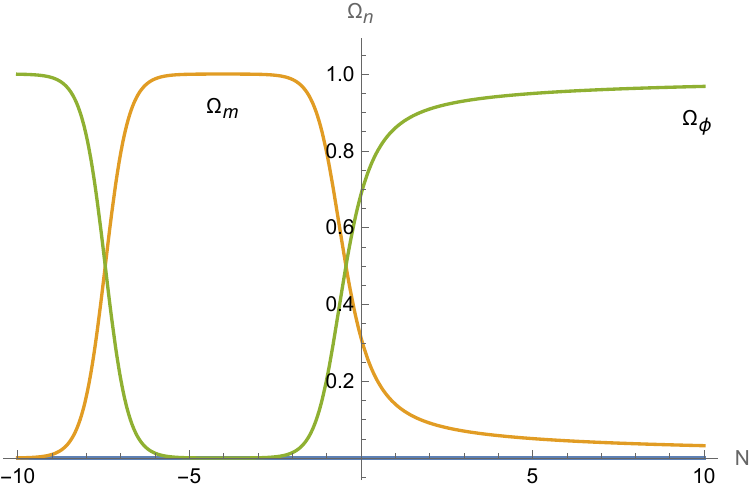}\caption{$\Omega_n$}\label{fig:SolmO}
\end{subfigure}\quad
\begin{subfigure}[H]{0.48\textwidth}
\includegraphics[width=\textwidth]{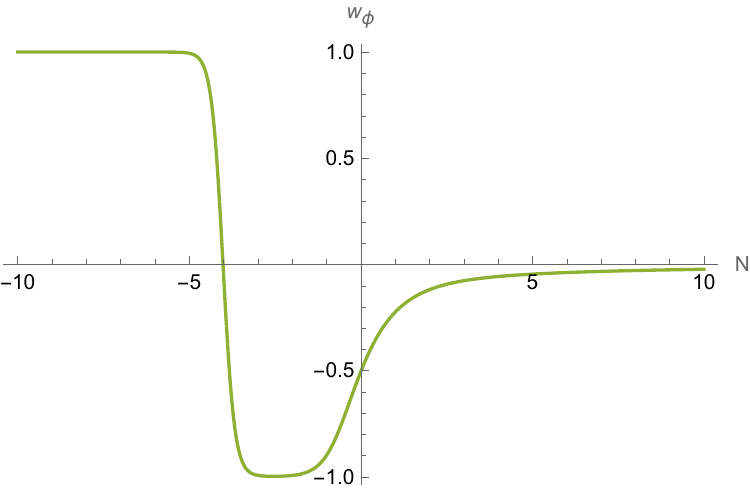}\caption{$w_{\phi}$}\label{fig:Solmwp}
\end{subfigure}
\caption{Two cosmological solutions (red curves) in the phase space of figure \ref{fig:PhaseSpace}, where radiation is turned off ($u=0$), for $\lambda= \sqrt{3}+0.01$. The two solutions are required to pass by a point corresponding to today's universe ($N=0$), defined by $\Omega_{\phi0}=0.685=1-\Omega_{m0}$, and by a certain value of $w_{\phi0}$. The (half) circle corresponding to today's value $x^2+y^2=\Omega_{\phi0}=0.685$ is depicted in orange in figure \ref{fig:Solm0}, while $w_{\phi0}$ corresponds to an angle on the $(x,y)$ plane; fixing $w_{\phi0}$ thus picks an orange point on that circle, through which the solutions are required to pass. The solution starting at $\Pkinp$ has $w_{\phi0} =-0.50410$ while the one starting at $\Pkinm$ has $w_{\phi0} =-0.50416$. Those values are tuned to obtain the start of matter domination at around $N\approx -8$, as can be seen in figure \ref{fig:SolmO}. The evolution of $w_{\phi}$ along the solutions are shown  in figure \ref{fig:Solmwp}. After matter domination (passing close to $\Pm$), the two solutions cannot be distinguished in figure \ref{fig:Solm0}. We refer to the main text for further comments.}\label{fig:Solm}
\end{center}
\end{figure}

Having established that the realistic solutions \eqref{realsol} inevitably have a matter dominated phase that reaches $\Omega_{m\, {\rm max}} \approx 1$, and that starts around $N\approx -8$, it will sometimes be convenient to use the subsystem where radiation is turned off ($u=0$) to describe those solutions during and after the matter dominated phase. Indeed, given the constraint \eqref{Omegaxyzu}, if $\Omega_m\approx 1$, then all other variables are negligible, in particular $u\approx0$. At this point of maximal matter domination, we could then cut and glue a solution where radiation is turned on in the past before that point, while neglecting it ($u=0$) after that point. Using this subsystem or invariant subspace allows us to use the phase space illustration of figure \ref{fig:PhaseSpace}, as done e.g. in figure \ref{fig:Solm}, to which we will shortly turn.\\

So far we have discussed the first important feature of the solutions characterized in \eqref{realsol}, namely their inevitable epoch of matter domination.  Let us now discuss the second important feature, illustrated in figures \ref{fig:Solr} and \ref{fig:Solm}, namely that the value of $w_{\phi0}$ is tuned (to the $4^{\rm th}$ digit for a given $\lambda$ and given $\Omega_{n0}$) by requiring the past epoch of radiation domination specified in \eqref{realsol}. The level of precision necessary in $w_{\phi0}$ to reach a past radiation domination, and thus a realistic solution, certainly goes beyond observational precision.  Nevertheless, some important observations can be made.  Figure \ref{fig:Solr} shows that the amplitude and duration of the radiation domination is increased by increasing the degree of fine tuning in $w_{\phi0}$: at a certain (unknown) critical value of $w_{\phi0}$ we expect the duration of radiation domination to become infinite, and this value can be approached from above or below.  Figure \ref{fig:Solm} illustrates how solutions with a value of $w_{\phi0}$ that is higher (less negative) than the critical value originated from $\Pkinp$, whilst solutions with a lower value of $w_{\phi0}$ start at $\Pkinm$. In addition, the fixing of $w_{\phi0}$ corresponds, in the phase space of figure \ref{fig:Solm}, to fixing the angle in the $(x,y)$ plane at which the solutions cross the orange circle ($x^2+y^2=\Omega_{\phi0}$): this point represents the universe today.\footnote{Strictly speaking, there is an ambiguity in the point for the universe today: indeed, $\Omega_{\phi0}$ and $w_{\phi0}$ only fix $x^2, y^2$. While we restrict to $y\geq0$ for an expanding universe, there is an ambiguity in the sign of $x$. As can be seen in figure \ref{fig:PhaseSpaceL} and \ref{fig:PhaseSpaceLbis}, solutions that pass close to $\Pm$, i.e.~matter domination then turn to $x\geq 0$, so choosing this sign for $x$ seems the right choice for a realistic present universe, and this is the sign we choose.}  Note that the two solutions shown are almost indistinguishable after matter domination, including their values for $w_{\phi0}$.

In the following, this will be a general feature; for a given $\lambda$ value, there are always two solutions with very close values of $w_{\phi0}$ (identical to the $4^{\rm th}$ digit),  which realise radiation and matter domination (more precisely, there is a continuum of solutions between those two; fixing the duration of the radiation domination phase breaks this degeneracy to only the two solutions).

\begin{figure}[H]
\begin{center}
\includegraphics[width=0.7\textwidth]{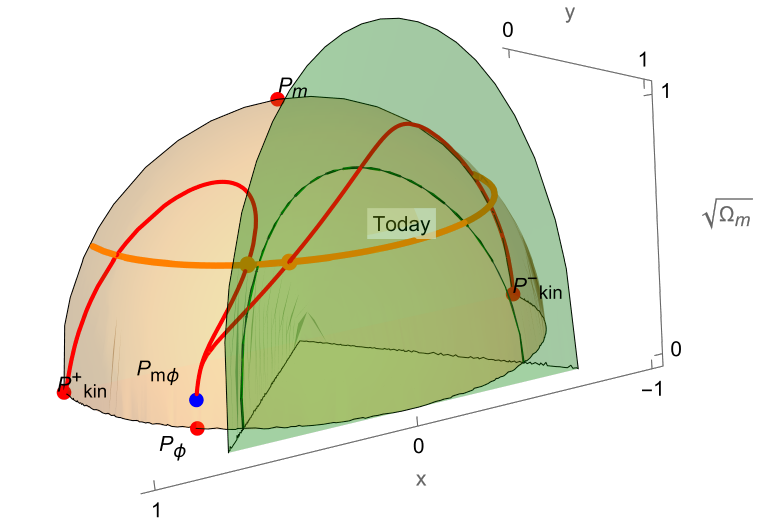}
\caption{Two cosmological solutions (red curves) in the phase space of figure \ref{fig:PhaseSpace}, where radiation is turned off ($u=0$), for $\lambda= \sqrt{3}+0.01$. As in figure \ref{fig:Solm}, the two solutions are required to pass by an orange point corresponding to today's universe with $\Omega_{\phi0}=0.685=1-\Omega_{m0}$ and a value of $w_{\phi0}$. The solution starting at $\Pkinp$ has $w_{\phi0} =-0.4$ while the one starting at $\Pkinm$ has $w_{\phi0} =-0.7$. Their matter domination is such that $\Omega_{m\, {\rm max}} < 0.8$ and lasts less than 2 e-folds. Those solutions correspond to the green ones in figure \ref{fig:Solr}. We refer to the main text for further comments.} \label{fig:Solmnotok}
\end{center}
\end{figure}

It is useful to contrast the above solutions in figure \ref{fig:Solm} to the solutions in figure \ref{fig:Solmnotok}, where we choose different values of $w_{\phi0}$. Such solutions exist (as anticipated in figure \ref{fig:PhaseSpaceL}) and pass by a point corresponding to today's density parameters $\Omega_{n0}$, but they do not pass close to $\Pm$. In other words, they do not achieve a past matter dominated phase with $\Omega_{m\, {\rm max}} \approx 1$ and an appropriate duration. As a consequence, such solutions cannot be continued to a solution with radiation domination in the past and they cannot be realistic.

Based on the above, let us now assume some benchmark values for $\lambda$ and give the corresponding values for $w_{\phi0}$, obtained by taking $\Omega_{\phi0}=0.685=1-\Omega_{m0}$ and requiring a matter dominated phase that starts no later than $N\approx -8$. For simplicity, we turned off radiation, but as argued, the requirement of matter domination is necessary for the possibility of radiation domination in the past: the values of $w_{\phi0}$ are then those of the candidate realistic solutions \eqref{realsol}. We obtain\footnote{The case $\lambda=0$ should be considered as a limit, since we assumed $\lambda>0$. For $\lambda=0$, if we pick the strict value $w_{\phi 0}=-1$, we cannot obtain a solution with $x\neq0$, in particular not starting at $\Pkin$; moving slightly away from $-1$ provides the same type of realistic solution as before.}
\bea
\lambda= 0:\quad & w_{\phi 0} \approx -1.0000 \ ,\nn\\
\lambda= 1:\quad & w_{\phi 0} \approx -0.8486 \ ,\nn\\
\lambda= \sqrt{2}:\quad & w_{\phi 0} \approx -0.6874 \ ,\nn\\
\lambda= \sqrt{8/3}:\quad & w_{\phi 0} \approx -0.5719 \ , \label{wphi0values}\\
\lambda= \sqrt{3}:\quad & w_{\phi 0} \approx -0.5107 \ ,\nn\\
\lambda=2:\quad & w_{\phi 0} \approx -0.3028 \ .\nn
\eea
Note that for $\lambda=2$, ensuring past matter domination requires $w_{\phi 0}$ to take a value that cannot drive an accelerated expansion today (indeed, $w_{\text{eff}\,0} \approx w_{\phi 0} \Omega_{\phi 0}$, so requiring acceleration today, $w_{\text{eff}\,0}<-\frac13$, whilst recalling that $0<\Omega_{\phi 0} < 1$, implies $w_{\phi 0} < -\frac13$). A related comment on figure \ref{fig:Solm}, where $\lambda=\sqrt{3}+0.01$, is that the point corresponding to the universe today turns out, accidentally, to be very close to the boundary of the acceleration region; we verify this by computing $w_{\rm eff} \approx -0.3453$. These observations will play an important role when studying higher $\lambda$ values, to which we now turn.

\subsection{An upper bound on $\lambda$}\label{sec:upperbound}

\begin{figure}[h]
\begin{center}
\includegraphics[width=0.7\textwidth]{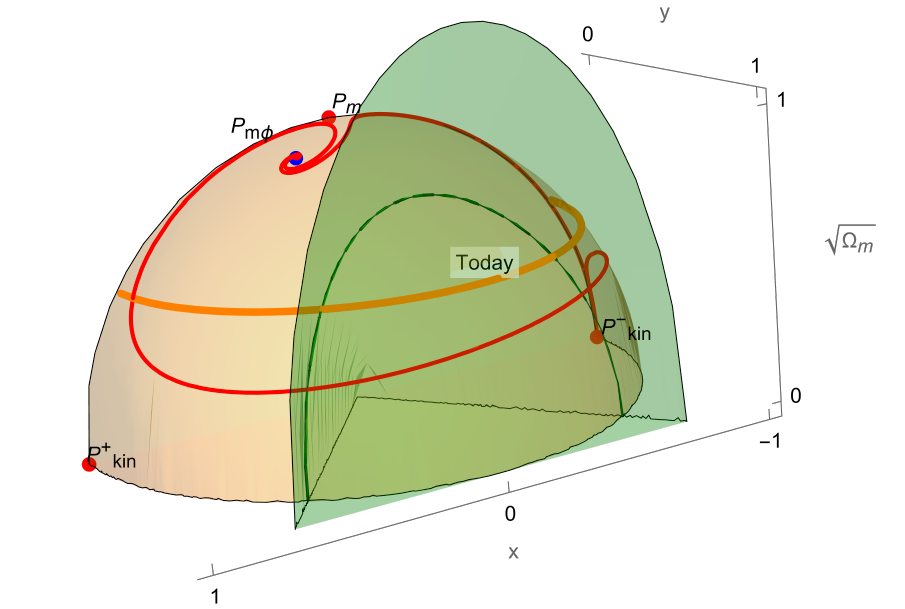}
\caption{Two cosmological solutions (red curves) in the phase space of figure \ref{fig:PhaseSpace}, where radiation is turned off ($u=0$), for $\lambda= 5$. The two solutions are required to pass by a point very close to matter domination ($\Omega_{\phi}=0.01=1-\Omega_{m}$, $w_{\phi}=-0.6$, giving two opposite values of $x$). The orange circle corresponding to today's universe is the same as in figure \ref{fig:Solm}. This circle is not crossed by the solution after matter domination. We refer to the main text for more comments.} \label{fig:Solhighlambda}
\end{center}
\end{figure}

When trying to find realistic solutions \eqref{realsol} for large values of $\lambda $  we fail, as we now explain. We looked numerically for an appropriate $w_{\phi 0}$ value that would provide solutions with a radiation dominated phase in the past and passes by appropriate $\Omega_{n0}$, as in figure \ref{fig:Solr}, for large $\lambda$ values: we could not find any such $w_{\phi 0}$ after $\lambda \gtrsim 2$. Considering again that the radiation dominated phase is inevitably followed by a matter dominated one, with $\Omega_{m\, {\rm max}} \approx 1$, we can use the phase space picture of figure \ref{fig:PhaseSpace}, and find an explanation to this failure by combining figures \ref{fig:PhaseSpaceLbis} and \ref{fig:Solm}. Indeed, as we increase $\lambda$, the (attractor) fixed point $\Pmp$ moves upwards on the sphere towards $\Pm$, crossing the circle $x^2+y^2=\Omega_{\phi0}$ when $\lambda = \sqrt{3/\Omega_{\phi 0}}$ (see table \ref{tab:fixedpoints}). Once $\Pmp$ is above this circle, any solution that passes very close to $\Pm$ will subsequently approach the attractor $\Pmp$ directly without passing again through the circle. In that case, today's universe cannot be reached after matter domination. We illustrate this situation in figure \ref{fig:Solhighlambda} with $\lambda=5$.

Requiring that $\Pmp$ stays below the circle, thus allowing matter domination to be followed by today's density parameters $\Omega_{n0}$, leads to the following upper bound for $\lambda$
\be
\lambda \leq \sqrt{\frac{3}{\Omega_{\phi0}}} \approx 2.093 \ , \label{lambdaup1}
\ee
corresponding indeed to the observed bound close to 2. Note that the inspiraling approach may allow a solution to still cross the circle for a slightly higher value of $\lambda$.\\

As observed already for solutions of figure \ref{fig:Solm}, or as can be seen in figure \ref{fig:Solmnotok}, there is no guarantee that the point at which the solution has today's energy densities (i.e. crosses the orange circle) be in the acceleration region. Actually, in the above limiting case where this point is exactly $\Pmp$, one has $x=y$, leading to $w_{\phi0}=0$, which clearly lies outside the acceleration region. Requiring that in a realistic solution, the universe today is accelerating, should thus give a stronger upper bound than the previous one: we now turn to this.

In the system considered here, turning off radiation for simplicity, we read from \eqref{weffO} that $w_{\rm eff} = w_{\phi} \Omega_{\phi}$. A bound for today's acceleration is thus easily obtained as
\be
w_{\phi0} \leq -\frac{1}{3\, \Omega_{\phi0}} \approx - 0.4866 \ .
\ee
The benchmark values \eqref{wphi0values} indicate a corresponding $\lambda$ slightly above $\sqrt{3}$; a more detailed empirical analysis on realistic solutions gives the upper bound\footnote{See \cite{Cline:2001nq} for a similar bound $1.4 < \lambda < 1.6$ where the lower bound comes from requiring no event horizon and the upper bound is derived from the observational constraint at the time, $\Omega_{\phi0} > 0.5$.}
\be
\lambda \leq \sqrt{3} + 0.0362 \approx 1.7683 \ .\label{lambdaboundacc}
\ee
Note that the precise value for this bound is dependent on the values taken for $\Omega_{n0}$. Taking for instance the lowest $\Lambda$CDM value of $\Omega_{\Lambda}$, using the error bars from \cite{Planck:2018vyg}, namely $\Omega_{\phi0} = 0.6847 - 0.0073$, we eventually obtain the upper bound $\lambda \leq \sqrt{3} + 0.0428 \approx 1.7748$.

We now turn to observational constraints on the quintessence model, which will lead to tighter bounds on $\lambda$ than the theoretical ones above.

\subsection{Summary and first observational constraints}\label{sec:obs}

Above, we have identified potentially realistic cosmological solutions to our quintessence model for given values of $\lambda$. These solutions are defined as in \eqref{realsol}, by requiring a past radiation domination, together with observationally appropriate values for $\Omega_{n0}$ in today's universe.  Observational constraints on $\Omega_{n0}$  are not expected to change much with the model or with $\lambda$, so we took fixed and probably admissible values for them. On the contrary, constraints on $w_{\phi0}$ seem to be so far very model and data dependent.

We have shown that the requirements in \eqref{realsol} are sufficient to guarantee a past matter domination phase and therefore provide candidate solutions with a realistic past. In addition, these requirements fix the value of $w_{\phi}$ today, given values for $\lambda$ and $\Omega_{n0}$, as detailed in the sample \eqref{wphi0values}.  Using the relation between $\lambda$ and $w_{\phi0}$ for the realistic cosmological solutions, and requiring moreover acceleration today, we finally obtained an upper bound \eqref{lambdaboundacc} on admissible $\lambda$ values, $\lambda \lesssim \sqrt{3}$. We do, however, expect stronger bounds from fitting the model to the wealth of observational data available.

In the literature, we find several works which have determined constraints on the exponential quintessence model from various observational data, without spatial curvature ($k=0$), as considered in this section. The following references find an upper bound on $\lambda$ at $95\%$ confidence level of 0.6 \cite{Agrawal:2018own}, 0.8 \cite{Akrami:2018ylq}, 0.5 \cite{Raveri:2018ddi} or ranging between 0.6-1.7 \cite{Schoneberg:2023lun} depending on the data used. As expected, these upper bounds on $\lambda$ are a little tighter than the one we obtained from minimally requiring past radiation and matter domination and acceleration today, $\lambda \lesssim \sqrt{3}$. So these works seem broadly consistent with our values of $\lambda$ and $w_{\phi0}$ for candidate realistic solutions.\footnote{We thank Nils Sch\"oneberg for helpful exchanges on the $w_{\phi0}$ values; it seems in particular that our values \eqref{wphi0values} can be reproduced using CLASS with an exponential potential, supporting again our identification of candidate realistic solutions.} On the other hand, observational fits do seem to make it difficult to reach $\lambda \geq \sqrt{2}$, which are the values of most interest for string theory models; in section \ref{sec:curvature}, we will investigate whether curvature can change this conclusion. Prior to this, we will explore one further set of observational constraints and  discuss quantitatively the acceleration epoch in the absence of curvature.

\subsection{$w_0 w_a$ parametrisation}\label{sec:w0wa}

Another set of observational constraints worth mentioning is that for the flat $w_0 w_a$CDM model, which assumes a dark energy equation of state parameter that varies linearly with $a$, that is, $w_\phi(a) = w_0+\left(1-\frac{a}{a_0}\right)w_a$; this is also known as the Chevallier-Polarski-Linder (CPL) parametrisation \cite{Chevallier:2000qy, Linder:2002et} and is commonly used by cosmologists as a fiducial model to fit to data.  The $w_0w_a$ parametrisation is clearly not a good approximation of the equation of state parameter for the exponential quintessence model throughout the cosmological history (see e.g. figure \ref{fig:Solmwp}); however, as is generally the case, it is a reasonable approximation at small redshifts.  We will apply the parametrisation in two ways.  First, we will provide an analytic expression that relates the theory parameter $\lambda$ to the model parameters $w_0, w_a$ and $\Omega_{\phi 0}$.  Secondly, we will, for given a $\lambda$ and candidate realistic solutions out to redshift $z \approx 4$, provide numerical fits for the parameters $w_0, w_a$, which could be compared with appropriate cosmological fits.

Let us start with the analytic approach: we show in the following how to relate $w_0, w_a$ to $\lambda$. We first reformulate the dynamical system \eqref{eq:system} as follows
\bea
{\Omega_k}'&=&(1+3w_{\text{eff}})\, \Omega_k\ ,\quad {\Omega_r}'=(-1+3w_{\text{eff}})\,\Omega_r \ , \quad {\Omega_{\phi}}'=3w_{\text{eff}}\, (\Omega_{\phi}-1)-\Omega_k+\Omega_r \ ,\nn\\
{w_{\text{eff}}}'&=&-3\Omega_{\phi}-\frac13(\Omega_k+\Omega_r)+3w_{\text{eff}}^2 \label{dynsysalt} \\
&&~~~~~~~~~\!+\sqrt{3}\,\lambda\, s_x \left|\Omega_{\phi}+w_{\text{eff}} +\frac13(\Omega_k-\Omega_r)\right|^{1/2}
\left(\Omega_{\phi}-w_{\text{eff}} -\frac13(\Omega_k-\Omega_r)\right)\ , \nn
\eea
where we traded the variables $x, y, z, u$ for $\Omega_{\varphi}, \Omega_k, \Omega_r$ and $w_{\text{eff}}$, using the various relations among those and the constraint $\Omega_m=1-\Omega_{\varphi}-\Omega_k-\Omega_r$. The sign of $x$, denoted $s_x$, is not fixed in terms of the new variables and needs to be added to the system. The last equation provides the following expression
\be
\lambda = \frac{3\Omega_{\phi}+\frac13(\Omega_k+\Omega_r)-3w_{\text{eff}}^2 + {w_{\text{eff}}}' }{\sqrt{3}\, s_x \left|\Omega_{\phi}+w_{\text{eff}} +\frac13(\Omega_k-\Omega_r)\right|^{1/2}
\left(\Omega_{\phi}-w_{\text{eff}} -\frac13(\Omega_k-\Omega_r)\right)} \ . \label{lambdaexprgen}
\ee
We can now apply this formula to the universe today, neglecting radiation and curvature.\footnote{For completeness, let us add that the threshold value $w_{\text{eff}} =-1/3$ gives $\lambda \approx 1.646 + 0.956\ {w_{\text{eff}}}'$, with $\Omega_{\phi}=0.685$.} Substituting ${w_{\text{eff}}}'\approx {\Omega_\phi}'\, w_\phi + \Omega_\phi\, {w_\phi}'$, using \eqref{dynsysalt} as well as the $w_0 w_a$ parametrisation of $w_{\phi}$, we can finally rewrite the expression \eqref{lambdaexprgen} as
\be
\lambda \approx \frac{3(1-w_0^2)-w_a}{\sqrt{3\,\Omega_{\phi 0}\, |1+w_0|} \, (1-w_0)}\,.\label{lambdaw0wa}
\ee

This expression relates analytically $\lambda, w_0$ and $w_a$ in a cosmological solution for a given $\Omega_{\phi 0}$. In particular, since we already determined in \eqref{wphi0values} the value $w_{\phi0}=w_0$ corresponding to each $\lambda$, for the candidate realistic solutions, it is straightforward to deduce from \eqref{lambdaw0wa} the  $w_a$ value. In fact, $-w_a$ computed in this way is equivalent to the gradient at $a=a_0$ of the curve $w_{\phi}(a)$ plotted against $a/a_0$ (i.e. $-w_a =e^{-N}\frac{dw_\phi}{dN}\vline_{N=0}$); the linear parametrisation $w_\phi(a) \approx w_0+w_a(1-\frac{a}{a_0})$ thus computed amounts to  trading the curve $w_{\phi}(a)$ against $a/a_0$ with its tangent at $a=a_0$.

In the second scheme, we use the $w_0 w_a$ parametrisation as an approximation of $w_{\phi}(a)$ over an extended period, say,  from today back to $z  =4$ or $\frac{a}{a_0}=0.2$. We  then determine numerically the (least-squares) best fit of this parametrisation to the curve for a given $\lambda$ and cosmological solution. Since the corresponding straight line is now an approximation of the curve over the period $0.2 \leq \frac{a}{a_0} \leq 1 $, the analytic expression \eqref{lambdaw0wa}, which is valid instead at a given point in time, is no longer exactly verified, but the results can be compared with cosmological fits using low-redshift data.

\begin{figure}[H]
\begin{center}
\includegraphics[width=0.6\textwidth]{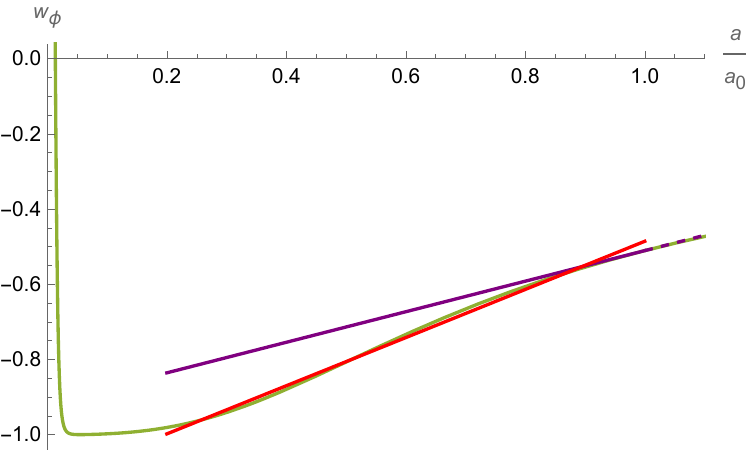}
\caption{Quintessence equation of state parameter, $w_{\phi}(a)$, (in green) plotted against the normalised scale factor, $\frac{a}{a_0}$, for $0 \leq \frac{a}{a_0} \leq 1.1$ where $a=a_0$ today. This curve is depicted for a candidate realistic solution with $\lambda= \sqrt{3}$, using $w_{\phi0}=-0.5107$ given in \eqref{wphi0values} and $\Omega_{m0}=0.315=1-\Omega_{\phi0}$. We display the tangent line to the curve at the point $a=a_0$ (in purple, plain in the past and dashed in the future); this can be equivalently computed either by applying equation \eqref{lambdaw0wa} to compute $w_a$ from $\lambda$, $w_{\phi0}$ and $\Omega_{\phi 0}$, or by computing numerically the gradient to the curve $w_{\phi}(a)$ against $\frac{a}{a_0}$ at $a=a_0$. Finally, we present the least-squares best fit of the curve $w_\phi(a)$ against $\frac{a}{a_0}$ over the period $0.2 \leq \frac{a}{a_0} \leq 1$ to the linear $w_0w_a$ parametrisation, $w_\phi(a) \approx w_0 + w_a(1-\frac{a}{a_0})$ (in red). We refer to the main text for more details.
}\label{fig:wphia}
\end{center}
\end{figure}

The results of these two schemes are illustrated in figure \ref{fig:wphia}, and the corresponding parameters for each $\lambda$ are given in table \ref{tab:w0wa}.\footnote{In the final stage of this work, the paper \cite{Shlivko:2024llw} appeared, and it has some overlap with the present subsection. In particular, the values presented in table \ref{tab:w0wa} seem to match those of \cite[Fig. 3]{Shlivko:2024llw}, up to the slightly different value of $\Omega_{\phi0}$.} As can be seen in figure \ref{fig:wphia}, the $w_0w_a$ parametrisation is indeed a reasonable approximation for the exponential quintessence model, at least at small redshifts, when applied as a ``best fit'' of the curve. We may then use the values of $w_0, w_a$ obtained in table \ref{tab:w0wa} to compare the model to current and future observational constraints. Intriguingly, there have been some recent preliminary hints in the cosmological data for so-called thawing quintessence models such as those we are considering, which have $w_0\gtrsim -1,\, w_a<0$, even if there is still much variation and uncertainty in the best fit parameters depending on the data sets used. For example, the DES collaboration finds $w_0= -0.773^{+0.075}_{-0.067}$ and $w_a  = -0.83^{+0.33}_{-0.42}$ (and $\Omega_m=0.325\pm 0.008$) using DES-SN5Y + Planck 2020 + SDSS BAO + DES Y3 3 $\times$ 2pt \cite[Tab. 2]{DES:2024tys}.  Looking at \eqref{wphi0values}, $w_0= -0.773$ would imply that $1< \lambda < \sqrt{2}$, a range again not favoured by asymptotic field space limits in string theory.  The first DESI results \cite[Tab. 3]{DESI:2024mwx} are also consistent with thawing quintessence, with e.g. $w_0 = -0.55^{+0.39}_{-0.21}$ and $w_a < -1.32$ (and $\Omega_m=0.344^{+0.047}_{-0.026}$), using low-redshift DESI data only; our $w_a$ (see table \ref{tab:w0wa}) might then be difficult to accommodate.   We will soon investigate whether curvature can help to improve these issues. Prior to this, we add some words on the acceleration phase of the solutions.

\begin{table}[H]
\begin{center}
\centering
\begin{tabular}{c||c|c||c|c}
 &  \multicolumn{2}{c}{Tangent} & \multicolumn{2}{c}{Best Fit}\\[-6pt]
$\ \lambda\ $ & \multicolumn{2}{c}{} & \multicolumn{2}{c}{} \\[-6pt]
 &  $\ w_0 \ $ &  $\ w_a \ $ & $\ w_0 \ $ &  $\ w_a \ $ \\
\hline
0 & -1.0000 & 0 & -1.0000 & 0 \\
1 & -0.8486 & -0.1915 & -0.8559 & -0.1914 \\
$\sqrt{2}$ & -0.6874 & -0.3302 & -0.6885 & -0.4034 \\
$\sqrt{8/3}$ & -0.5719 & -0.3888 & -0.5584 & -0.5586 \\
$\sqrt{3}$ & -0.5107 & -0.4063 & -0.4859 & -0.6402 \\
2 & -0.3028 & -0.3939 & -0.2217 &  -0.8932
\end{tabular}
\end{center}
\caption{Values of the parameters $w_0$ and $w_a$  in the different uses of this $w_0w_a$ parametrisation, described in the main text. In the ``tangent'' version, $w_0=w_{\phi0}$ from \eqref{wphi0values}, and $w_a$ is obtained analytically via \eqref{lambdaw0wa}, for $\Omega_{\phi0}=0.685$; $-w_a$ corresponds to the slope of the tangent to the curve $w_{\phi}(a)$ against $\frac{a}{a_0}$ today, which can also be computed numerically. In the ``best fit'' approach, the values are obtained numerically by obtaining the least-squares best fit of the curve $w_{\phi}(a)$ against $\frac{a}{a_0}$ to the linear parametrisation $w_{\phi}(a) \approx w_0 + w_a(1-\frac{a}{a_0})$, for the period $0.2 \leq \frac{a}{a_0} \leq 1$, i.e.~between redshift $z=4$ and today. These two schemes are illustrated in figure \ref{fig:wphia}.}
\label{tab:w0wa}
\end{table}

\subsection{Acceleration phase}\label{sec:accphase}

\begin{figure}[H]
\begin{center}
\begin{subfigure}[H]{0.48\textwidth}
\includegraphics[width=\textwidth]{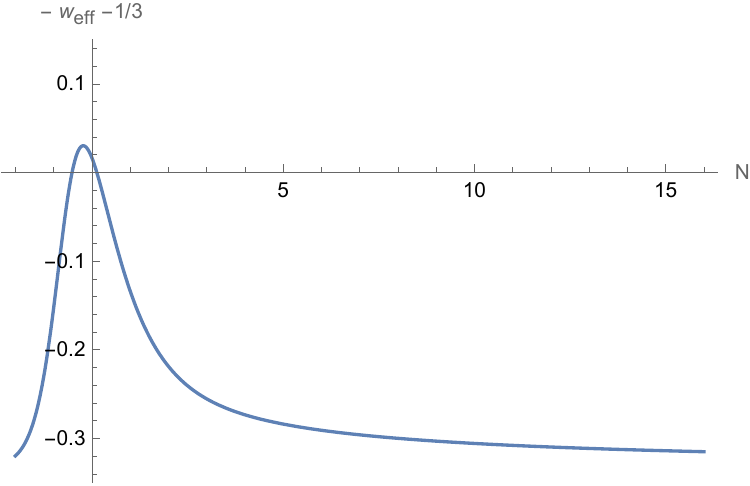}\caption{$\lambda=\sqrt{3}$\\$-0.52 \leq N \leq  0.13$}\label{fig:Acc1}
\end{subfigure}\quad
\begin{subfigure}[H]{0.48\textwidth}
\includegraphics[width=\textwidth]{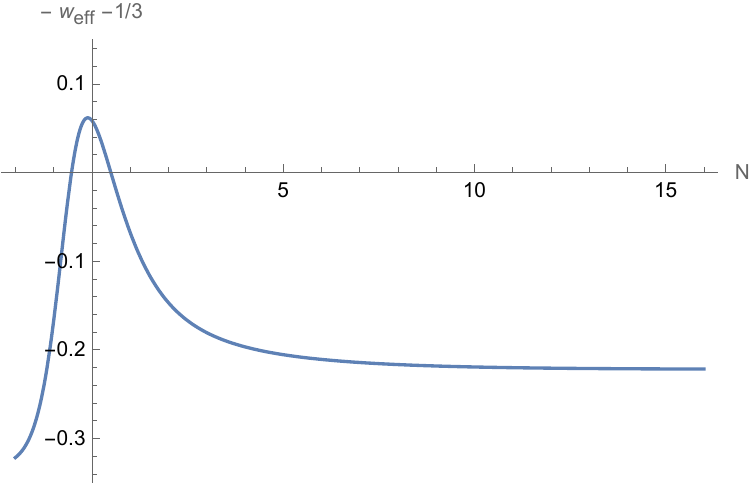}\caption{$\lambda=\sqrt{\frac{8}{3}}$\\ $-0.53 \leq N \leq 0.50$}\label{fig:Acc2}
\end{subfigure}\\
\begin{subfigure}[H]{0.48\textwidth}
\includegraphics[width=\textwidth]{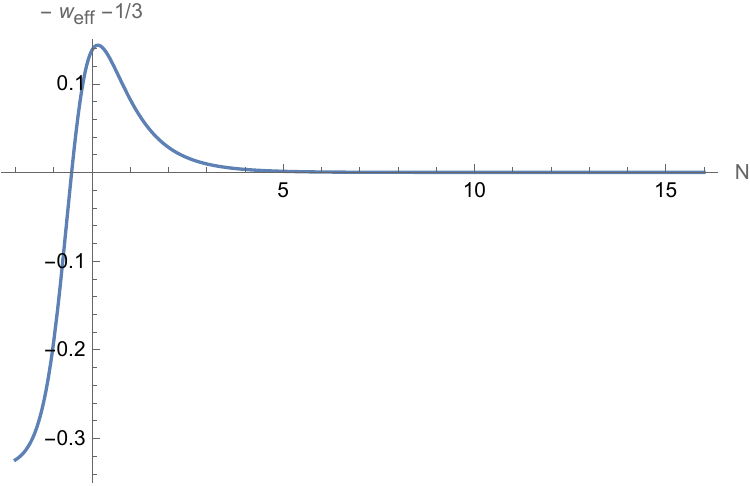}\caption{$\lambda=\sqrt{2}$\\$-0.53 \leq N $}\label{fig:Acc3}
\end{subfigure}\quad
\begin{subfigure}[H]{0.48\textwidth}
\includegraphics[width=\textwidth]{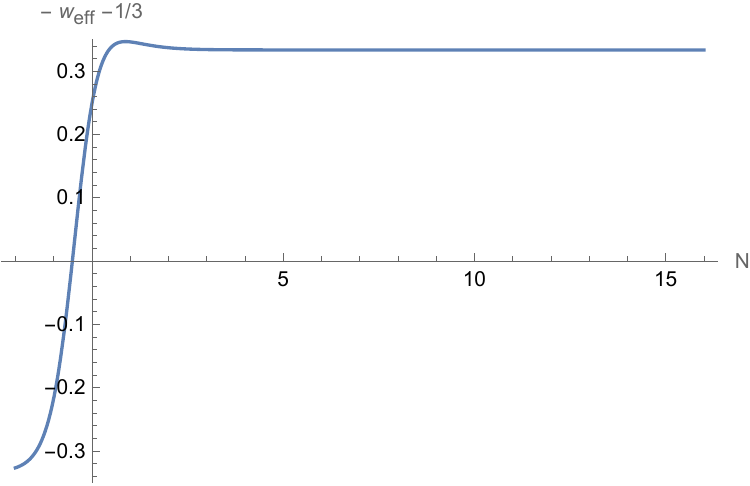}\caption{$\lambda=1$\\$ -0.51 \leq N $}\label{fig:Acc4}
\end{subfigure}
\caption{Evaluation of $-w_{{\rm eff}}-\frac{1}{3}$ in candidate realistic solutions for e-folds between $-2 \leq N \leq 16$, using $w_{\phi0}$ given in \eqref{wphi0values} for each $\lambda$ value, and $\Omega_{m0}=0.315=1-\Omega_{\phi0}$. When the quantity $-w_{{\rm eff}}-\frac{1}{3}$ is positive, there is acceleration: for each $\lambda$ value, we indicate the corresponding e-fold range. As expected, we find transient acceleration phases for $\lambda > \sqrt{2}$, and semi-eternal ones for $\lambda \leq \sqrt{2}$. We note the asymptotics to the constant value $\frac{2-\lambda^2}{3}$, corresponding to the value of $-w_{{\rm eff}}-\frac{1}{3}$ at $\Pp$. The universe today stands at $N=0$, and we see the bound on acceleration today to be slightly above $\lambda=\sqrt{3}$, as discussed around~\eqref{lambdaboundacc}.}\label{fig:Acc}
\end{center}
\end{figure}

Let us now focus on the acceleration phase in the candidate realistic solutions of the quintessence models we are considering; this phase includes today's universe. We illustrate this matter in figure \ref{fig:Acc}.

A first question is to know when the phase starts. To offer a comparison, the moment at which acceleration starts for $\Lambda$CDM can be computed, neglecting radiation and using $\Omega_{m0}=0.3150=1-\Omega_{\phi0}$: we obtain
\be
N = -\frac{1}{3} \ln \left( \frac{2\, \Omega_{\phi0}}{\Omega_{m0}} \right) \approx -0.49 \ .
\ee
We see in figure \ref{fig:Acc}, for a sample of $\lambda$ values, that the quintessence solutions have their acceleration phase starting slightly earlier.

A second question is the duration of the acceleration epoch; in particular, for $\lambda > \sqrt{2}$ the acceleration phase is transient, whereas it is semi-eternal\footnote{Semi-eternal acceleration refers to eternal acceleration in the future, but not in the past (see e.g. \cite{Andriot:2023wvg} for examples of eternal acceleration in both past and future).} for $\lambda \leq \sqrt{2}$.  This is also illustrated in figure \ref{fig:Acc}, where we give the duration for the same sample of $\lambda$ values. Having a transient acceleration instead of a semi-eternal one affects future cosmology, but has little impact regarding observations; it remains, however, a conceptually important difference.

\section{Including curvature}\label{sec:curvature}

We  now  consider cosmological solutions with a non-zero spatial curvature, in particular, we assume an open universe ($k=-1$), with $\Omega_k=z^2 \neq 0$. We have seen above that the dynamical system for a flat universe has a stable attractor at $\Pp$ (when $\lambda \leq \sqrt{3}$) or $\Pmp$ (when $\lambda > \sqrt{3}$).  It allows for asymptotic acceleration, a.k.a. (semi-)eternal acceleration, only when $\lambda \leq \sqrt{2}$, which is at best difficult to realise in string theory.  Intriguingly, as recently discussed in \cite{Marconnet:2022fmx, Andriot:2023wvg}, for an open universe the dynamical system has a new stable attractor, $\Pkp$, precisely when $\lambda$ takes values within the range favoured by string theory, $\lambda > \sqrt{2}$.  This fixed point, whilst not corresponding to acceleration itself, allows for (semi-)eternal acceleration as it is approached, with the cosmological event horizon going out to infinity \cite{Andriot:2023wvg}.  It is also interesting to note that open universes may be favoured by the tunnelling processes that are expected between vacua in the string landscape during the early universe \cite{Freivogel:2005vv} (see also, \cite{Buniy:2006ed, Horn:2017kmv, Cespedes:2020xpn, Cespedes:2023jdk}).  We will now investigate whether solutions with $k=-1$ can be used to model the observed dark energy in our universe, with a particular interest in those with $\lambda > \sqrt{2}$ and, possibly, asymptotic acceleration.  In particular, this requires us to extend the analyses of  \cite{Marconnet:2022fmx, Andriot:2023wvg} to include matter and radiation.

A first question to ask is how much curvature  we can allow for in our observed universe, that is how large can $\Omega_{k0}$ be? The answer is unfortunately model dependent, and to our knowledge observational constraints for the exponential quintessence model including curvature are not available, but see forthcoming work \cite{BBMPTZ, Alestas:2024gxe}. We can nevertheless read a few values from the literature for related models. To start with, consider $\Lambda$CDM with curvature, denoted $k\Lambda$CDM, which assumes $w_{\rm DE}=-1$. The strongest constraints come from combining Planck data with BAO, which leads to $\Omega_{k0}=0.0007\pm 0.0019$ \cite{Planck:2018vyg}. CMB independent constraints can be found in \cite[Tab. 5]{Bel:2022iuf}, e.g. using clustering, BBN and BAO leads to $\Omega_{k0} = 0.0484^{+0.0748}_{-0.0756}$ (with $\Omega_{m0} =  0.3299^{+0.0157}_{-0.0158}$). Similarly, the latest value of $\Omega_{k0}$ for $k\Lambda$CDM from DESI \cite{DESI:2024mwx} varies between $\Omega_{k0}=0.065^{+0.068}_{-0.078}$ (with $\Omega_{m0} = 0.284 \pm 0.020$) and $\Omega_{k0}=0.0003^{+0.0048}_{-0.0054}$ (with $\Omega_{m0} = 0.296 \pm 0.014$), depending on the data used. All these results are compatible with a spatially flat universe, but allow for curvature up to $\mathcal{O}(10^{-3}-10^{-1})$.

Turning to dynamical dark energy models, a first set of values can be obtained from \cite{Aurich:2003it}, which used WMAP data \cite{WMAP:2003ivt} to constrain a quintessence model with $w_\phi=\text{constant}$ (this corresponds to a rather specific potential different to the exponential) in the presence of curvature. The best fit for this quintessence model (assuming $n_s=1$) gave $\Omega_{k0}=0.05$, $\Omega_{m0}=0.34$ and $w_{\phi0} = -0.58$, with a value for $\chi_{\text{eff}}^2=975$, compared to $\chi_{\text{eff}}^2=978$ for flat $\Lambda$CDM. More recently, DESI \cite{DESI:2024mwx} provides constraints on the $w_0w_a$CDM model with curvature, with the values of $\Omega_{k0}$ for an open universe varying between $\Omega_{k0}=0.087^{+0.1}_{-0.085}$ (with $\Omega_{m0} = 0.313 \pm 0.049$, $w_{0} = -0.70^{+0.49}_{-0.25}$, $w_{a} < -1.21$) and $\Omega_{k0}=0.0003 \pm 0.0018$ (with $\Omega_{m0} =0.3084 \pm 0.0067$, $w_{0} = -0.831\pm 0.066$, $w_{a}=-0.73^{+0.32}_{-0.28}$) (and some data combinations yielding instead a best fit to a closed universe). Again, these results are compatible with a spatially flat universe, but allow for curvature up to $\mathcal{O}(10^{-3}-10^{-1})$.

\begin{figure}[H]
\begin{center}
\begin{subfigure}[H]{0.40\textwidth}
\includegraphics[width=\textwidth]{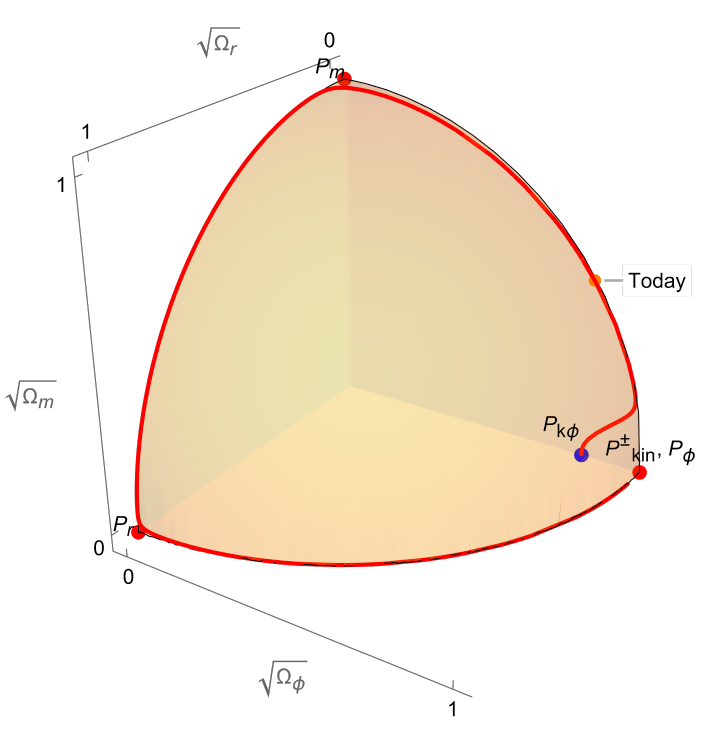}\caption{}\label{fig:Solrk10}
\end{subfigure}\quad\quad
\begin{subfigure}[H]{0.48\textwidth}
\includegraphics[width=\textwidth]{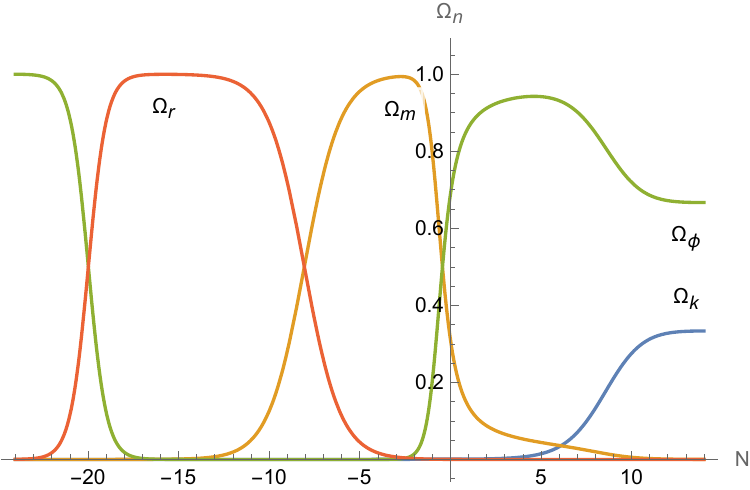}\caption{$\Omega_{k0}=0.0007$}\label{fig:Solrk1O}
\end{subfigure}\\
\begin{subfigure}[H]{0.40\textwidth}
\includegraphics[width=\textwidth]{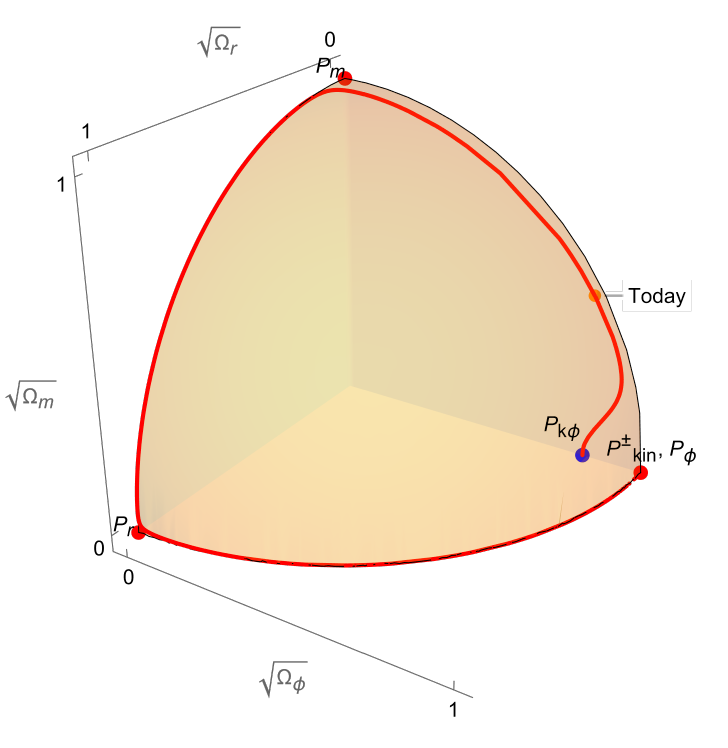}\caption{}\label{fig:Solrk20}
\end{subfigure}\quad\quad
\begin{subfigure}[H]{0.48\textwidth}
\includegraphics[width=\textwidth]{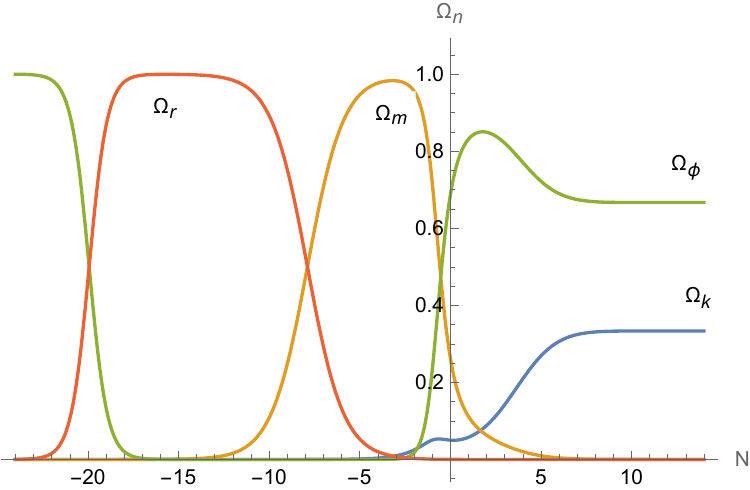}\caption{$\Omega_{k0}=0.05$}\label{fig:Solrk2O}
\end{subfigure}
\caption{Cosmological solutions in the phase space illustration of figure \ref{fig:PhaseSpacer} for $\lambda=\sqrt{3}$, together with their $\Omega_n$. The solutions start at $\Pkin$ and end at $\Pkp$. We require in addition the solutions to pass by the orange point corresponding to today's universe. The latter is defined by $\Omega_{\phi0}=0.685$, $\Omega_{r0}=0.0001$ and $\Omega_{m0}=0.3149-\Omega_{k0}$. The value of $w_{\phi0}$ is tuned to obtain past radiation domination, as well as the equality $\Omega_{\phi} = \Omega_r$ around $N_{\phi=r} \approx -20$, motivated by BBN. In figure \ref{fig:Solrk10} and \ref{fig:Solrk1O}, we take $\Omega_{k0}=0.0007$, giving $w_{\phi0} = -0.51054453650$ for $N_{\phi=r} = -19.97$. In figure \ref{fig:Solrk20} and \ref{fig:Solrk2O}, we take $\Omega_{k0}=0.05$, giving $w_{\phi0} = -0.49618615823$ for $N_{\phi=r} = -19.93$. This is to be compared to figure \ref{fig:Solr} where $\Omega_{k0}=0$.}\label{fig:Solrk}
\end{center}
\end{figure}

\begin{figure}[H]
\begin{center}
\begin{subfigure}[H]{0.48\textwidth}
\includegraphics[width=\textwidth]{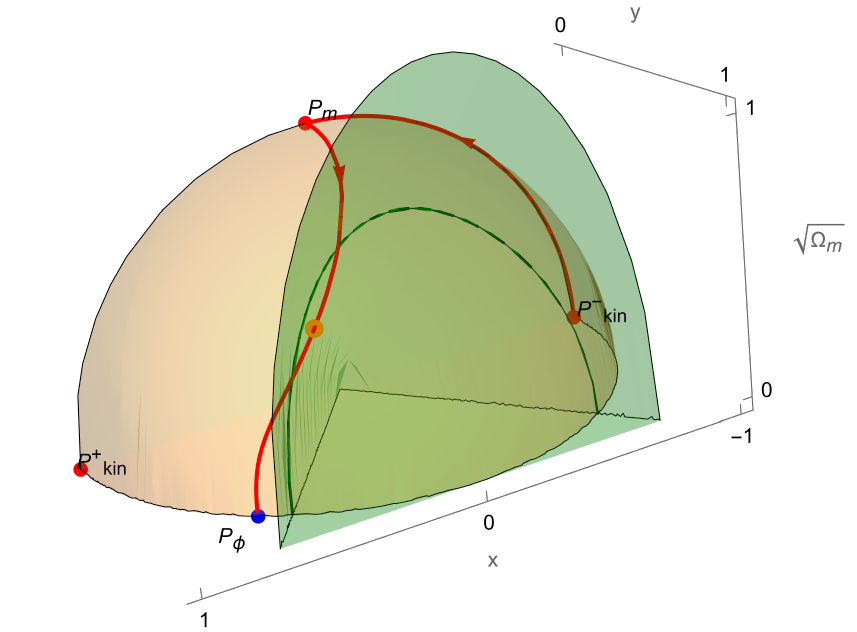}\caption{}\label{fig:Solmk10}
\end{subfigure}\quad\
\begin{subfigure}[H]{0.48\textwidth}
\includegraphics[width=\textwidth]{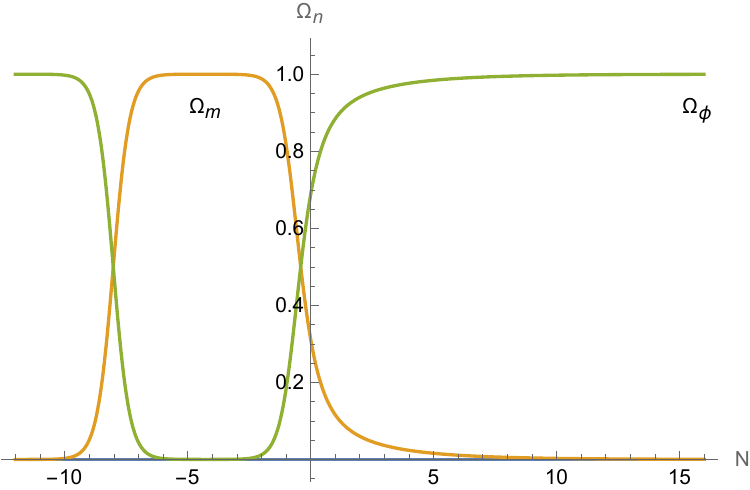}\caption{$\Omega_{k0}=0$}\label{fig:Solmk1O}
\end{subfigure}\\
\begin{subfigure}[H]{0.48\textwidth}
\includegraphics[width=\textwidth]{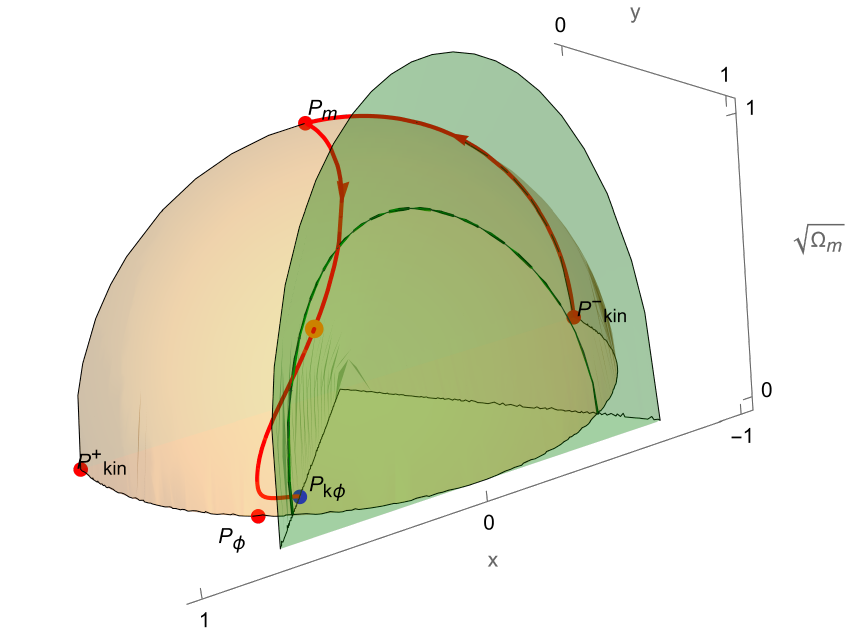}\caption{}\label{fig:Solmk20}
\end{subfigure}\quad\
\begin{subfigure}[H]{0.48\textwidth}
\includegraphics[width=\textwidth]{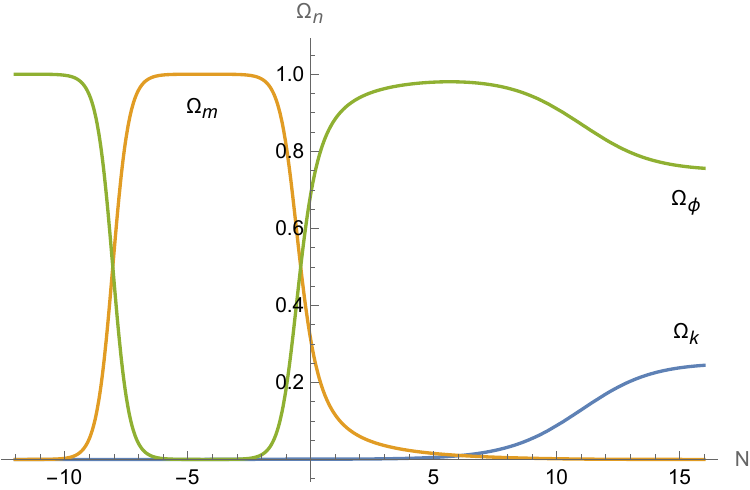}\caption{$\Omega_{k0}=0.0007$}\label{fig:Solmk2O}
\end{subfigure}
\caption{Cosmological solutions in the phase space illustration of figure \ref{fig:PhaseSpace} for $\lambda=\sqrt{\frac{8}{3}}$, together with their $\Omega_n$. We require the solutions to pass by the orange point corresponding to today's universe. The latter is defined by $\Omega_{\phi0}=0.685$ and $\Omega_{m0}=0.315-\Omega_{k0}$. The value of $w_{\phi0}$ is tuned to obtain past matter domination starting at $N_{m} \approx -8.05$. In figure \ref{fig:Solmk10} and \ref{fig:Solmk1O}, we take $\Omega_{k0}=0$, giving $w_{\phi0} = -0.571933$ for $N_{m} = -8.02$. In figure \ref{fig:Solmk20} and \ref{fig:Solmk2O}, we take $\Omega_{k0}=0.0007$, giving $w_{\phi0} = -0.571775$ for $N_{m} = -8.04$. The former is similar to figure \ref{fig:Solm} where $\lambda=\sqrt{3}$.}\label{fig:Solmk}
\end{center}
\end{figure}

In summary, the central value of $\Omega_{k0}$ for an open universe varies between $0.09$ and $0.0001$ depending on the model and data, with $\Omega_{m0}$ varying by a few percent around the values \eqref{OmegaLCDM} previously considered. In other words, even though we do not have precise observational constraints for the model we are considering, we expect the curvature contribution $\Omega_{k0}$ to be small. This will have important consequences, as we now summarize. We will first argue that the past universe is not qualitatively altered by the addition of curvature within the expected limits, and all previously considered solutions and their properties will remain; the future of the universe on the contrary can change drastically. Finally, few changes occur in today's universe, but we will study in detail the quantitative changes. Before entering into these discussions, we depict in figure \ref{fig:Solrk} and \ref{fig:Solmk} different candidate realistic cosmological solutions for various values of $\lambda$ and $\Omega_{k0}$; we will comment on these figures throughout this section.

\subsection{Impact on the past and on the future}
\label{sec:pastfuture}

Let us start by examining the impact of curvature on the past of the candidate realistic cosmological solutions. We have argued above that $\Omega_{k0}<\Omega_{m0}$. Using the dependence of the $\Omega_n$ on $a$ in \eqref{Omegas}, it is straightforward to see that
\be
\frac{\Omega_k}{\Omega_m} = \frac{\Omega_{k0}}{\Omega_{m0}}\ \frac{a}{a_0} < \frac{\Omega_{k0}}{\Omega_{m0}} < 1 \ \ \text{in the past} \ .
\ee
In other words, if the curvature contribution is small today, it will be even more subdominant in the past, as a direct consequence of the scaling with $a$ of the different energy densities:
\be
\rho_{{\rm kin}} \propto a^{-6} ,\  \rho_r \propto a^{-4}, \ \rho_m \propto a^{-3},\ \rho_k \propto a^{-2} \ ,\label{rhoscalings}
\ee
(see table \ref{tab:n} and footnote \ref{foot:kin}).\footnote{One can also see the behaviour of $\rho_{{\rm kin}}$ by considering the solution close to $\Pkin$, using as an approximation the expression given in table \ref{tab:fixedpointsfields}, from which we read $\rho_{{\rm kin}} = 3 H^2 x^2 = \tfrac{1}{2} \dot{\phi}^2 = \tfrac{1}{3} \left( a/a_0 \right)^{-6}$. In the same way, we find for $\Pkin$ that $\rho_V = 3 H^2 y^2 = V \propto a^{\mp \lambda \sqrt{6}}$, assuming a solution close to these points with $y\neq0$. For $\Pkinm$, this $\rho_V$ is always subdominant to $\rho_{{\rm kin}} $ for $a \rightarrow 0$; for $\Pkinp$ however, it becomes dominant for $\lambda >  \sqrt{6}$. In other words, one has $y\gg x$ in solutions close to $\Pkinp$ for $\lambda >  \sqrt{6}$ (as long as $y\neq0$). Since $\Pkinp$ has $x=1, y=0$, this explains why this fixed point is a saddle for $\lambda >  \sqrt{6}$: it cannot be reached by solutions with $y\neq0$ since they have $y\gg x$. This is also related to solutions described in appendix \ref{ap:analytic} and \ref{ap:asymptotics}.} That curvature is (very) subdominant in the past until today, in realistic solutions, can be verified explicitly in the evolution of the $\Omega_n$ depicted in figures \ref{fig:Solrk1O}, \ref{fig:Solrk2O} and \ref{fig:Solmk2O}. It can also be seen in figures \ref{fig:PhaseSpaceL} and \ref{fig:PhaseSpaceLbis}; there, we already noticed that any solution having a small curvature contribution ($0 \leq \Omega_k \leq 0.1$) at some point in time before approaching close to $\Pp$ or $\Pmp$, always had a small $\Omega_k$ in the past, because the solutions stayed within this slice of $\Omega_k$ values. This is in particular true in the past of realistic solutions: indeed, today's universe has to be placed in the acceleration region, and this is met before the points $\Pp$ or $\Pmp$.\\

If $\Omega_k$ is negligible in the past of realistic solutions, it is expected to have negligible impact in that past.\footnote{Note that setting $z=0$, i.e.~no curvature, defines an invariant subspace of solutions, see appendix \ref{ap:fixedsubsets}. Adding a very small amount of curvature is thus not expected to change the solution, as long as the evolution keeps this contribution small.} All properties drawn from the past in absence of curvature should then still hold. We defined previously realistic cosmological solutions in \eqref{realsol}, as those having a past radiation domination, and today's values $\Omega_{n0}$ close to \eqref{OmegaLCDM}. As argued at the beginning of this section, in the presence of curvature, the values of the $\Omega_{n0}$ are expected to change at most by a few percent. And as just argued, a past radiation domination is not expected to be affected by curvature. We also showed in the absence of curvature that matter domination would automatically be obtained in the past, with fixed amplitude and duration, after radiation domination: we do not expect any change in this in the presence of small curvature. We verify this explicitly in figure \ref{fig:Solrk}. Finally, the property that $w_{\phi0}$ is fixed to the 4th digit in those solutions for a given $\lambda$ is also verified, e.g.~in figure \ref{fig:Solrk} and \ref{fig:Solmk}. The precise value, as a quantitative property of today's universe, may however differ due to curvature, as we will see below.

The reasonings that led to upper bounds on $\lambda$ also remain valid. Beyond explicit attempts to find solutions, arguments for the bounds were based on the phase space illustrations as well as on the past of the solutions. As can be seen in figure \ref{fig:PhaseSpaceL} and \ref{fig:PhaseSpaceLbis}, the solutions behave in the same way in the phase space for small $\Omega_k>0$, so those arguments remain valid. We will work out the precise bound values in the following subsection.\\

We now turn to the impact of curvature for the future of the universe. While the scaling in $a$ of energy densities makes curvature subdominant in the past, it also makes curvature dominant in the future over matter and radiation. The only contribution whose scaling is ambiguous, and could therefore be competing with curvature, is that of the quintessence scalar field, both through its kinetic and potential energy. In fact, the behaviour of $\Omega_{\phi}$ in the future is dictated by the stable fixed point that the solutions eventually reach.

\begin{figure}[H]
\begin{center}
\begin{subfigure}[H]{0.48\textwidth}
\includegraphics[width=\textwidth]{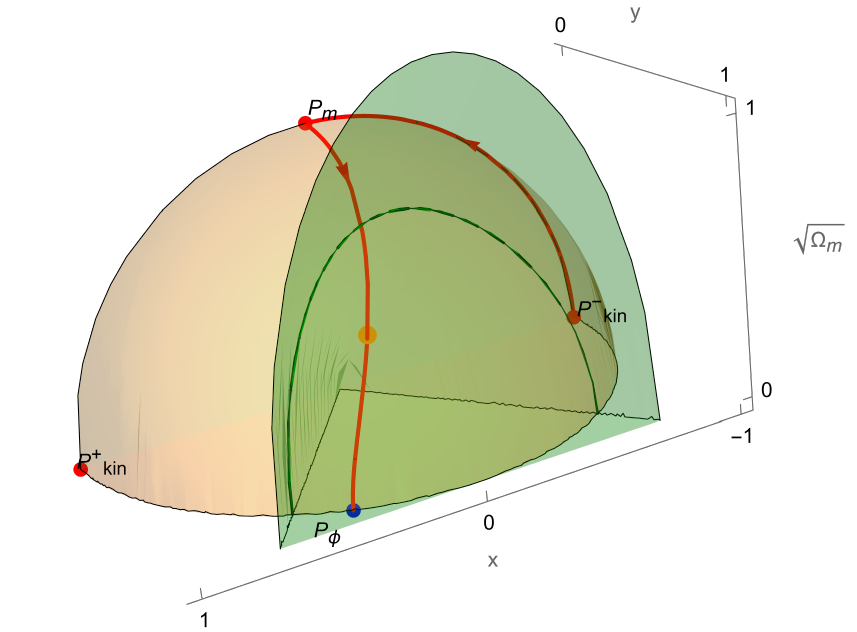}\caption{}\label{fig:Solmk30}
\end{subfigure}\quad\
\begin{subfigure}[H]{0.48\textwidth}
\includegraphics[width=\textwidth]{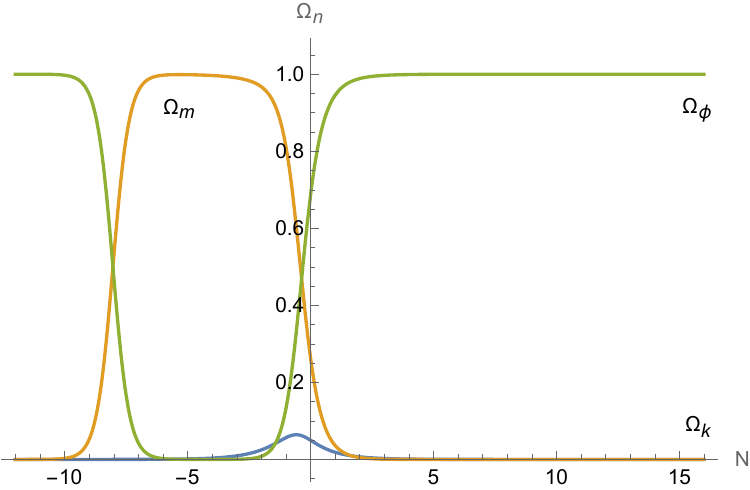}\caption{}\label{fig:Solmk3O}
\end{subfigure}
\caption{Cosmological solution in the phase space illustration of figure \ref{fig:PhaseSpace} for $\lambda=1$, together with its $\Omega_n$. We require the solution to pass by the orange point corresponding to today's universe. The latter is defined by $\Omega_{\phi0}=0.685$, $\Omega_{k0}=0.05$ and $\Omega_{m0}=0.315-\Omega_{k0}$. The value of $w_{\phi0}=-0.845258$ is tuned to obtain past matter domination starting at $N_{m} = -8.03$. This can be compared to figure \ref{fig:Solmk}.}\label{fig:Solmk3}
\end{center}
\end{figure}

As studied in section \ref{sec:stab}, for $\lambda < \sqrt{2}$, the final fixed point is $\Pp$, with or without curvature. It can be verified that this fixed point has $\Omega_{\phi}=1$ and $\Omega_k=0$, thus making $\Omega_k$  necessarily subdominant to dark energy in the future. We can also verify this property by considering solutions close to $\Pp$, using as an approximation the expressions of table \ref{tab:fixedpointsfields}: we read from there that $\dot{\phi}^2 \sim V \sim a^{-\lambda^2}$. In the future, this is thus dominant over $\rho_k$ for $\lambda < \sqrt{2}$, as expected. We finally see this subdominance explicitly in figure \ref{fig:Solmk3} for $\lambda=1$. We conclude that curvature has no impact in the future of realistic solutions for $\lambda < \sqrt{2}$.

For $\lambda>\sqrt{2}$, the situation with or without curvature becomes distinct: with curvature, the attractor becomes $\Pkp$, while without curvature, it remains $\Pp$ for $\sqrt{2} < \lambda \leq \sqrt{3}$, and becomes $\Pmp$ for $\lambda>\sqrt{3}$ (see section \ref{sec:stab}). $\Pkp$ gives a very different future, since it allows for curvature to compete with dark energy. One verifies for instance that $\dot{\phi}^2 \sim V \sim a^{-2} \sim \rho_k$ at $\Pkp$. Also, it has $\Omega_{\phi} = \frac{2}{\lambda^2}, \, \Omega_k = \frac{\lambda^2 - 2}{\lambda^2}$. This competition in the future is obvious in the evolution of the $\Omega_n$ in figure \ref{fig:Solrk1O}, \ref{fig:Solrk2O} and \ref{fig:Solmk2O}. The difference in final fixed points, with or without curvature, can be seen in the phase space illustrations of the solutions by comparing figure \ref{fig:Solrk10} or  \ref{fig:Solrk20} (with curvature) to \ref{fig:Solr0} (without curvature). Another comparison is that of figure \ref{fig:Solmk20} (with curvature) to \ref{fig:Solmk10} (without curvature); for those we also see the difference in the $\Omega_n$ in the future by comparing figures  \ref{fig:Solmk1O} to \ref{fig:Solmk2O}. Including curvature therefore has an important impact on the future of candidate realistic cosmological solutions for $\lambda>\sqrt{2}$, which interestingly, is the favorable range of values for string theory models.

Alongside the differences in the future density parameters' evolution, a consequence of the change of final fixed point between curvature and no curvature for $\lambda>\sqrt{2}$ can also be seen in the behaviour of the asymptotic deceleration or acceleration. In the absence of curvature, recall from figure \ref{fig:Acc} that realistic solutions for $\lambda>\sqrt{2}$ admit only a transient phase of acceleration: it ends at some finite time after today, with the solutions then entering a phase of deceleration. Indeed, the final fixed point, $\Pp$ or $\Pmp$, has a constant $w_{\rm eff}> - \frac{1}{3}$ (see table \ref{tab:fixedpoints}) and the universe will reach this value while decelerating. With curvature, the final fixed point is $\Pkp$ which has $\ddot{a}=0$. As can be seen in figure \ref{fig:Acck}, in the solution, $\ddot{a}$ will first reach a (negative) minimum, i.e.~a ``maximal deceleration'', before turning back to zero. So again the universe has only a transient acceleration and reaches the final fixed point while decelerating, but now the fixed point has $w_{\text{eff}}=-\frac13$ and no deceleration or acceleration.\footnote{The approach to zero acceleration is actually achieved by successive phases of deceleration and acceleration if the approach to $\Pkp$ is a spiral. The amplitude of these phases is however smaller than that of the first deceleration.} This is to be contrasted to the case without matter and radiation, studied in \cite{Marconnet:2022fmx, Andriot:2023wvg}, where solutions with (semi-)eternal acceleration can be found for $\lambda>\sqrt{2}$ (as well as solutions with transient acceleration).

Whilst differences in the future with or without curvature are important, it is fair to note that this has little chance to ever be observationally verified by humankind, given the large time scales; e.g. using $\int \d t = \int \d N/H = \int \d N\, a_0\, e^N\, z$, where $N=\ln(a/a_0)$ and $z$ is the variable in \eqref{eq:variables} with $k=-1$, we find that for $\lambda=\sqrt{\frac{8}{3}}$ the duration between today and the time at which $w_{\rm eff}$ starts to turn downwards towards $w_{\rm eff}=-\frac13$ is around 17.4 times the duration between BBN at $N\approx-20$ and today (see figure  \ref{fig:Acck}).\footnote{Note that to determine that time in years we would need to infer a value of $H_0$ from observations, which would in turn fix $a_0$ via $z_0 = 1/(a_0 H_0)$.} To discover any observable impact of curvature on realistic quintessence solutions, we are then left with today's universe, to which we now turn.

\subsection{Impact on today's universe}\label{sec:today}

The first impact of curvature on today's universe is the fact that $\Omega_{k0}\neq0$, as made visible in figure \ref{fig:Solrk2O} or \ref{fig:Solmk3O}. Since $\Omega_k = 1- \Omega_T$, this curvature contribution has to be taken off from the others, namely $\Omega_{m0}$ or $\Omega_{\phi0}$. As argued previously, this does not lead to qualitative changes, but we would like here to evaluate the quantitative changes. To that end, recall that current observational constraints allow for curvature up to $\mathcal{O}(10^{-3}-10^{-1})$, e.g. the recent data from DESI results in best fits for the curved $w_0w_a$CDM model parameters to be $\Omega_{k0}=0.087^{+0.010}_{-0.085}, \Omega_{m0}=0.313\pm 0.049$, $w_0 = -0.70^{+0.49}_{-0.25}$, and $w_a<-1.21$ \cite{DESI:2024mwx}. For simplicity and to offer a better comparison to the case without curvature (see \eqref{OmegaLCDM}), we will use the fiducial values
\be
\Omega_{k0}=0.085,\quad \Omega_{m0}=0.315 ,\quad \Omega_{\phi0}= 0.600 \ ,\label{OmegaCurv}
\ee
neglecting radiation today.

We have argued that including curvature had no  significant impact on the past of realistic solutions. As a consequence, the requirement of radiation domination leads again to matter domination, and eventually fixes $w_{\phi_0}$ today. We now evaluate how much the value of $w_{\phi_0}$ changes in presence of curvature, as compared to \eqref{wphi0values}. This is done again by requiring that matter domination starts no later than $N \approx -8$, which is consistent with having past radiation domination. We obtain the following results:
\be
\begin{tabular}{c||c|c}
 & \multicolumn{2}{c}{$w_{\phi0}$ for}\\[-6pt]
$\ \lambda\ $ & \multicolumn{2}{c}{ } \\[-6pt]
 & $\ \Omega_{k0}=0\ $ & $\Omega_{k0}=0.085$ \\
\hline
0 & -1.0000 & -1.0000 \\
1 & -0.8486 & -0.8720 \\
$\sqrt{2}$ & -0.6874 & -0.7363 \\
$\sqrt{8/3}$ & -0.5719 & -0.6400 \\
$\sqrt{3}$ & -0.5107 & -0.5894 \\
2 & -0.3028 & -0.4231
\end{tabular}   \label{wphi0valuesk}
\ee
Including curvature, and correspondingly lowering the dark energy density parameter, leads to more negative values of $w_{\phi0}$ at a given $\lambda$. This is probably observationally favorable. In turn, at a given $w_{\phi0}$, one reaches higher $\lambda$ values, which is favorable for string theory models. In particular, the above values of $\Omega_{n0}$  were found \cite{DESI:2024mwx} to correspond to a central value of $w_{\phi0} = -0.70$. According to \eqref{wphi0valuesk}, this would correspond to $\sqrt{2}< \lambda < \sqrt{8/3}$, again favorable to string model realisations.

Note that including curvature in the opposite way, namely taking it off the matter density contribution leads to the opposite effect: for $\Omega_{k0}=0.085, \Omega_{\phi0}= 0.685, \Omega_{m0}=0.230$, and $\lambda=\sqrt{8/3}$, we obtain $w_{\phi0} = -0.5507$. However, on this choice of the universe budget, we were guided by the values of \cite{DESI:2024mwx}, which rather preserve $\Omega_{m0}$ when including curvature.\\

Next, we evaluate the change in the upper bounds on $\lambda$. The first bound was obtained from the crossing of the circle $x^2+y^2=\Omega_{\phi0}$ by $\Pmp$. In presence of curvature, $\Pmp$ becomes a saddle point: the solutions go close to it and then inside the phase space ball towards $\Pkp$. In addition, the point corresponding to the universe today is now slightly inside the ball, away from the spherical boundary. These two differences make the crossing criterion less exact. We can still use it in a first approximation to obtain the following upper bound
\be
\lambda \leq \sqrt{\frac{3}{\Omega_{\phi0}}} \approx 2.236 \ ,
\ee
which, as above, gives a higher $\lambda$ value than in the case without curvature ($\lambda \leq 2.093$ in \eqref{lambdaup1}).

A more stringent value is obtained by requiring acceleration today. From \eqref{weffO}, we have $w_{\rm eff} = w_{\phi} \Omega_{\phi} - \frac{1}{3} \Omega_k \leq - \frac{1}{3} $, giving here
\be
w_{\phi0} \leq - \frac{1}{3 \Omega_{\phi0} }  (1 - \Omega_{k0}) \approx -0.5083 \ ,
\ee
slightly more negative than the case without curvature. Requiring acceleration today together with matter domination in the past leads to the upper bound
\be
\lambda \leq \sqrt{3} + 0.1407 \approx 1.8727 \ .\label{lambdaboundacccurv}
\ee
This is higher than the one without curvature given in \eqref{lambdaboundacc}, $\lambda \leq 1.7683$. These results go again in a favorable direction for string theory models.

We expect observational constraints to give rise to a tighter upper bound on $\lambda$ than the one derived above. As mentioned, we are not aware of any paper where this has been evaluated in the presence of curvature. However, it appears from our results that observational upper bounds on $\lambda$ with curvature will be higher than the ones without curvature and it will be extremely interesting to check if string models with $\lambda \geq \sqrt{2}$ could be observationally viable.\\

For completeness, we now consider the CPL or $w_0 w_a$ parametrisation that was also discussed in section \ref{sec:w0wa}: it assumes the linear dependence $w_\phi(a) = w_0+\left(1-\frac{a}{a_0}\right)w_a$. Let us evaluate how much the results of this parametrisation change in the presence of curvature. As in section \ref{sec:w0wa}, we use this parametrisation in two ways. Firstly, we derive an analytical formula that relates $\lambda$ to the model parameters; correspondingly, we trade the $w_\phi(a)$ against $\frac{a}{a_0}$ curve for its tangent at the point $a=a_0$, namely today's universe. Secondly, we take the candidate realistic solutions for given $\lambda$ over some small redshift interval and perform a (least-squares) best fit to the $w_0w_a$ parametrisation.

Using \eqref{dynsysalt} and \eqref{lambdaexprgen}, we compute the relation between $\lambda$, $w_0$ and $w_a$, in the presence of curvature but still neglecting radiation. In this case  $w_{\text{eff}}=w_{\phi} \Omega_{\phi} - \frac{1}{3} \Omega_k$. Eventually, it turns out that expression \eqref{lambdaw0wa} remains valid, namely
\be
\lambda \approx \frac{3(1-w_0^2)-w_a}{\sqrt{3\,\Omega_{\phi 0}\, |1+w_0|} \, (1-w_0)}\,.\label{lambdaw0wak}
\ee
Using the pairs $\lambda, w_{\phi0}=w_0$ determined in \eqref{wphi0valuesk} with curvature, we deduce from \eqref{lambdaw0wak} the value of $w_a$. We report the results in table \ref{tab:w0wak}, together with the best fits to the $w_0w_a$ parametrisation over the redshift interval from $z=4$ to today.  These results are to be compared with  the case without curvature, given in table \ref{tab:w0wa}; we see that the values of $w_0, w_a$ differ somewhat with or without curvature. It is interesting to compare our predictions for the case with curvature in table \ref{tab:w0wak} with the first results from DESI \cite[Tab. 3]{DESI:2024mwx}: the latter give $w_0 = -0.70^{+0.49}_{-0.25}$ and $w_a < -1.21$ (with $\Omega_m=0.313 \pm 0.049$ and $\Omega_k = 0.087^{+0.100}_{-0.085}$) for the low-redshift DESI-data-only fits.  As in the case without curvature, it seems difficult to accommodate the current observational bounds on $w_a$. It will, however, be interesting to keep in mind the best fit values of $w_0, w_a$ obtained here, when future observational data becomes available.

\begin{table}[H]
\begin{center}
\centering
\begin{tabular}{c||c|c||c|c}
 &  \multicolumn{2}{c}{Tangent} & \multicolumn{2}{c}{Best Fit}\\[-6pt]
$\ \lambda\ $ & \multicolumn{2}{c}{} & \multicolumn{2}{c}{} \\[-6pt]
 &  $\ w_0 \ $ &  $\ w_a \ $ & $\ w_0 \ $ &  $\ w_a \ $ \\
\hline
0 & -1.0000 & 0 & -1.0000 & 0 \\
1 & -0.8720 & -0.1797 & -0.8811 & -0.1589 \\
$\sqrt{2}$ & -0.7363 & -0.3182 & -0.7451 & -0.3344 \\
$\sqrt{8/3}$ & -0.6400 & -0.3846 & -0.6414 & -0.4629 \\
$\sqrt{3}$ & -0.5894 & -0.4088 & -0.5844 & -0.5310 \\
2 & -0.4231 & -0.4374 & -0.3840 & -0.7526
\end{tabular}
\end{center}
\caption{Values of the parameters $w_0$ and $w_a$ in the different uses of the $w_0w_a$ parametrisation, described in the main text and in section \ref{sec:w0wa}, now with $\Omega_{\phi0}=0.600$, $\Omega_{k0}=0.085$. In the ``tangent'' version, $w_0=w_{\phi0}$ from \eqref{wphi0valuesk}, and $w_a$ is obtained analytically via \eqref{lambdaw0wak}; $-w_a$ corresponds to the slope of the tangent, to the curve $w_{\phi}(a)$ against $\frac{a}{a_0}$, today. In the ``best fit'' approach, the values are obtained numerically by obtaining the least-squares best fit of the curve $w_{\phi}(a)$ to $w_{\phi}(a) \approx w_0 + w_a(1-\frac{a}{a_0})$, for the period $0.2 \leq \frac{a}{a_0} \leq 1$, i.e.~between redshift $z=4$ and today. This table can be compared to table \ref{tab:w0wa}, without curvature.}
\label{tab:w0wak}
\end{table}

\begin{figure}[H]
\begin{center}
\begin{subfigure}[H]{0.48\textwidth}
\includegraphics[width=\textwidth]{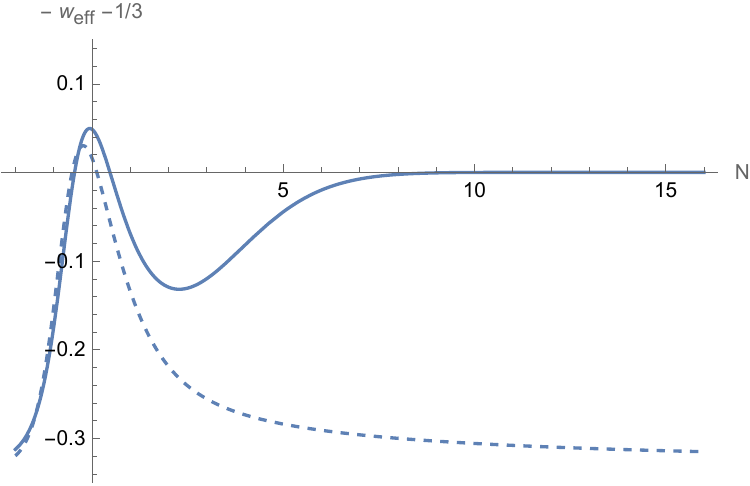}\caption{$\lambda=\sqrt{3}$\\ $-0.45 \leq N \leq 0.47$}\label{fig:Acck1}
\end{subfigure}\quad
\begin{subfigure}[H]{0.48\textwidth}
\includegraphics[width=\textwidth]{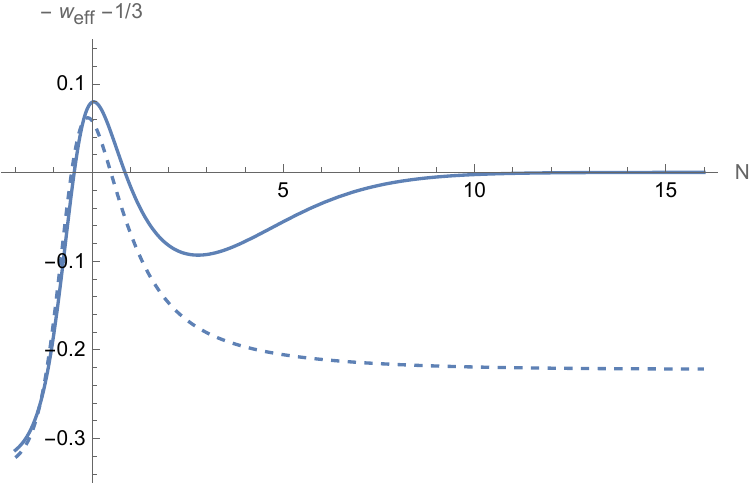}\caption{$\lambda=\sqrt{\frac{8}{3}}$\\ $-0.46 \leq N \leq 0.87$}\label{fig:Acck2}
\end{subfigure}\\
\begin{subfigure}[H]{0.48\textwidth}
\includegraphics[width=\textwidth]{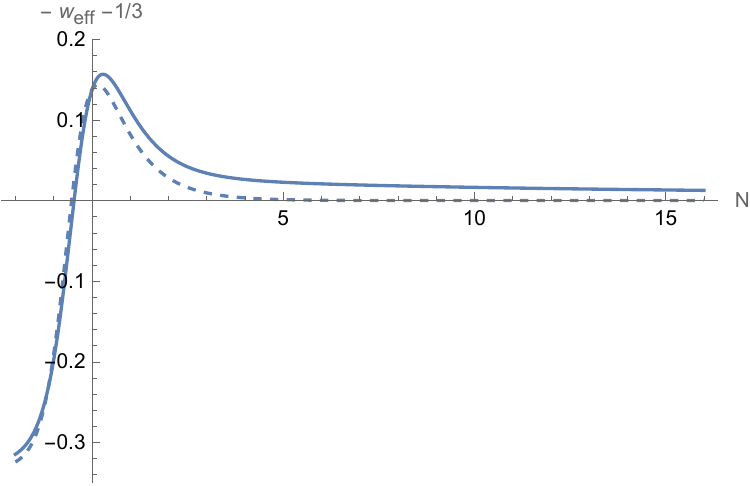}\caption{$\lambda=\sqrt{2}$\\ $-0.47 \leq N $}\label{fig:Acck3}
\end{subfigure}\quad
\begin{subfigure}[H]{0.48\textwidth}
\includegraphics[width=\textwidth]{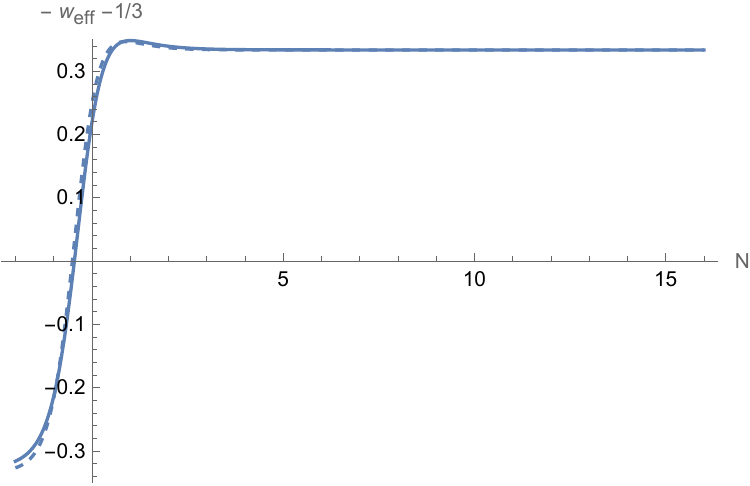}\caption{$\lambda=1$\\ $-0.46 \leq N $}\label{fig:Acck4}
\end{subfigure}
\caption{Evaluation of $-w_{{\rm eff}}-\frac{1}{3}$ in candidate realistic solutions (plain curves), using $w_{\phi0}$ given in \eqref{wphi0valuesk} for each $\lambda$ value, and $\Omega_{k0}=0.085,\, \Omega_{m0}=0.315 ,\, \Omega_{\phi0}=0.600$. This figure is to be compared to figure \ref{fig:Acc}, where $\Omega_{k0}=0$: for convenience, we reproduce here the corresponding curves as dashed. When the quantity $-w_{{\rm eff}}-\frac{1}{3}$ is positive, we have acceleration: for each $\lambda$ value, we indicate the corresponding e-fold range. As expected, we find  transient acceleration phases for $\lambda > \sqrt{2}$, and semi-eternal ones for $\lambda \leq \sqrt{2}$. We also note the change in asymptotic behaviours, according to $\lambda$ and to the corresponding fixed point. For $\lambda > \sqrt{2}$, we obtain asymptotically here no acceleration as expected for $\Pkp$. In addition, for $\lambda=\sqrt{3}$ a zoom in at the tail (not displayed here) reveals an oscillation between deceleration and acceleration phases, as expected from the inspiraling approach to $\Pkp$ for this value of $\lambda$.}\label{fig:Acck}
\end{center}
\end{figure}

Finally, we compare the acceleration phase of the solutions in presence of curvature to the results of section \ref{sec:accphase} without curvature. As shown in figure \ref{fig:Acck}, we see small differences in its start and duration; in particular curvature (with a correspondingly lower dark energy density parameter) leads to a later start of the acceleration phase.

\section{String theory realisations}\label{sec:stringtheory}

Considering a quintessence model with a single (canonically normalised) scalar field and an   exponential potential \eqref{potexp}, together with matter and radiation contributions, we have identified cosmological solutions that could provide a realistic past and present universe. Theoretical and observational constraints nevertheless set an upper bound on the exponential rate $\lambda$. We have argued that spatial curvature could alleviate these constraints by raising the  upper bound, possibly allowing for $\lambda \geq \sqrt{2}$. If this is confirmed by a more advanced cosmological analysis, it opens the door to simple string theoretical realisations of those solutions. While including matter and radiation would remain a challenge for string theory models in cosmological settings, obtaining an exponential scalar potential with $\lambda \geq \sqrt{2}$ is possible. We discuss in this section stringy constructions of such potentially viable quintessence scenarios.

Several compactifications of 10d string theory to 4d models with an exponential potential were presented in \cite{Marconnet:2022fmx}, as consistent truncations of 10d type II supergravities. Two of them were detailed in \cite[Sec. 4]{Andriot:2023wvg}: one with $\lambda =\sqrt{\frac{8}{3}}\approx 1.63$ and one with $\lambda =\sqrt{\frac{26}{3}}\approx 2.94$. The latter is larger than the upper bounds found theoretically in section \ref{sec:today}, from the requirements of past radiation and matter domination and acceleration today, so such a model is not expected to provide a realistic quintessence scenario; instead, the former may stand a chance. This compactification with $\lambda =\sqrt{\frac{8}{3}}$ and $k=-1$ was first considered in \cite{Andersson:2006du} and \cite[Sec. 6.1]{Marconnet:2022fmx}: it consists in a negatively curved 6d compact Einstein manifold, with all 10d fluxes vanishing, and a constant dilaton. The quintessence 4d scalar field corresponds to the 6d volume, which grows when rolling down the potential. The constant dilaton can be fixed in such a way that one has a small string coupling. As argued in \cite[Sec. 4.3]{Andriot:2023wvg}, this allows the solution to be in a classical regime of string theory, neglecting $\alpha'$ and loop corrections. In addition, the 6d Einstein manifold can be chosen in such a way that the 6d volume is the only geometric modulus. Finally, the universe's expansion is faster than the growth of the 6d volume, such that the 4d cosmology remains scale separated. We refer to \cite[Sec. 4.3]{Andriot:2023wvg} for more detail on the control of this string compactification.

While appearing trustable, this string construction still exhibits an evolving internal volume; this is a general feature of the time dependent compactifications of \cite{Marconnet:2022fmx}. The volume modulus generally couples with gravitational strength to all fields in the 4d theory, including the visible matter; moreover, due to its special role, it generally appears in the masses and couplings of the 4d effective field theory. A light and evolving volume could therefore lead to time varying fundamental constants and fifth forces, challenges that are difficult to address without a more concrete model.  With respect to fifth forces, it would be interesting to explore possible screening mechanisms within a concrete string construction that includes both quintessence and matter, e.g. via interactions with its axionic partner \cite{Brax:2023qyp}.

\begin{figure}[H]
\begin{center}
\begin{subfigure}[H]{0.48\textwidth}
\includegraphics[width=\textwidth]{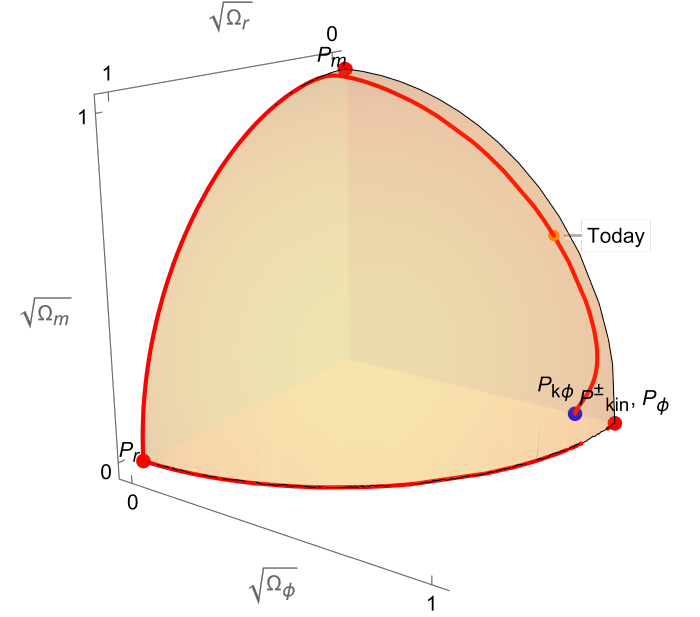}\caption{Solutions}\label{fig:Solphi0}
\end{subfigure}\quad
\begin{subfigure}[H]{0.48\textwidth}
\includegraphics[width=\textwidth]{SolphiO.pdf}\caption{$\Omega_n$}\label{fig:SolphiO}
\end{subfigure}\\
\begin{subfigure}[H]{0.48\textwidth}
\includegraphics[width=\textwidth]{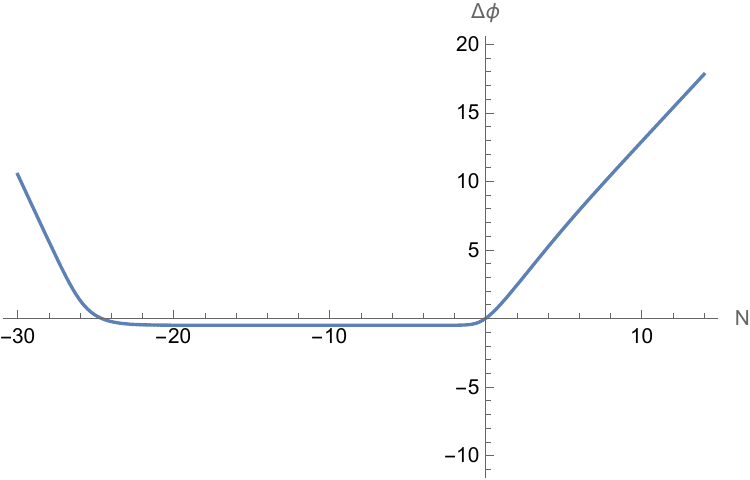}\caption{$\Delta \phi,\, \Pkinm$}\label{fig:Solphiphim}
\end{subfigure}\quad
\begin{subfigure}[H]{0.48\textwidth}
\includegraphics[width=\textwidth]{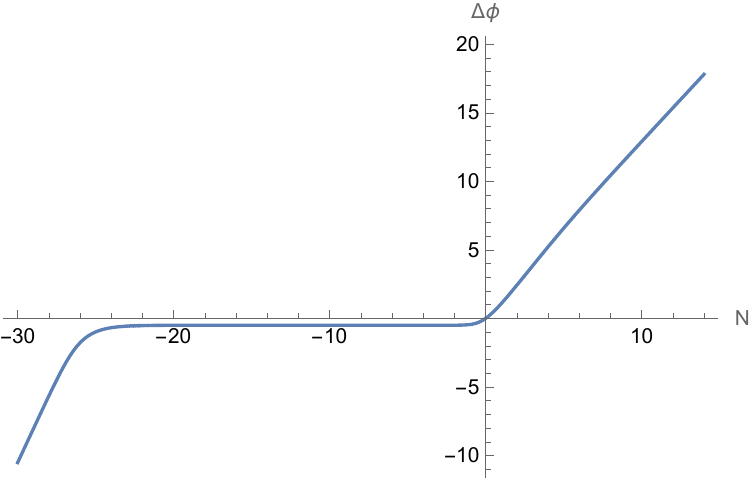}\caption{$\Delta \phi,\, \Pkinp$}\label{fig:Solphiphip}
\end{subfigure}
\caption{Cosmological solutions for $\lambda=\sqrt{\frac{8}{3}}$, starting at $\Pkin$ and passing through today's universe given by $\Omega_{\phi0}=0.600,\, \Omega_{k_0}=0.085,\, \Omega_{r0}=0.0001,\, \Omega_{m0}=0.3149$, and $w_{\phi_0}=-0.6399875086151$ for $\Pkinm$, $w_{\phi_0}=-0.6399875086149$ for $\Pkinp$. The latter tuning allows to have radiation domination starting in both cases at $N_r \approx -26$. We present the evolution of the $\Omega_n$, in agreement with descriptions of section \ref{sec:curvature}; $w_{\phi}(N)$ is depicted in figure \ref{fig:Solphiwphiintro}. In figure \ref{fig:Solphiphim} and \ref{fig:Solphiphip}, we present, for the solution starting at $\Pkinm$ and $\Pkinp$ respectively, the field evolution of $\Delta\phi(N) \equiv \phi(N)-\phi(0)$. This is obtained using the following expression for the field: $-\frac{1}{\lambda} \ln \left(\frac{y^2}{z^2}\, e^{-2\, N} \right)$. In these curves, we first observe the decreasing or growing field in the initial kination phase, depending on the starting point $\Pkin$. We then observe an approximately constant field for $-20 \lesssim N\lesssim -2$; in particular we find $\Delta\phi(-20)=-0.495$ (for $\Pkinm$) or $\Delta\phi(-20)=-0.503$ (for $\Pkinp$), and $\Delta\phi(-2)=-0.497$ in both cases. We interpret this constant field, during radiation and matter domination, as being due to a high Hubble friction; we also verify numerically the large values $H(N)/H(0)$ for negative $N$. The end of the field evolution follows the growth expected from \eqref{solfixev}.}\label{fig:Solphi}
\end{center}
\end{figure}

As a first look at the potential variation of fundamental constants we can examine the evolution of the scalar field and determine its field range. This evolution can be partially understood analytically by considering the different fixed points that the solutions of interest pass close to: the  solutions presented in table \ref{tab:fixedpointsfields} may then serve as local approximations to the solutions we consider. For $k=-1$ and $\sqrt{2} <  \lambda < \sqrt{3}$, as is the case here, the candidate realistic solutions evolve as follows along the fixed points
\be
\hspace{-0.25in} \xymatrix{
\text{Solution:}\hspace{-0.2in} & \Pkin\ \ar[r] & \ \Prr \ \ar[r] & \ \Pm \ \ar[r]^{\text{(today)}} & \ \Pp \ \ar[r] & \ \Pkp \\
\phi(N):\hspace{-0.2in} &  \phi_{0,{\rm kin}} \pm \sqrt{6}\, N \ \ar[r] & \ \phi_{0,r} \ \ar[r]  & \ \phi_{0,m} \ \ar[r] & \ \phi_{0,\phi} +\lambda \, N \ \ar[r] & \ \phi_{0,k\,\phi} +\frac{2}{\lambda} \, N
} \label{solfixev}
\ee
where the constant field values $\phi_{0,n}$ may differ at each point, though constrained by continuity of the solution. We see that close to maximal radiation domination, and to maximal matter domination, the field is approximately constant (taking possibly different values $\phi_{0,r},\ \phi_{0,m}$); what happens in between is a priori unclear. However, we verify numerically for the solution in figure \ref{fig:Solphi} that the field remains almost constant throughout this whole period, for $-20 \lesssim N\lesssim -2$, with $\Delta\phi\approx 0.5\, M_p$ between $-20 \leq N \leq 0$ (restoring units).

This behaviour of the scalar field helps towards  circumventing the issues pointed out above:  at least while the field is constant -- as ensured by Hubble friction -- there is no issue of time varying fundamental constants.   Moreover, the scalar being frozen for most of the cosmological history helps to avoid the implications of the swampland distance conjecture, namely that when the field traverses super-Planckian distances a tower of light states emerges with an associated breakdown of the effective field theory \cite{Ooguri:2006in, Baume:2016psm, Klaewer:2016kiy}. Understanding how much the field is allowed to vary once it starts thawing until today, whilst still being consistent with observations and control of the effective theory, would require a complete model.

On the other hand, it is also interesting to note that all the solutions that we have presented start at the one of the fully unstable fixed points, $\Pkin$, which correspond to a kination epoch.\footnote{This is a consequence of quintessence, $\forall \lambda >0$: the strict $\Lambda$CDM limit where $\lambda=0, w_{\phi}=-1$ does not allow for such an initial phase.} Indeed, early epochs of kination followed by a braking Hubble friction due to radiation, have recently been discussed \cite{Conlon:2022pnx, Apers:2022cyl, Cicoli:2023opf, Apers:2024ffe} as a class of non-standard cosmology that appears very naturally in string cosmology (and which solves an otherwise common ``overshoot problem'' in string cosmology). Clearly, during kination, the scalar evolves significantly, traversing super-Planckian distances; the effective field theory may therefore actually break down before $\Pkin$ is reached in the past. However, a complete model, patching to an early inflation phase, would be needed before reaching any such conclusion; this work provides some motivation for constructing it.

\vspace{0.4in}

\subsection*{Acknowledgements}

We would like to thank A.~Moradinezhad, V.~Poulin, N.~Sch\"oneberg and G.~Tringas for very helpful discussions. We also thank the organisers of the String Phenomenology 2023 Conference, that allowed  interactions leading to this work. The work of S.~L.~P.~is partially supported by the UK Science and Technology Facilities Council grant ST/X000699/1. The work of T.~W.~is supported in part by the NSF grant PHY-2210271. I.~Z.~is partially funded by the STFC grants ST/T000813/1 and ST/X000648/1. This research was supported in part by grant NSF PHY-2309135 to the Kavli Institute for Theoretical Physics (KITP). For the purpose of
open access, the authors have applied a Creative Commons Attribution licence to any Author Accepted Manuscript version arising.

\newpage

\begin{appendix}

\section{$d$-dimensional dynamical system}\label{ap:dynsys}

In this appendix, we generalize the dynamical system analysis of section \ref{sec:dynsys} to a $d$-dimensional spacetime, $d\geq 3$. Beyond the scalar field and the spatial curvature, we allow for simplicity for only one perfect fluid component, leaving it however free (e.g.~matter or radiation).

\subsection{Cosmology}

In $d$ dimensions, the FLRW metric is given by
\be\label{metricd}
\d s^2 = -\d t^2 +a^2(t) \left( \frac{\d r^2}{1-kr^2}+r^2 \d\Omega_{d-2}^2\right)\,,
\ee
with $a>0$ and $k=0,\pm1$ characterizing the spatial curvature. The equations of motion \eqref{eq:Fried}-\eqref{eq:phi} generalize to the following ones
\begin{subequations}
 \begin{align}
      H^2 &= \frac{2}{(d-1)(d-2)}\ \sum_n\rho_n \,,\label{eq:Friedd}\\
    \dot H &= -\frac{1}{d-2}\ \sum_n\rho_n +p_n \,,\label{eq:Hdotd}\\
    \ddot \phi &= -(d-1)H\dot \phi - \del_\phi V \,,  \label{eq:phid}
 \end{align}
\end{subequations}
and we  assume $H\neq 0$. The sum over $n$ includes all possible effective fluid components, for instance radiation, matter, the scalar field $\phi$ and the spatial curvature $k$. For each component, we introduce the equation of state parameter $w_n$ such that $p_n= w_n \, \rho_n$. The components to be considered in the following are listed in table \ref{tab:nd}: for simplicity, we consider only one perfect fluid component, but we keep it generic, capturing the possibility of having e.g.~either radiation or matter (see table \ref{tab:n} for the 4d study which includes both radiation and matter).

\begin{table}[H]
\begin{center}
\centering
\begin{tabular}{| l | c || c | c | c |}
\hline
\cellcolor[gray]{0.9} & \cellcolor[gray]{0.9} & \cellcolor[gray]{0.9} & \cellcolor[gray]{0.9} & \cellcolor[gray]{0.9}\\[-8pt]
\cellcolor[gray]{0.9} $\!\! n$ & \cellcolor[gray]{0.9} component &  \cellcolor[gray]{0.9} $\rho_n$ & \cellcolor[gray]{0.9} $p_n$ & \cellcolor[gray]{0.9} $w_n$ \\[5pt]
\hline
&&&&\\[-8pt]
 & generic & $\rho$  & $p$ & $w$ \\[3pt]
\hline
&&&&\\[-8pt]
$k$ & curvature & $-(d-1)(d-2)\, \frac{k}{2a^2}$  & $(d-3) (d-2)\, \frac{k}{2a^2}$ & $-\frac{d-3}{d-1}$ \\[6pt]
\hline
&&&&\\[-8pt]
$\phi$ & scalar field & $\frac{\dot\phi^2}{2} + V(\phi)$  & $\frac{\dot\phi^2}{2} - V(\phi)$ & $w_{\phi}$ \\[6pt]
\hline
\end{tabular}
\end{center}
\caption{Perfect fluid notations in $d$ dimensions with, for each component $n$, the energy density $\rho_n$, the pressure $p_n$, and the equation of state parameter $w_n=\frac{p_n}{\rho_n}$.}
\label{tab:nd}
\end{table}

For each component, we define the density ratio as
\begin{equation}
\Omega_n=\frac{2\rho_n}{(d-1)(d-2)H^2}\,,
\end{equation}
allowing one to write the first Friedmann equation \eqref{eq:Friedd} as in \eqref{eq:friO}
\be\label{eq:friOd}
1 = \sum_n\Omega_n \quad \Leftrightarrow \quad 1-\Omega_k =\Omega_T \,.
\ee
As in 4d, we also define an effective equation of state of the full system, $p_{\rm eff}=w_{\rm eff} \rho_{\rm eff}$, where $p_{\rm eff} = \sum_n p_n$, $\rho_{\rm eff} = \sum_n \rho_n$. Similarly to \eqref{weffO}, we obtain here
\be
w_{\rm eff} = \sum_n w_n \Omega_n = w_{\phi} \,\Omega_{\phi} -\frac{d-3}{d-1}\Omega_k + w \Omega \ .\label{weffOd}
\ee
Finally, while $\epsilon$ is defined as in \eqref{epsilon}, we now rewrite it with \eqref{eq:Hdotd} as
\be
\epsilon = \frac{d-1}{2} (1 + w_{\rm eff} ) \,,
\ee
leading to the following condition for acceleration
\be
\ddot{a}>0 \ \Leftrightarrow \ \epsilon<1 \ \Leftrightarrow \ w_{\rm eff} <-\frac{d-3}{d-1} \,. \label{accd}
\ee

\subsection{Dynamical system}\label{ap:dynsysintro}

Proceeding as in section \ref{sec:dynsysintro}, we focus on the case $k \leq 0$, with a single perfect fluid, and define the following dynamical variables in $d$ dimensions
\be\label{eq:variablesd}
   x= \frac{\phi'}{\sqrt{(d-1)(d-2)}} \,, \qquad
   y = \frac{\sqrt{2V}}{\sqrt{(d-1)(d-2)}\, H}\,,    \qquad
   z = \frac{\sqrt{-k}}{a H}\,, \qquad
   \lambda= -\frac{\del_\phi V}{V} \,,
\ee
with the prime $'$ denoting the derivative with respect to the number of e-folds $\d N= H \d t$. We assume again $a> 0,\ H\neq0,\ V\geq 0$. We have $\Omega_\phi=x^2+y^2$ and $\Omega_k=z^2$. The first Friedmann equation \eqref{eq:friOd} can be rewritten as
\be
 \Omega=1-  x^2- y^2- z^2 \,, \label{Omegaxyzd}
\ee
and will be used as an extra constraint on the system. We also rewrite \eqref{weffOd} as
\be
w_{\rm eff}  = x^2 -y^2 -\frac{d-3}{d-1}z^2 +w\,\Omega\,.\label{weffd}
\ee

Using the second Friedmann equation \eqref{eq:Hdotd} and the scalar field e.o.m.~\eqref{eq:phid}, the 4d system is generalized as follows
\begin{subequations}\label{eq:systemd}
  \begin{empheq}[box=\fbox]{align}
 \ \  x'&= \frac{\sqrt{(d-1)(d-2)}}{2}\, y^2\lambda +  x \left( (d-1) (x^2-1)  + z^2 +\frac{d-1}{2}(1+w) \,\Omega \right) \,,\
    \label{eq:xpd}\\
  y' &= y\left( -\frac{\sqrt{(d-1)(d-2)}}{2}\, x \,\lambda + (d-1) x^2 + z^2 +\frac{d-1}{2}(1+w)\, \Omega  \right) \,, \label{eq:ypd}\\
        z' &= z\left( z^2-1 +(d-1) x^2 + \frac{d-1}{2}(1+w)\,\Omega \right) \,, \label{eq:zpd}\\
        \lambda' &= -\sqrt{(d-1)(d-2)}\,x\, \left( \frac{\del^2_{\phi} V}{V} -\frac{(\del_{\phi} V)^2}{V^2}\right)\,,\label{eq:lpd}
\end{empheq}
\end{subequations}
together with the constraint \eqref{Omegaxyzd}.

In the following, we restrict ourselves to an exponential scalar potential
\be
V(\phi) = V_0\ e^{-\lambda \phi} \ , \label{potexpd}
\ee
where $\lambda$ now matches the exponential rate; we take $\lambda>0$ and $V_0\geq0$. As in 4d, the above equations and variables can then also be considered in the case where $V = V_0 = 0$. As $\lambda$ is constant, its equation \eqref{eq:lpd} is trivially satisfied, so we can ignore this variable and equation in the following.

\subsection{Fixed points}\label{ap:fixedpointsd}

We now look for the fixed points to the system \eqref{eq:systemd} with constraint \eqref{Omegaxyzd} and exponential potential \eqref{potexpd}. To that end, let us first consider the possibility of having $z\neq0$. The system boils down to the following three equations at the fixed points
\bea
0 &=& \frac{\sqrt{(d-1)(d-2)}}{2}\, y^2\lambda - (d-2) x \nn\\
0 &=& y \left(1- \frac{\sqrt{(d-1)(d-2)}}{2} \, x \lambda \right) \\
0 &=& z^2-1 +(d-1) x^2 + \frac{d-1}{2}(1+w)\,\Omega \,.\nn
\eea
For $y=0$, one finds $x=0$. One is then left to solve the last equation together with the constraint \eqref{Omegaxyzd}. There are two solutions in terms of $(x,y,z)$:
\be
(x,y,z)= (0,0,\pm 1) \ {\rm with}\ \Omega=0
\ee
and
\be
(x,y,z)= (0,0,\pm \sqrt{1-\Omega}) \ {\rm with}\ w=-\frac{d-3}{d-1} \,,
\ee
where the latter is not a fixed point, but a fixed line (segment). For $y\neq0$, one obtains again two solutions to the system with the constraint: the fixed point
\be
(x,y,z)= \left(\frac{2}{\lambda \sqrt{(d-1)(d-2)}} , \pm \frac{2}{\lambda \sqrt{d-1}} , \pm \sqrt{1-\frac{4}{\lambda^2 (d-2)}}\right)  \ {\rm with}\ \Omega=0 \,,
\ee
and the fixed line
\be
(x,y,z)= \left(\frac{2}{\lambda \sqrt{(d-1)(d-2)}} , \pm \frac{2}{\lambda \sqrt{d-1}} , \pm \sqrt{1-\frac{4}{\lambda^2 (d-2)} - \Omega}\right)  \ {\rm with}\ w=-\frac{d-3}{d-1} \,.\nn
\ee

We now turn to the case of $z=0$. For $y=0$, we obtain three solutions to the system with constraint: two fixed points
\bea
& (x,y,z)= (0,0,0) \ {\rm with}\ \Omega=1 \,,\\
& (x,y,z)= (\pm 1,0,0) \ {\rm with}\ \Omega=0 \,,
\eea
and the fixed line
\be
(x,y,z)= (\pm \sqrt{1-\Omega},0,0) \ {\rm with}\ w=1 \,.
\ee
Finally, for $z=0$ and $y\neq0$, using the constraint, the system can be rewritten as
\bea
\!\!\! y^2 &=& \frac{2}{\lambda}\, \sqrt{\frac{d-1}{d-2}} \ x - x^2 \,,\\
\!\!\! \frac{d-1}{2}(1+w)\, y^2 &=& \frac{d-1}{2}(1+w) -\frac{\sqrt{(d-1)(d-2)}}{2}\, x \,\lambda + \frac{d-1}{2}(1-w)\, x^2  \,.
\eea
This can be solved and gives two fixed point solutions:
\be
(x,y,z)= \left(\frac{1+w}{\lambda} \sqrt{\frac{d-1}{d-2}},\pm \frac{\sqrt{1-w^2}}{\lambda} \sqrt{\frac{d-1}{d-2}} , 0\right) \ {\rm with}\ \Omega=1- \frac{2(1+w)}{\lambda^2} \frac{d-1}{d-2} \,,\ |w|< 1 \,,
\ee
and
\be
(x,y,z)= \left(\frac{\lambda}{2} \sqrt{\frac{d-2}{d-1}},\pm\sqrt{1-\frac{\lambda^2}{4} \frac{d-2}{d-1}},0 \right) \ {\rm with}\ \Omega=0 \,.
\ee

We summarize in table \ref{tab:fixedpointsd} the fixed points found for the system \eqref{eq:systemd} with constraint \eqref{Omegaxyzd} for the exponential potential \eqref{potexpd}. We do not summarize the three fixed lines found above since the corresponding $w$ values generally correspond to neither matter nor radiation (see however appendix \ref{ap:analytic} for a related discussion of fixed lines and fixed planes). For each fixed point, we compute $w_{\rm eff}$ using \eqref{weffd}; from the latter, one determines whether the fixed point solution is accelerating, as indicated in \eqref{accd}. For $\Pp$, we recover the known condition for acceleration $\lambda < 2/\sqrt{d-2}$ \cite{Rudelius:2022gbz, Andriot:2022xjh}.

\begin{table}[H]
\begin{center}
\centering
\begin{tabular}{| l | c | c | c | }
\hline
\cellcolor[gray]{0.9} & \cellcolor[gray]{0.9} & \cellcolor[gray]{0.9} & \cellcolor[gray]{0.9} \\[-8pt]
\cellcolor[gray]{0.9} \hskip 1.4cm $(x,y,z)$ & \cellcolor[gray]{0.9} $\Omega$ &  \cellcolor[gray]{0.9} Existence & \cellcolor[gray]{0.9} $w_{\rm eff}$ \\[5pt]
\hline
&&&\\[-8pt]
 $\Pkin= $ $(\pm 1,0,0) $ & $0$  & $ \forall\, \lambda, w$ & $1$ \\[3pt]
\hline
&&&\\[-8pt]
$\Pk= $ $(0,0,\pm1) $ & $ 0$  & $ \forall\, \lambda, w$ & $-\frac{d-3}{d-1}$ \\[3pt]
\hline
&&&\\[-10pt]
 $\Pkp= $ $\lp\frac{2}{\lambda \sqrt{(d-2)(d-1)}},\pm\frac{2}{\lambda \sqrt{d-1}},\pm \sqrt{1-\frac{4}{\lambda^2 (d-2)}}\rp $ & $0$  & $ \lambda>\frac{2}{\sqrt{d-2}}$ & $ -\frac{d-3}{d-1}$ \\[8pt]
\hline
&&&\\[-10pt]
$\Pp= $ $\lp \frac{\lambda}{2}\sqrt{\frac{d-2}{d-1}}, \pm \sqrt{1-\frac{\lambda^2}{4}\frac{d-2}{d-1}}, 0 \rp $ & $ 0$  & $  \lambda < 2\sqrt{\frac{d-1}{d-2}}$ & $\frac{\lambda^2}{2}\frac{d-2}{d-1}-1$ \\[8pt]
\hline
&&&\\[-10pt]
$P_{w\phi}=$ $\lp \frac{1+w}{\lambda} \sqrt{\frac{d-1}{d-2}},\pm \frac{\sqrt{1-w^2}}{\lambda} \sqrt{\frac{d-1}{d-2}},0\rp $ & $ 1-\frac{2(1+w)}{\lambda^2} \frac{d-1}{d-2}$  & $\lambda^2 > 2(1+w) \frac{d-1}{d-2} \ ,$   & $w $\\[4pt]
&& $|w| < 1$ &\\[8pt]
\hline
&&&\\[-8pt]
$P_{w}=(0,0,0) $ & $1$  & $ \forall\, \lambda, w$ & $w$ \\[3pt]
\hline
\end{tabular}
\end{center}
\caption {Fixed points for the system \eqref{eq:systemd} with the constraint \eqref{Omegaxyzd} and the exponential potential \eqref{potexpd}, for a given value of $w$. The existence condition on $\lambda$ for $P_{w\phi}$ comes from the requirement $\Omega \geq 0$ that we impose here; we make the inequality strict to distinguish it from $\Pp$. The points obtained generalize those of table \ref{tab:fixedpoints}, and we give them here the corresponding names; $P_{w\phi}$ could stand e.g.~for $\Pmp$ or $\Prp$ , and $P_{w}$ for $\Pm$ or $\Prr$.}
\label{tab:fixedpointsd}
\end{table}

\subsection{Fixed points, including radiation and matter}\label{ap:radmat}

Let us now derive the fixed points for the case $d=4$, including both radiation and matter, as presented in table \ref{tab:fixedpoints} of section \ref{sec:fixedpoints4d}.  In fact, if we set $d=4$ and $w=0$ in the system \eqref{eq:systemd} and constraint \eqref{Omegaxyzd}, we obtain the same system and constraint as considered in 4d, \eqref{eq:system} and \eqref{Omegaxyzu}, provided one neglects the radiation there by setting $u=0$. The way we found the fixed points above therefore applies to section \ref{sec:fixedpoints4d} for $u=0$.  It remains to find the fixed points for the 4d system that have $u\neq 0$. Plugging the latter into \eqref{eq:system} implies that $z=0$, so the system boils down to
\be
0 = \sqrt{\frac{3}{2}}\, y^2\,\lambda - x\,,\quad 0 = y \left(- \sqrt{\frac{3}{2}}\, x \,\lambda + 2\right) \,, \quad \frac32 \Omega_m +2u^2 = \, 2- 3 \, x^2  \,,
\ee
with the constraint
 \be
 \Omega_m= 2 x^2 - 4 y^2 \,.
\ee
For $y=0$, one finds $x=0$, implying with the constraint that $\Omega_m=0$ and $u^2=1$. In other words, we obtain for $(x,y,z,u)$ the fixed point
\be
(x,y,z,u)= (0,0,0,\pm 1) \ {\rm with}\ \Omega_m=0 \,.
\ee
For $y\neq 0$, the system can also be solved and one finds the following fixed point
\be
(x,y,z,u)= \left( \frac{1}{\lambda} \sqrt{\frac{8}{3}},\pm \frac{2}{\lambda\, \sqrt{3}},0, \pm \sqrt{1-\frac{4}{\lambda^2}} \right) \ {\rm with}\ \Omega_m=0 \,.
\ee
These are the results summarized in table \ref{tab:fixedpoints} for $u\neq0$.

\subsection{Stability}\label{ap:stab}

We end the $d$-dimensional analysis by outlining the stability of the fixed points listed in table \ref{tab:fixedpointsd}. We proceed as described in section \ref{sec:stab}, for the system \eqref{eq:systemd} with constraint \eqref{Omegaxyzd}, and list the results in table \ref{tab:stabd}. We refrain from commenting on the corresponding stability behaviour, since it depends on the general parameters $d,w$.

\begin{table}[H]
\begin{center}
\centering
\begin{tabular}{| l | c | }
\hline
\cellcolor[gray]{0.9} & \cellcolor[gray]{0.9} \\[-8pt]
\cellcolor[gray]{0.9} Point & \cellcolor[gray]{0.9} Eigenvalues \\[5pt]
\hline
&\\[-10pt]
$\Pkin $ & $ \lp (d-1)(1-w),\ d-1 \mp \frac{\lambda}{2} \sqrt{(d-1)(d-2)},\ d-2 \rp$ \\[8pt]
\hline
&\\[-8pt]
$\Pk$ & $\lp 2-d,\ 1 ,\ 2+ (1-d)(1+w)  \rp $ \\[4pt]
\hline
&\\[-10pt]
$\Pkp$ & $w=0:\ \lp 3-d , \ \frac{1}{2}\left(2 - d \pm \sqrt{(10-d)(2-d) + \frac{32}{\lambda^2}}\right)  \rp$  \\[8pt]
\hline
&\\[-10pt]
$\Pp$ & $\lp \frac{\lambda^2}{4} (d-2) -1 ,\ \frac{\lambda^2}{4} (d-2) +  1-d ,\ (1-d)(1+w) + \frac{\lambda^2}{2} (d-2)    \rp$  \\[8pt]
\hline
&\\[-10pt]
$P_{w\phi}$ & $\lp \frac{d-1}{4} (w - 1), \left( 1 \pm \sqrt{ \frac{ 16 (1-d) (1+ w)^2 + (d-2) \lambda^2 (7+9w) }{(d-2)(w - 1) \lambda^2 }  } \right) ,\ \frac{1}{2} \left( -2 + (d-1) (1+w) \right) \rp$  \\[8pt]
\hline
&\\[-8pt]
$P_{w}$ & $\lp \frac{1}{2}(d-1)(w-1),\ \frac{1}{2}(d-1)(1+w), \ \frac{1}{2}(d-1)(1+w)-1  \rp$ \\[4pt]
\hline
\end{tabular}
\end{center}
\caption {Eigenvalues of the system Jacobian, that determine the stability of the fixed points of table \ref{tab:fixedpointsd}. For $\Pkp$ and $P_{w\phi}$ we give for simplicity two eigenvalues together, as they only differ by a sign. For $\Pkp$, we only determined the eigenvalues for $w=0$.}
\label{tab:stabd}
\end{table}

\section{Fixed subspaces and analytical solutions}\label{ap:analytic}
In this appendix we further study the fixed loci, namely fixed curves and surfaces, of the dynamical system in equation \eqref{eq:systemd} with constraint \eqref{Omegaxyzd} and present corresponding analytical solutions. As above, we restrict to an exponential potential $V(\phi) = V_0\ e^{-\lambda \phi}$ that automatically solves the equation \eqref{eq:lpd} for $\lambda$ since we take the latter to be a positive constant. We also take $d=4$ to allow easier comparison with the  solutions in the vicinity of the fixed points, described in appendix \ref{ap:asymptotics}.

\subsection{Fixed curves and surfaces}\label{ap:fixedsubsets}
The dynamical system in equation \eqref{eq:systemd} admits two fixed planes given by
\be
\mathcal{P}_y=\{(x,y,z)~|~y=0\}~, \qquad \mathcal{P}_z=\{(x,y,z)~|~z=0\}~,
\ee
corresponding to vanishing potential or curvature, respectively. It follows  that the intersection,
$$\mathcal{L}_x=\mathcal{P}_y\cap \mathcal{P}_z~,$$
is a fixed line. It can also be seen that the $z$-axis,
$$\mathcal{L}_z=\{(x,y,z)~|~x=y=0\}~,$$
is a fixed line. The unit sphere corresponding to $\Omega=0$ (see \eqref{Omegaxyzd}),
$$\mathcal{S}=\{(x,y,z)~|~x^2+y^2+z^2=1\}~,$$
is a fixed surface, as follows from,
\eq{
(x^2+y^2+z^2)'=3\Omega\left[(w-1)x^2+(w+1)y^2+(w+\frac13)z^2\right]
~.}
It also follows from the above that the intersections,
$$\mathcal{C}_y=\mathcal{S}\cap \mathcal{P}_y~,~~~\mathcal{C}_z=\mathcal{S}\cap \mathcal{P}_z~,$$
are fixed circles.

The fixed surface $\mathcal{S}$ corresponds to a model of the universe without matter, radiation or any other extra fluid, $\Omega=0$. In particular the dynamical system and associated solutions of \cite{Andriot:2023wvg} are obtained by restriction of the system \eqref{eq:systemd} to this surface ${\cal S}$ of the unit sphere. Indeed the fixed points $\Pkin, \Pk, \Pkp,\Pp$ all lie on this surface and have been examined in~\cite{Andriot:2023wvg}.

The dynamical system in equation \eqref{eq:systemd} admits two fixed planes in addition to the ones above, for the following special cases:
\begin{itemize}
\item Curvature: $w=-\frac13$ and $\lambda=\sqrt{\frac{2}{3}}$.
In this case we have the two additional fixed planes $x\pm\sqrt{2}y+1=0$, as can be seen from the equations (following from \eqref{eq:systemd}),
\eq{
x'\pm\sqrt{2}y'=(x\pm\sqrt{2}y+1)(2x^2-y^2\pm \sqrt{2}y-2x)\,.
}
\item Radiation: $w=\frac13$ and $\lambda=2\sqrt{\frac{2}{3}}$.
In this case we have the two additional fixed planes $x\pm z-1=0$, as can be seen from,
\eq{
x'\pm z'=(x\pm z-1)(x^2-2y^2-z^2+x\mp z)\,.
}
\end{itemize}

\subsection{Analytic solutions}

In this subsection we find analytic solutions for some of the fixed subspaces discussed in the previous subsection.

\subsubsection{Analytic solutions on $\mathcal{L}_x$}\label{sec:analyticI}
On $\mathcal{L}_x$ the dynamical system reduces to the equation,
\eq{
x'=-\frac32(1-w)x(1-x^2)
~,}
which can be solved to give
\eq{
x=\pm\frac{1}{\sqrt{1+c~\!a^{3(1-w)}}}
~,}
where $c$ is a positive constant. This describes a system with vanishing potential ($V=0$) and curvature ($k=0$).
Integrating the above using equation \eqref{eq:variablesd}, we obtain the expression for the scalar,
\eq{
\phi=\phi_0\mp2\sqrt{\frac23}\frac{1}{1-w}\text{arctanh}\left(\sqrt{1+c~\!a^{3(1-w)}}\right)~,
}
with $\phi_0$ a real constant. To find the expression of the scale factor in terms of cosmological time we proceed as follows. From the conservation law for the fluid we find
\eq{\rho=\rho_{0}\, a^{-3(1+w)}~.}
Inserting this in $\Omega=1-x^2=\rho/3H^2$ and using the above, we obtain
\eq{\dot{a}^2=c_1a^{-1-3w}+c_2a^{-4}
~,}
with $c=c_1/c_2$ and $\rho_{0}=3c_1$. The equation above can be solved analytically in the case of matter ($w=0$) to give
\eq{
a(t)=\left(
\frac34t(4\sqrt{c_2} +3c_1t)
\right)^{\frac13}~,
}
where without loss of generality we have chosen the origin of time  so that  the scale factor vanishes at $t=0$. Moreover we can verify that the solution above interpolates between the two asymptotic solutions corresponding to the fixed points (see table \ref{tab:fixedpointsfields}):
\eq{\label{230}
\Pkin~\text{:}~~~a(t)=\left(3\sqrt{c_2} ~\!t\right)^{\frac13}\,,\qquad \phi(t)=\phi_0\pm\sqrt{6}~\!\ln a(t)\,, \qquad \rho=0~,
}
in the limit $t\rightarrow0$, and
\eq{\label{231}
\Pm~\text{:}~~~a(t)=\left(\frac32 \sqrt{c_1} ~\!t\right)^{\frac23}\,,\qquad \phi=\phi_0\,,\qquad \rho=\frac{4}{3t^2}~,
}
in the limit $t\rightarrow\infty$.

\subsubsection{Analytic solutions on $\mathcal{L}_z$} \label{sec:analyticII}
On $\mathcal{L}_z$ the dynamical system reduces to,
\eq{
z'=\frac12(1+3w)z(1-z^2)
~,}
which can be solved to give,
\eq{\label{zsol}
z=\pm\frac{1}{\sqrt{1+c~\!a^{-(1+3w)}}}
~,}
where $c$ is a positive constant. This describes a system with vanishing potential ($V=0$) and constant scalar,
\eq{
\phi=\phi_0~,
}
with $\phi_0$ a real constant.
For the case of matter, $w=0$, integrating \eqref{zsol} using \eqref{eq:variablesd} gives
\eq{\label{tsol}
t=\pm \sqrt{a(a+c)}\mp c~\!
\text{arcsinh}\left(\sqrt{ \frac{a}{c}}\right)~,
}
where we have chosen the origin of time so that $t=0$ corresponds to vanishing scale factor. In the following we will choose the upper sign, which corresponds to an expanding universe.

Expanding around $t=0$, corresponds to an asymptotic expansion around the fixed point $\Pm$. In this case  we can invert \eqref{tsol} to obtain,
\eq{
a(t)=c\left[
\left( \frac{3t}{2c}\right)^{\frac23}
+\frac15
\left( \frac{3t}{2c}\right)^{\frac43}+\dots
\right]
~.
}
The first term in the expansion corresponds to the exact scaling solution associated with $\Pm$, given in equation \eqref{231} above.

Expanding around $t=\infty$, corresponds to an asymptotic expansion around the fixed point $\Pk$. In this case  we can invert \eqref{tsol} to obtain,
\eq{\label{ai0}
a(t)=
t+\frac{c}{2}\ln t
+\dots
~.
}
The first term in the expansion above  corresponds to the  exact scaling solution associated to $\Pk$ as given in table \ref{tab:fixedpointsfields} above.

\subsubsection{Analytic solutions on $\mathcal{C}_z$} \label{sec:Analytic solutions III}
On $\mathcal{C}_z$ the dynamical system reduces to
\eq{
\theta'=3(\cos\theta-c)\sin\theta
~,}
where we have defined $x=\cos\theta$, $y=\sin\theta$, and $c=\frac{\lambda}{\sqrt{6}}$. This can be integrated to give
\eq{\label{57}
N=\frac{1}{3(c^2-1)}\ln\left|
\frac{\cos\theta-c}{\sin\theta}\left(
\cot\frac{\theta}{2}
\right)^c
\right|
~,}
where we have dropped an integration constant by suitably choosing the origin of e-folds $N$.
We distinguish the following two cases:
\begin{itemize}
\item $~\lambda>\sqrt{6}$ (equivalent to $c> 1$): in this case \eqref{57} describes a flow from $\Pkinm$ (corresponding to $N \rightarrow -\infty$ and $\theta\rightarrow\pi$)
to $\Pkinp$ (corresponding to $N\rightarrow +\infty$ and $\theta\rightarrow 0$). Note that indeed in the case $\lambda>\sqrt{6}$, $\Pkinp$ has one negative eigenvalue (see table \ref{tab:fixedpointstab}), which corresponds precisely to a direction of approach along the eigenvector $(0,1,0)$, i.e. tangent to $\mathcal{C}_z$ at $\Pkinp$. In other words, $\Pkinp$ is a stable node for trajectories on $\mathcal{C}_z$, in agreement with the flow from $\Pkinm$ to $\Pkinp$.
\item $\lambda<\sqrt{6}$ (equivalent to $c< 1$): let us set $\cos\theta_0=c$. In this case the critical point $\Pp$ exists and it is situated precisely on $\mathcal{C}_z$ at angle $\theta=\theta_0$  (see table \ref{tab:fixedpoints}). Now \eqref{57} describes a flow from $\Pkin$ (corresponding to $N \rightarrow -\infty$ and $\theta\rightarrow0$ or $\pi$) to $\Pp$ (corresponding to $N \rightarrow +\infty$ and $\theta\rightarrow \theta_0$). Note that indeed in the case $\lambda<\sqrt{6}$, $\Pkin$ are both fully unstable nodes (see table \ref{tab:fixedpointstab}), so they have eigenvectors tangent to $\mathcal{C}_z$ allowing to have a flow originating from either of them. One can also show the existence of appropriate eigenvectors that drive either flow towards $\Pp$ along $\mathcal{C}_z$; those  correspond to $\Pp$'s negative eigenvalue in table \ref{tab:fixedpointstab}.
\end{itemize}

\section{Asymptotic solutions near the fixed points}\label{ap:asymptotics}

In this appendix we solve the equations of motion for $a(t)$ and $\phi(t)$ perturbatively around the different fixed points. We do so in $d=4$ spacetime dimensions and we restrict to matter with $w=0$. We thus solve equations \eqref{eq:Friedd}-\eqref{eq:phid} for $\rho_m=\rho_{m,0}/a(t)^3$ and $p_m=0$. Since we are working in $d=4$ all  expansions around the fixed points asymptote to the results given in table \ref{tab:fixedpointsfields}. However, since we only allow for matter, we work in the subspace with $u=0$ as compared to section \ref{sec:dynsys}. We will frequently refer to the further restricted subspaces defined in appendix \ref{ap:fixedsubsets} above.

\paragraph{The points $\Pkin$:} From the dynamical system \eqref{eq:systemd} we see that the trajectories asymptote to $\Pkin = (x,y,z) =(\pm1,0,0)$ along the directions $\vec{v}_1=(1,0,0)$, $\vec{v}_2=(0,1,0)$, or $\vec{v}_3=(0,0,1)$ in the $(x,y,z)$ space. This follows since the Jacobian (similar to equation \eqref{Jacobian}) for the system \eqref{eq:systemd} has these three eigenvectors at this fixed point. The corresponding asymptotic solutions of the equations \eqref{eq:Friedd}-\eqref{eq:phid} for $d=4$ read
\be\label{c1}
a(t)=a_0t^{\frac13}\left(1+a_1 t^p+\mathcal{O}(t^{2p})\right)\,,\qquad \phi(t)=\phi_0\pm\sqrt{\frac{2}{3}}\ln t+\phi_1 t^p+\mathcal{O}(t^{2p})\,,
\ee
where $p=1$, $2\mp\sqrt{\frac{2}{3}}\lambda$, $\frac43$ and $\phi_1/a_1=\mp\sqrt{6}, \left(\lambda\mp\sqrt{\frac32}\right), \mp\left(\frac32\right)^{3/2}$ for $\vec{v}_1, \vec{v}_2, \vec{v}_3$, respectively.

Trajectories approaching $\Pkin$ along $\vec{v}_1$, lie entirely on the line $\mathcal{L}_x$ defined in appendix \ref{ap:fixedsubsets} and one can actually find a fully analytic solution. As we discussed in appendix \ref{sec:analyticI} it has $k=0$, $V_0=0$, and $\rho_0 = 4 a_0^3 a_1$. Trajectories along $\vec{v}_2$ lie entirely in $\mathcal{C}_z$ and correspond to the analytic solution of appendix \ref{sec:Analytic solutions III}. It has $k=0$, $\rho_0=0$, and $a_1=\frac{3V_0e^{-\lambda \phi_0}}{2(\lambda \mp 3 \sqrt{\frac32}) (\lambda \mp \sqrt{6})}$. Lastly, trajectories along $\vec{v}_3$ lie entirely in $\mathcal{C}_y$; they have $k=-1$, $V_0=0$, $\rho_0=0$, and $a_1 = \frac{9}{14 a_0^2}$.

The above trajectories asymptote to $\Pkin$ in the past, as $t\rightarrow 0$, except for $\Pkin$ in the direction of approach along $\vec{v}_{2}$ with $\lambda>\sqrt{6}$: in the latter case the exponent $p$ above is negative, and the trajectories asymptote to $\Pkinp$ in the future, as $t\rightarrow \infty$.  This is consistent with $\Pkinp$ being a stable node specifically along the $\vec{v}_{2}$ direction provided that $\lambda>\sqrt{6}$, see table \ref{tab:fixedpointstab} and figure \ref{fig:Solhighlambda}.

\paragraph{The point $\Pk$:} The point $\Pk$ corresponds to a regular Milne universe with constant scalar field, $a(t)=t$, $\phi=\phi_0$. There are again three distinct directions of approach in the $(x,y,z)$ space: $\vec{v}_1=(1,0,0)$, $\vec{v}_2=(0,1,0)$, or $\vec{v}_3=(0,0,1)$.
Trajectories approaching $\Pk$ along the $\vec{v}_1$ direction, lie entirely in $\mathcal{C}_y$ and thus have $V_0=0$, $\rho_0=0$. The corresponding asymptotic solution reads
\eq{
a(t)=t+a_1 t^{-3}+\mathcal{O}(t^{-4})\,, \qquad\phi=\phi_0+\phi_1 t^{-2}+\mathcal{O}(t^{-4}) ~,}
with $a_1=-\frac19 \phi_1^2$, and it asymptotes to $\Pk$ as $t\rightarrow\infty$.

Trajectories approaching $\Pk$ along the $\vec{v}_2$ direction, lie entirely on the sphere  $\mathcal{S}$ and thus have $\rho_0=0$. The corresponding asymptotic solutions read
\eq{
a(t)=t+a_1 t^{3}+\mathcal{O}(t^{4}) \,, \qquad \phi=\phi_0+\phi_1 t^{2}+\mathcal{O}(t^{4}) ~,}
with $\phi_1/a_1=\frac94\lambda$, $a_1 = \frac{1}{18} e^{-\lambda \phi_0} V_0$, and they asymptote to $\Pk$ as $t\rightarrow 0$.

Trajectories approaching $\Pk$ along the $\vec{v}_3$ direction, lie entirely in $\mathcal{L}_z$ and thus have $V_0=0$, $\phi=\text{const}$. The corresponding asymptotic solutions read
\eq{
a(t)=t+a_1 \ln t+\mathcal{O}(t^{-1}) \,, \qquad \phi=\phi_0 ~,}
with $a_1=\rho_0/6$ with $\rho_0$ arbitrary. They asymptote to $\Pk$ as $t\rightarrow \infty$, in accordance with the analytic solutions of appendix \ref{sec:analyticII}.

\paragraph{The point $\Pkp$:} The point $\Pkp$ corresponds to a Milne universe with angular defect,
$a(t)= a_0 t\,, \phi=\phi_0+\frac{2}{\lambda}\ln t$, with $a_0=\frac{\lambda}{\sqrt{\lambda^2-2}}$, $\phi_0=\frac{1}{\lambda}\ln\frac{\lambda^2V_0}{4}$ and $\rho_0=0$. There is one direction of approach, given by $\vec{v}_1=(-\sqrt{\frac23}, 0, \sqrt{\lambda^2-2})$, which exists for all $\lambda>\sqrt{2}$. The corresponding solution asymptotes to $\Pkp$ as $t\rightarrow \infty$,
\eq{
a(t)= a_0 t+a_l\ln t+\mathcal{O}(t^{-1})\,, \qquad \phi=\phi_0+\frac{2}{\lambda}\ln t+t^{-1}(\phi_1+\phi_{1l}\ln t)+\mathcal{O}(t^{-2})
~,}
where
\eq{
a_l=   \frac{(\lambda^2-2)}{2(3 \lambda^2-8)} \rho_0\,, \qquad
\phi_{1l} = -\phi_{1} = \frac{(\lambda^2-2)^{\frac32}}{\lambda^2(3 \lambda^2-8)} \rho_0
~,}
with $\rho_0$ unconstrained. For $\rho_0>0$ and $\lambda>\sqrt{\frac{8}{3}}$, these solutions are semi-eternally decelerating, since
\eq{
\ddot{a}= -\frac{a_l}{t^2}+ \mathcal{O}(t^{-3})
~.}
For $\sqrt{2}<\lambda<\sqrt{\frac83}$, $\Pkp$ is a stable node and there are two additional distinct directions of approach, $\vec{v}_2, \vec{v}_3$, which are tangent to the unit sphere: these trajectories thus lie entirely on the sphere $\mathcal{S}$ and correspond to the asymptotic solutions given in equations (3.5) and (3.6) of \cite{Andriot:2023wvg}.

The case $\lambda=\sqrt{\frac83}$ is degenerate, so that $\vec{v}_i=(-1,0,1)$, for $i=1,2,3$. The asymptotic solutions in this case are given in equations (3.9) and (3.10) of \cite{Andriot:2023wvg}. For $\lambda>\sqrt{\frac83}$, $\Pkp$ becomes a stable spiral for trajectories on the sphere $\mathcal{S}$. The corresponding oscillatory asymptotic solutions are given in equations (3.13) and (3.14) of \cite{Andriot:2023wvg}.

\paragraph{The point $\Pp$:} The point $\Pp$ corresponds to the scaling solution $a(t)= a_0 t^{\frac{2}{\lambda^2}}\,, \phi=\phi_0+\frac{2}{\lambda}\ln t$, with $k=0$, $\rho_0=0$, $\phi_0=\frac{1}{\lambda}\ln\frac{\lambda^4V_0}{2(6-\lambda^2)}$, and $a_0$ arbitrary. There are three directions of approach in the $(x,y,z)$ space, along $\vec{v}_1=(-\sqrt{6 - \lambda^2}, \lambda, 0)$, $\vec{v}_2=(1,0,0)$, or $\vec{v}_3=(0,0,1)$. Trajectories approaching $\Pp$ along $\vec{v}_1$, lie entirely in $\mathcal{C}_z$ and correspond to the analytic solutions of appendix \ref{sec:Analytic solutions III}; they have $k=0$, $\rho_0=0$. Trajectories along $\vec{v}_2$ lie entirely in $\mathcal{P}_z$ and they have $k=0$, $\rho_0=\frac{12 (\lambda^2-3) (\lambda^2-2)}{\lambda^4}a_1a_0^3$. Trajectories along $\vec{v}_3$ lie entirely in $\mathcal{S}$ and they have $k=-1$, $\rho_0=0$.

The corresponding asymptotic solutions read,
\eq{
a(t)= a_0 t^{\frac{2}{\lambda^2}} \left(1+a_1 t^p+\mathcal{O}(t^{2p})\right) \,, \qquad \phi=\phi_0+\frac{2}{\lambda}\ln t+\phi_1 t^p+\mathcal{O}(t^{2p})
~,}
with $p=\frac{(\lambda^2-6)}{\lambda^2}$, $\frac{2(\lambda^2-3)}{\lambda^2}$,
$\frac{2(\lambda^2-2)}{\lambda^2}$ and $\phi_1/a_1=\frac{3}{\lambda}$, $\frac{2(3-\lambda^2)}{\lambda}$, $\frac{3\lambda(2-\lambda^2)}{3\lambda^2-2}$, for $\vec{v}_1, \vec{v}_2, \vec{v}_3$, respectively, with $a_1$ arbitrary.   For the trajectories approaching  along $\vec{v}_{1}$, the exponent $p$ is always negative and $\Pp$ is reached at future infinity, $t\rightarrow\infty$. For the other two directions of approach along  $\vec{v}_2, \vec{v}_3$, if $p$ is positive $\Pp$ is reached in the past as $t\rightarrow 0$, or if $p$ is negative $\Pp$  is reached asymptotically in the future as $t\rightarrow \infty$. The case for which all three $p$-exponents are negative thus corresponds to the condition of stability of $\Pp$: $\lambda<\sqrt{2}$.

\paragraph{The point $\Pmp$:} The point $\Pmp$ corresponds to the scaling solution $a(t)= a_0 t^{\frac{2}{3}}\,, \phi=\phi_0+\frac{2}{\lambda}\ln t$, with $k=0$, $\rho_0=\frac{4a_0^3(\lambda^2-3)}{3\lambda^2}$, $\phi_0=\frac{1}{\lambda}\ln\frac{\lambda^2V_0}{2}$. If $\lambda>\sqrt{3}$ then $\Pmp$ lies in the interior of $\mathcal{S}$, i.e., it satisfies $x^2+y^2+z^2<1$. It is a node for $2 \sqrt{\frac67}>\lambda>\sqrt{3}$, in which case it has three directions of asymptotic approach: $\vec{v}_1=(0,0,1)$ and $\vec{v}_2, \vec{v}_3=( \lambda^2-6 \mp\sqrt{24 \lambda^2 - 7 \lambda^4},2( \lambda^2-3),0)$, corresponding to the three eigenvalues of table \ref{tab:fixedpointstabwithoutr}.
The directions of approach along $\vec{v}_2, \vec{v}_3$ correspond to trajectories lying entirely in $\mathcal{P}_z$. $\Pmp$ is repulsive along  $\vec{v}_1$,  while it is attractive along $\vec{v}_2, \vec{v}_3$. In other words, $\Pmp$ is stable for trajectories in $\mathcal{P}_z$ (corresponding to flat cosmologies, $k=0$), while it is repulsive otherwise (corresponding to open cosmologies, $k=-1$). For $\lambda>2 \sqrt{\frac67}$, $\Pmp$ becomes a stable spiral for trajectories in $\mathcal{P}_z$.

For trajectories approaching $\Pmp$ along the $\vec{v}_1$ direction, the solution reads
\eq{
a(t)= a_0 t^{\frac{2}{3}}\left(
1+a_1t^{\frac{2}{3}}+\mathcal{O}(t^{\frac{4}{3}})
\right)\,,\qquad \phi=\phi_0+\frac{2}{\lambda}\ln t + \phi_1t^{\frac{2}{3}}+\mathcal{O}(t^{\frac{4}{3}})
~,}
with $k=-1$, ${\phi_1}/{a_1}=-\frac{9}{7 \lambda}$, $a_1 = \frac{63 \lambda^2}{10 a_0^2 (14 \lambda^2-27 )}$, and reaches $\Pmp$ in the past, as $t\rightarrow 0$.

For $2 \sqrt{\frac67}>\lambda>\sqrt{3}$ and trajectories approaching $\Pmp$ along $\vec{v}_2,\vec{v}_3$, the solution reads
\eq{
a(t)= a_0 t^{\frac{2}{3}}\left(
1+a_1t^{p}+\mathcal{O}(t^{2p})
\right)
 ~;~~~\phi=\phi_0+\frac{2}{\lambda}\ln t + \phi_1 t^{p}+\mathcal{O}(t^{2p})
~,}
with $k=0$, $p=- \frac{ \lambda \pm \sqrt{24  - 7 \lambda^2} }{ 2 \lambda}$, ${\phi_1}/{a_1}=-   \lambda~\! p$ with $a_1$ arbitrary, and reaches $\Pmp$ in the future, as $t\rightarrow \infty$.

For  $\lambda>2 \sqrt{\frac67}$ and trajectories approaching $\Pmp$ in $\mathcal{P}_z$, so that $k=0$, the solution reads
\eq{\spl{
a(t)&= a_0 t^{\frac{2}{3}}\left(
1
+  t^{-\frac12}\left(
 a_{1c}\cos\left[q\ln t\right]+ a_{1s}\sin\left[q\ln t\right]
 \right)
 +\mathcal{O}(t^{-1})
\right)~,\\
 \phi&=\phi_0+\frac{2}{\lambda}\ln t +  t^{-\frac12}\left(
 \phi_{1c}\cos\left[q\ln t\right]+ \phi_{1s}\sin\left[q\ln t\right]
 \right)
 +\mathcal{O}(t^{-1})
~,}}
where
\eq{
q=\frac{  \sqrt{7 \lambda^2-24} }{ 2 \lambda}~;~~~
\phi_{1c}=
 \frac12  (a_{1c} \lambda - a_{1s} \sqrt{7 \lambda^2-24})~, ~~~
 \phi_{1s}=
 \frac12
  (a_{1s} \lambda + a_{1c} \sqrt{ 7 \lambda^2-24})
  ~,}
with $a_{1c,s}$ arbitrary,
and $\Pmp$ is reached in the future, as $t\rightarrow \infty$.

\paragraph{The point $\Pm$:}
The point $\Pm$ corresponds to the scaling solution $a(t)= a_0 t^{\frac{2}{3}}$, $\phi=\phi_0$, with $k=0$, $V_0=0$, $\phi_0$ arbitrary, and $\rho_0=\frac43 a_0^3$. $\Pm$ has one attractive direction of asymptotic approach: $\vec{v}_1=(1,0,0)$ and two repulsive ones: $\vec{v}_{2}=(0,1,0)$ and $\vec{v}_{3}=(0,0,1)$. The trajectories approaching along $\vec{v}_1$ lie entirely in $\mathcal{L}_x$ so that $k=0$, $V_0=0$. They correspond to the analytic solutions given in appendix \ref{sec:analyticI}. Asymptotically we have
\eq{
a(t)= a_0 t^{\frac{2}{3}}\left(
1+a_1t^{-1}+\mathcal{O}(t^{-2})
\right)
 \,,\qquad \phi=\phi_0+ \phi_1 t^{-1}+\mathcal{O}(t^{-2})
~,}
with $\rho_0=\frac43 a_0^3$, and $a_1$, $\phi_1$ arbitrary. $\Pm$ is reached in the future, as $t\rightarrow \infty$.

Trajectories approaching along $\vec{v}_2$ lie all in $\mathcal{P}_z$, so that $k=0$. Asymptotically we have
\eq{
a(t)= a_0 t^{\frac{2}{3}}\left( 1+a_1t^{2}+\mathcal{O}(t^{4}) \right) \,, \qquad \phi=\phi_0+ \phi_1 t^{2}+\mathcal{O}(t^{4})
~,}
with $\rho_0=\frac43 a_0^3$, $\phi_1/a_1=2\lambda$, $a_1=\frac{1}{12} e^{-\lambda \phi_0} V_0$. $\Pm$ is reached in the past, as $t\rightarrow 0$.

The trajectories approaching along $\vec{v}_3$ lie entirely in $\mathcal{L}_z$, so that $V_0=0$, $\phi=\text{const}$. They correspond to the analytic solutions given in appendix \ref{sec:analyticII}. Asymptotically we have
\eq{
a(t)= a_0 t^{\frac{2}{3}}\left( 1+a_1t^{\frac23}+\mathcal{O}(t^{\frac43}) \right) \,, \qquad \phi=\phi_0
~,}
with $k=-1$, $\rho_0=\frac43 a_0^3$, $a_1=\frac{9}{20} a_0^2$, and $\phi_0$ arbitrary. $\Pm$ is reached in the past, as $t\rightarrow 0$.

\end{appendix}

\newpage

\addcontentsline{toc}{section}{References}

\providecommand{\href}[2]{#2}\begingroup\raggedright\endgroup

\end{document}